\documentclass[fleqn,usenatbib]{mnras}

\usepackage{newtxtext,newtxmath}
\usepackage{nicefrac, xfrac}

\usepackage[T1]{fontenc}

\DeclareRobustCommand{\VAN}[3]{#2}
\let\VANthebibliography\thebibliography
\def\thebibliography{\DeclareRobustCommand{\VAN}[3]{##3}\VANthebibliography}

\usepackage{graphicx}	% Including figure files
\usepackage{amsmath}	% Advanced maths commands

\usepackage{pdflscape}

\usepackage{longtable}

\usepackage[table]{xcolor}
\title[Stochastic low-frequency variability]{Stochastic low-frequency variability of 50 massive stars in the Cygnus OB associations and the Small Magellanic Cloud}

\author[Pedersen and Bildsten]{
May G. Pedersen$^{1,2}$\thanks{E-mail: may.pedersen@sydney.edu.au}
and Lars Bildsten$^{2,3}$
\\
$^{1}$Sydney Institute for Astronomy, School of Physics, University of Sydney NSW 2006, Australia\\
$^{2}$Kavli Institute for Theoretical Physics, Kohn Hall, University of California, 
Santa Barbara, CA 93106, USA\\
$^{3}$Department of Physics, University of California, Santa Barbara, CA 93106, USA
}

\date{Accepted 2025 April 14. Received 2025 April 10; in original form 2024 December 31}

\pubyear{2025}

\begin{document}
\label{firstpage}
\pagerange{\pageref{firstpage}--\pageref{lastpage}}
\maketitle

% Abstract of the paper
\begin{abstract}
In recent years, high-precision high-cadence space photometry has revealed that stochastic low frequency (SLF) variability is common in the light curves of massive stars. We use the data from the \emph{Transiting Exoplanet Survey Satellite} (TESS) to study and characterize the SLF variability found in a sample of 49 O- and B-type main-sequence stars across six Cygnus OB~associations and one low-metallicity SMC star AV~232. We compare these results to 53 previously studied SLF variables. We adopt two different methods for characterizing the signal. In the first, we follow earlier work and fit a Lorentzian-like profile to the power density spectrum of the residual light curve to derive the amplitude $\alpha_0$, characteristic frequency $\nu_{\rm char}$, and slope $\gamma$ of the variability. In our second model-independent method, we calculate the root-mean-square (RMS) of the photometric variability as well as the frequency at 50\% of the accumulated power spectral density, $\nu_{50\%}$, and the width of the cumulative integrated power density, $w$. For the full sample of 103 SLF variables, we find that $\alpha_0$, $\gamma$, RMS, $\nu_{50\%}$, and $w$ correlate with the spectroscopic luminosity of the stars. Both $\alpha_0$ and RMS appear to increase for more evolved stars whereas $\nu_{\rm char}$ and $\nu_{50\%}$ both decrease. Finally, we compare our results to 2-D and 3-D simulations of subsurface convection, core-generated internal gravity waves, and surface stellar winds, and find good agreement between the observed $\nu_{\rm char}$ of our sample and predictions from sub-surface convection.
\end{abstract}

% Select between one and six entries from the list of approved keywords.
% Don't make up new ones.
%https://static.primary.prod.gcms.the-infra.com/static/site/mnras/document/MNRAS%20Keywords_November%202022.pdf?node=d966aa6b52fb4ac858b7

%%%%%%%%%%%%%%%%%%%%%%%%%%%%%%%%%%%%%%%%%%%%%%%%%%

%%%%%%%%%%%%%%%%% BODY OF PAPER %%%%%%%%%%%%%%%%%%

\section{Introduction}\label{Sect:Intro}

Stochastic low-frequency (SLF) variability is ubiquitous in massive stars, from main-sequence O- and B-type stars to the more evolved blue, yellow and red supergiants \citep[e.g.][]{Balona1992,Kiss2006,Blomme2011,Tkachenko2014,Aerts2015,Aerts2018,Ramiaramanantsoa2018,Pedersen2019,Dorn-Wallenstein2019,Bowman2019a,Bowman2019b,Bowman2020,Burssens2020,Bowman2022,Elliott2022,Ma2024,Shen2024,Zhang2024,Bowman2024,Kourniotis2025} as well as the more exotic Wolf-Rayet stars \citep[e.g.][]{Lamontagne1987,Lepine1999,Moffat2008,Chene2011,Ramiaramanantsoa2019,Naze2021,Lenoir-Craig2022}. In the frequency domain, such variability is seen as an increase in power towards low frequencies as characteristic of red noise, and has been speculated to be caused by surface granulation \citep[e.g.][]{Kiss2006}, sub-surface convection in the Fe peak opacity zone \citep[e.g.][]{Cantiello2021,Schultz2022,Schultz2023}, internal gravity waves excited by core convection \citep[e.g.][]{Aerts2015,Edelmann2019,Thompson2024}, or stellar winds \citep[e.g.][]{Krticka2018,Krticka2021}.

SLF variability has been characterised in different ways. In early studies of SLF variability detected in the space photometry of massive stars, the signal was generally characterised by fitting a Lorentzian-like profile to the amplitude or power density spectrum \citep[e.g.][]{Blomme2011,Tkachenko2014,Bowman2018,Bowman2019a,Bowman2019b,Bowman2020,Naze2021,Shen2024,Kourniotis2025}. Some more recent works have adopted Gaussian process regression to characterise the SLF variability in the time domain \citep{DornWallenstein2020,Bowman2022,Zhang2024,Bowman2024}. \citet{Ma2024} used a modified Lorentzian-like profile to fit the amplitude spectrum of a sample of stochastically variable blue supergiants in the Large Magellanic Cloud (LMC). Yet other studies suggest that a broken power law may be more appropriate to characterise the SLF variability depending on the origin of the signal \citep[e.g.][]{Krticka2021}. 

The goal of this work is two-fold. Firstly, using data from the Transiting Exoplanet Survey Satellite \citep[TESS,][]{TESS} we aim to identify and characterise the SLF variability of a sample of O- and B-type stars found in six different Cygnus OB associations containing 655 Galactic OB stars \citep{Quintana2021,Quintana2022}, and compare these to a smaller well-studied sample of 13 O-type stars in the lower metallicity Small Magellanic Cloud (SMC) \citep{Bouret2021}. Secondly, we test a new model-independent method for characterizing the SLF variability which aims to both simplify the SLF characterization process as well as circumvent having to make different choices in which model to use to characterize the variability. We discuss the sample selection in Sect.~\ref{sec:method} as well our new method for characterizing the SLF variability. For comparison, we also choose to fit a Lorentzian-like profile to the power density spectrum and leave a similar comparison to the Gaussian process regression method as future work. We repeat the analysis for a sample of 70 SLF variables previously studied by \citet{Bowman2020} in Sect.~\ref{sec:comparison_stars}. We present our results in Sect.~\ref{sec:results} and compare them to predictions for surface granulation, sub-surface convection, internal gravity waves excited by core convection, and stellar winds in Sect.~\ref{Sec:discussion}. Our final conclusions are provided in Sect.~\ref{Sec:conclusions}.

\section{Method}
\label{sec:method}

\subsection{Target selection}
\label{sec:target_selection}

\subsubsection{Cygnus OB association sample}

\begin{figure}%[ht!]
\begin{center}
\includegraphics[width=0.9\linewidth]{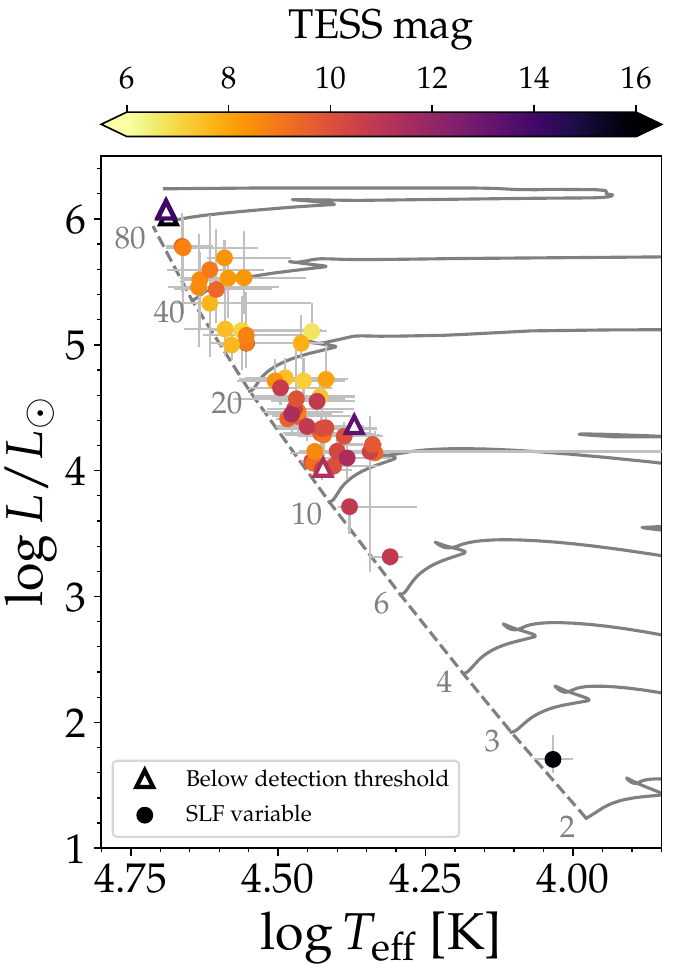}
\caption{HR diagram showing the position of the 49 SLF variables (circles) and the five stars with the SLF signal below the detection threshold (triangles) from the Cyg~OB sample. The symbol of the fifth star is overlapping with one of the other two close $\approx 80\,{\rm M}_\odot$ stars. The colour of the symbols indicate the TESS magnitude of the star. The zero-age-main-sequence is indicated by the dashed line, while non-rotating MIST evolutionary tracks \citep{Dotter2016,Choi2016} calculated with the stellar structure and evolution code \texttt{MESA} \citep{Paxton2011,Paxton2013,Paxton2015} are shown in grey full lines for eight different initial stellar masses in the range $2-80\ {\rm M}_\odot$ at solar metallicity.}\label{fig:SLF_sample}
\end{center}
\end{figure}

Starting with the initial sample of 655 O- and B-type stars in the six Cygnus OB associations identified by \cite{Quintana2021,Quintana2022} (group A-F), we include all stars with $\log L/L_\odot \geq 4$. Amongst these we exclude contaminated stars, as determined from a visual inspection of the TESS pixel data, and stars where the dominant variability is due to coherent mode pulsations ($\beta$~Cep and SPB stars), eclipsing binaries, or rotational variability. Finally, we add three additional stars with $\log L/L_\odot < 4$ where the dominant variability is clearly SLF. The resulting sample of 54 stars is shown in the HR diagram in Fig.~\ref{fig:SLF_sample}. We further reduce the sample to 49 O- and B-type stars after excluding stars where the signal-to-noise for the SLF variability is too low (see Sect.~\ref{sec:fitting}). The five stars below the detection threshold are shown as triangles in the figure. These are thus the only non-variable stars in the six Cygnus OB associations with $\log L/L_\odot \geq 4$, demonstrating that the SLF variability is a common feature among massive stars in line with previous observations (see Sect.~\ref{Sect:Intro}). For the rest of this paper we will refer to this sample as the Cyg~OB sample. 

\subsubsection{SMC sample}

To compare our results to findings in low-metallicity environments, we consider also an initial sample of 13 O-type stars in the SMC, whose UV and optical spectra have recently been analysed in detail by \cite{Bouret2021}. Stars observed by TESS in the SMC have a higher risk of contamination due to the higher crowding combined with the large $21'' \times 21''$ TESS pixel sizes. The \texttt{tess\_localize} \texttt{python} package \citep{Higgins2023} has recently been used successfully to localize the source of variability in TESS light curves of evolved massive stars in the Large and Small Magellanic Clouds \citep{Pedersen2023}. However, this tool takes as input a list of frequencies of coherent signals to determine the location where the amplitude of these signals is highest and therefore by design cannot be used to localize the source of SLF variability. Instead, we extract, background correct, and normalise the light curves of the nearby pixels surrounding the target following the method outlined in Sect.~\ref{sec:tess_data} and study these light curves in combination with their corresponding power density spectra (PDS) to determine if the detected variability originates from the target or a nearby star.

Based on this, we find one of the stars (AV~232) to be an SLF variable with an additional three (AV~43, AV~83 and AV~327) candidate SLF variables. Inspecting the pixel data of AV~83 shows that the star does appear to be an SLF variable, however, due to the high crowding the SLF signal is very likely diluted which would impact its characterization. AV~327 has an adjacent similar brightness star that is always in the same pixel as the target and it is therefore unclear if the signal is coming from AV~327 or its neighbour. AV~43 shows a small amount of SLF variability but is contaminated by a nearby Cepheid variable in some of its sectors. We therefore choose to label these three stars as SLF candidates and only characterize the signal of AV~232. Of the remaining 10 stars, two are found to be eclipsing binaries (AV~25 and AV~75), one is contaminated by a much brighter nearby star and another by a nearby Cepheid variable. The remaining two stars only contain white noise with (some) occasional contamination from nearby Cepheid variables.

\subsection{The TESS data}
\label{sec:tess_data}

The members of the Cygnus OB associations were observed by TESS in its observing cycle 2, 4, a few in cycle 5, and in cycle 6. Out of the 49 stars discussed in this work, 20 have 2-min cadence available, whereas the rest only have Full Frame Image (FFI) data available. Over these four TESS cycles, the cadence of the FFI data changed from 30-min, to 10-min and 200~sec, corresponding to a change in Nyquist frequency from $277\,\mu$Hz to $833\,\mu$Hz and $2499\,\mu$Hz. An overview of the sample and the available TESS data is given in Table~\ref{tab:sample_overview}. 

The SMC was observed by TESS in cycle 1, 3, and 5, with the FFI data of cycle 1 and 3 sharing the same observing cadence as cycle 2 and 4, respectively. AV~232 was observed in sector 1, 28, 67, and 68, and only has 2-min cadence data available for the last two sectors. An overview of the SMC O-type sample falling in the category of SLF variable, SLF candidate, and not variable (i.e. white noise only) is provided in Table~\ref{tab:sample_overview_smc}.

The light curves were extracted from the TESS FFI data and 2-min cadence target pixel files using the python packages \texttt{lightkurve} \citep{Lightkurve2018} and \texttt{tesscut} \citep{tesscut2019} using custom target pixel masks. To correct for background signals and systematics, a background pixel mask was created from which a time series of the background signals were extracted from each background pixel and a principal component analysis subsequently carried out to reduce the background signals in the time domain to their first one to seven principal components depending on the target. These principal components were used to remove the background signals from the target light curve using an adapted version of the \texttt{RegressionCorrector} functionality of \texttt{lightkurve}. The target light curves were then normalised by dividing by a low (typically zeroth or first) order polynomial fit to the first and second half of each sector (corresponding to two different TESS orbits) and the flux units changed to parts-per-million (ppm). Finally, the time stamps of the TESS FFI data were corrected following the methodology adapted in the TESS Asteroseismic Science Operations Centre (TASOC) pipeline developed by TESS Data for Asteroseismology (T'DA) group under the TESS Asteroseismic Science Consortium (TASC) \citep{Handberg2021}. For more information on the light curve extraction and preparation, please see Pedersen et al. (in prep).

\subsection{Iterative prewhitening}
\label{sec:prewhitening}

Before a study of the red noise, i.e. the stochastic low-frequency variability, can be carried out we first remove any remaining periodic coherent signals from the light curves, which are unlikely to be related to the dominant SLF variability. We use an iterative prewhitening scheme, where sinusoidal signals are removed one at a time until the signal-to-noise (S/N) ratio of an extracted signal goes below a predefined S/N limit. The optimal choice of this limit depends on the probability that the extracted signal may be a random noise peak instead of a genuine intrinsic signal of the extracted light curve. For one sector of TESS data, Pedersen et al. (in prep.) have shown that a S/N limit of 3.4, 4.0, and 4.6, where the noise is calculated using a $11.57\,\mu{\rm Hz}$ window around the extracted signal, corresponds to a probability of 50\%, 75\%, and 90\% that a signal of the given S/N value is not a random noise peak. In this work, we adopt $S/N = 4$ as the limit to be consistent with previous studies of SLF variability \citep{Bowman2018,Bowman2019a,Bowman2019b,Bowman2020}. To carry out the iterative prewhitening, we extract signals one by one from the observed light curve by calculating its corresponding Lomb-Scargle periodogram \citep{Lomb1976,Scargle1982} to identify the frequency $\nu_{\rm peak}$ and amplitude $A_{\rm peak}$ of the highest S/N peak at each step of the iteration. Using the identified $\nu_{\rm peak}$ and $A_{\rm peak}$ as initial guesses, we fit the equation 

\begin{equation}
    F(t) = C + \sum_i^N A_i \sin\left[2 \pi \left(t\nu_i + \phi_i \right)\right],
    \label{eq:lc_model}
\end{equation}

to the light curve using the non-linear least squares fitting algorithm provided by the \texttt{lmfit} python package \citep{lmfit}, allowing $\nu_i$, $A_i$, and $\phi_i$ of all the extracted $i=1,\dots, N$ signals to be optimized at each step of the iteration. Here $A_i$ is the amplitude of signal $i$, $\nu_i$ is the frequency, $\phi_i$ is the phase, and $C$ is a constant. The derived model in Eq.~\ref{eq:lc_model} is subsequently subtracted from the original light curve and a periodogram calculated for the resulting residual light curve to determine the next $\nu_{\rm peak}$ and $A_{\rm peak}$. This iterative process continues until the S/N of the next signal drops below four.

Following \cite{Papics2017}, we carry out the iterative prewhitening in the order of the highest to lowest significant signals rather than highest to lowest amplitude as for data with a significant amount of red noise the latter carries the risk of reaching the stopping criterion before all significant signals have been extracted, as lower-amplitude higher-S/N signals at higher frequencies may exist. By prewhitening according to peak significance, we effectively remove twice as many coherent signals compared to prewhitening according to the amplitudes.

\subsection{Fitting the power density spectrum}
\label{sec:fitting}

Following earlier studies of SLF variability in massive stars \citep{Blomme2011,Tkachenko2014,Bowman2018,Bowman2019a,Bowman2019b,Bowman2020}, we model the red noise in the power density spectrum of the residual light curve using the Lorentzian-like function

\begin{equation}
    \mathcal{M} (\nu) = \frac{\eta(\nu) \alpha_0 }{1 + \left( \frac{\nu}{\nu_{\rm char}}\right)^\gamma} + C_W,
    \label{eq:model_L}
\end{equation}

where $\nu$ is the frequency, $\alpha_0$ is the PDS at $0\,\mu{\rm Hz}$, $\gamma$ defines the slope of the variability, $\nu_{\rm char}$ is the characteristic frequency, and $C_w$ is the white noise contribution to the PDS. The spectrum below $\nu_0 = 1.157\,\mu{\rm Hz}$ is influenced by how the light curve is detrended, and therefore we do not model the spectrum for frequencies below this value. To convert the power to power-density we multiply the power by the effective length of the TESS sector, calculated as the inverse of the area under the spectral window \citep{Kjeldsen2005}.

\begin{figure}%[ht!]
\begin{center}
\includegraphics[width=\linewidth]{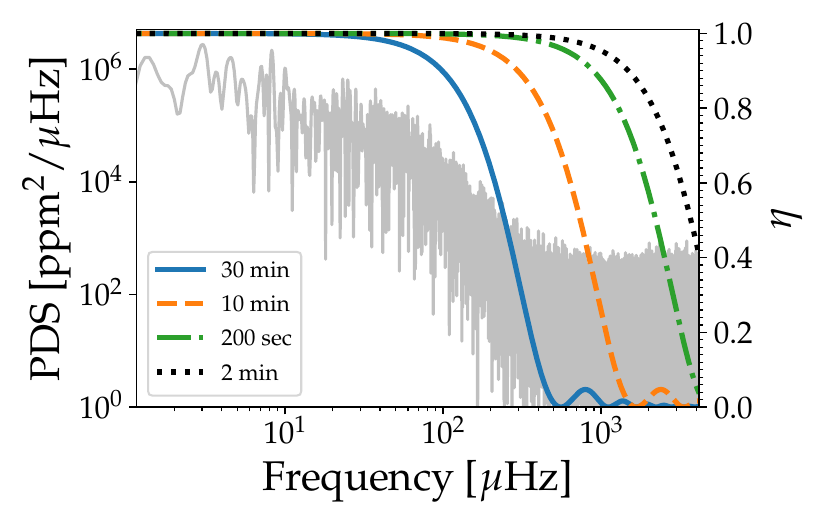}
\caption{Fractional power attenuation as a function of frequency for 30-min (full blue), 10-min (dashed orange), 200-sec (dot-dashed green), and 2-min (dotted black) cadence sampling. The power density spectrum of the 2-min cadence sector 41 light curve of Gaia~EDR3~2059070135632404992 is shown in grey for comparison.}\label{fig:attenuation}
\end{center}
\end{figure}

The first term of Eq.~\ref{eq:model_L} has been multiplied by the power attenuation \citep{Chaplin2011,Huber2022} 

\begin{equation}
    \eta \left(\nu \right) = {\rm sinc}^2 \left(\frac{\pi}{2}\frac{\nu}{\nu_{\rm Nyquist}} \right),
    \label{eq:P_att}
\end{equation}

to account for the time averaging of high frequency signals caused by long exposure times of the earlier TESS FFI data (see Fig.~\ref{fig:attenuation}). If the true power of a signal is $P_{\rm true}$, then the observed power is modulated by the power attenuation  and corresponds to $P_{\rm obs} = \eta P_{\rm true}$. As attenuation will have the highest impact on the results obtained from the 30-min cadence data, we choose to exclude data from TESS Cycle~1 and 2 for the remained of this paper. See Appendix~\ref{App:cadence_dependence} for further discussion of cadence dependence of the parameter estimates.

To fit Eq.~\ref{eq:model_L} to the PDS we derive posterior probability distributions and Bayesian evidence with the nested sampling Monte Carlo algorithm \texttt{MLFriends} \citep{Buchner2016,Buchner2019} using the \texttt{UltraNest} \texttt{python} package \citep{Buchner2021}. We adopt 400 live points to sample the parameter space assuming that $\alpha_0$, $\nu_{\rm char}$, $\gamma$, and $C_W$ all have uniform priors. At each iteration a model is calculated and evaluated using a log-likelihood function. Under the assumption of Gaussian noise in the time domain, which translates to a $\chi^2$ distribution in power with two degrees of freedom in the frequency domain, the corresponding log-likelihood function can be shown to be \citep{DuvallHarvey1986,Toutain1994} 

\begin{equation}
    \ln \mathcal{L} (\boldsymbol{\theta}) = - \sum_i \left[ \ln \mathcal{M}_i (\boldsymbol{\theta}) + \frac{S_i}{\mathcal{M}_i (\boldsymbol{\theta})} \right], 
    \label{eq:loglike}
\end{equation}

and has commonly been used when fitting Lorentzian functions \citep[or Harvey profiles;][]{Harvey1985} to the granulation background in solar-like stars as well as their oscillation frequencies \citep[e.g.][]{Davies2016,Lund2017,LiY2020,Nielsen2021}. We adopt the same log-likelihood function in Eq.~\ref{eq:loglike} to evaluate Eq.~\ref{eq:model_L} at each iteration.

\begin{figure*}
\begin{center}
\includegraphics[width=\linewidth]{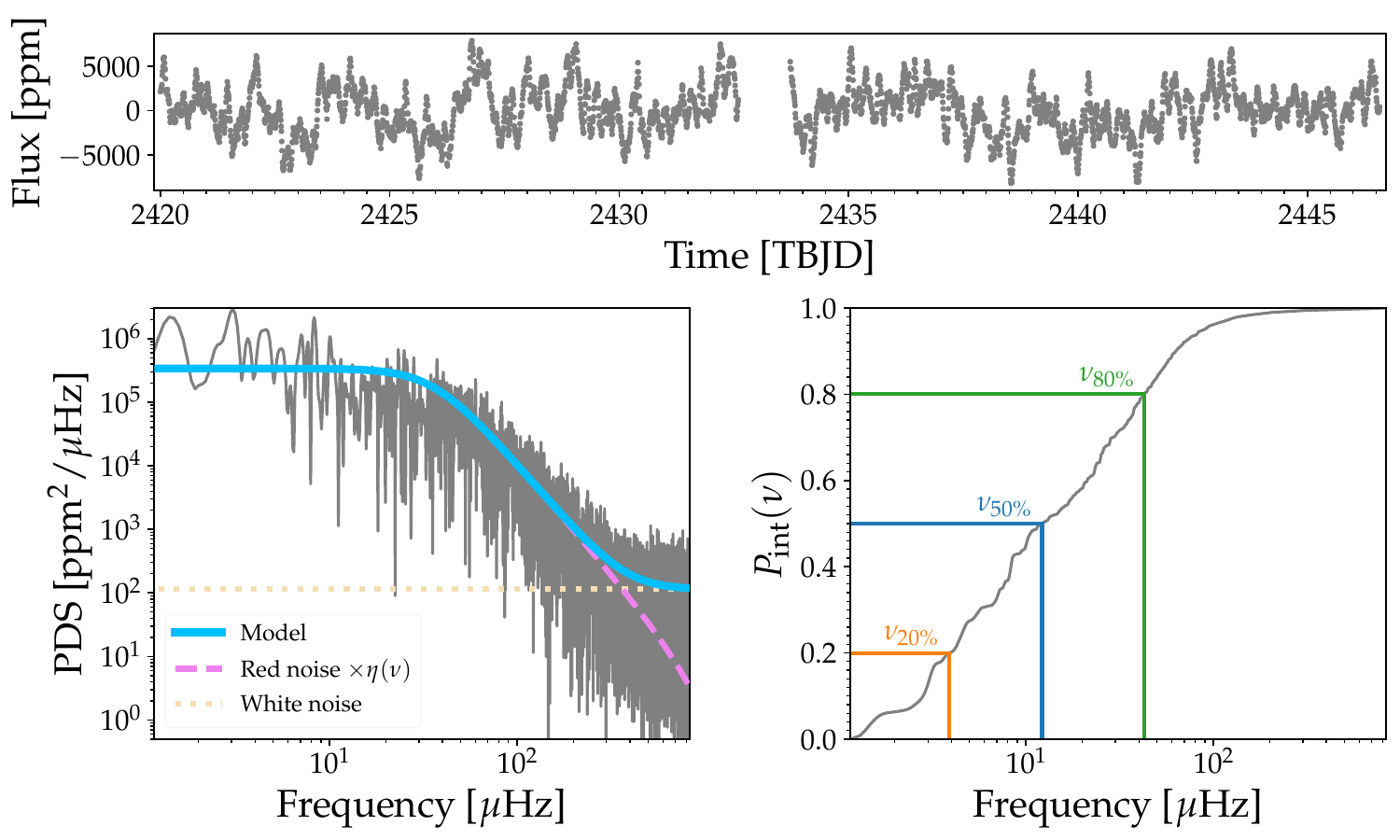}
\caption{\emph{Top:} Example residual 10-min cadence FFI light curve of Gaia~EDR3~2059070135632404992 from TESS sector 41. \emph{Bottom left:} Power density spectrum of the residual light curve (grey) and the corresponding best fitting Lorentzian-like model (blue). The red and white noise components are shown in pink and yellow, respectively. \emph{Bottom right:} Cumulative integrated power density spectrum (grey). An illustration of how the frequencies at 20\%, 50\%, and 80\% of the power are defined is shown in orange, blue, and green, respectively.}\label{fig:fit_example}
\end{center}
\end{figure*}

The final parameter estimates are taken as the 50th percentiles of the samples and the 16th and 84th percentiles their corresponding lower and upper values. For the SLF variability detection to be considered to be significant, we require that $\alpha_0/C_w > 10$ for more than half of the sectors that a given star was observed in. Five stars in our initial Cyg~OB sample of 54 O- and B-type stars failed this criterion, leaving us with the final sample of 49 stars reported in Table~\ref{tab:sample_overview}. An example residual FFI light curve (top panel) and its corresponding fitted PDS (bottom left) is shown in Fig.~\ref{fig:fit_example}.

\subsection{RMS and integrated power}
\label{sec:rms_int_power}

\begin{figure*}
\begin{center}
\includegraphics[width=\linewidth]{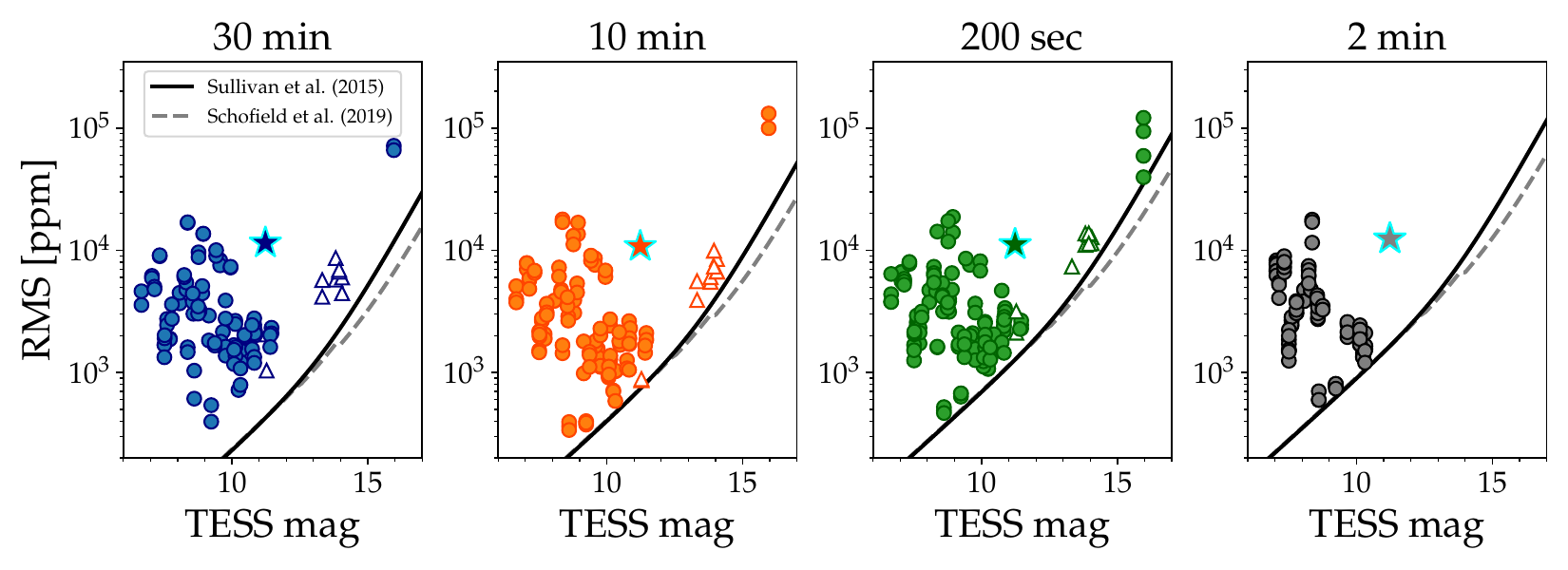}
\caption{Measured RMS of the residual light curves as a function of the TESS magnitude. The RMS is shown independently for each star and each individual TESS sector with available data as filled circles and triangles for the Cyg~OB sample. The star symbol represents the SMC star AV~232. The subplot corresponds to a different observing cadence, with the first three coming from the TESS FFI data for cycle 1+2, 3+4, and 5+6. The last subplot is the RMS for the available 2-min cadence light curves spread across all observing cycles. Open triangles are the measurements belonging to the five Cyg~OB stars which were excluded based on their derived $\alpha_0/C_w$ parameters. The black full line shows the corresponding TESS noise model from \citet{Sullivan2015}, while the grey dashed line is the updated TESS noise model from \citet{Schofield2019} where the number of pixels in the target mask is adjusted based on the TESS magnitude.}\label{fig:rms_vs_mag}
\end{center}
\end{figure*}

As an alternative to using the Lorentzian-like model in Eq.~(\ref{eq:model_L}) to characterise the SLF variability, we also calculate the root-mean-squared (RMS) variability of the residual light curve as a separate measure of the amplitude of the SLF variability, as well as the frequencies at which 20\%, 50\%, and 80\% of the power is located. This approach has the benefit that it does not rely on any model fits, but instead the quantities are calculated directly from the data. The calculated RMS is shown against the TESS magnitude of the 49 Cyg~OB stars in Fig.~\ref{fig:rms_vs_mag} (circles), with the TESS noise models from \citet{Sullivan2015} and \citet{Schofield2019} shown for comparison. The results for the five excluded stars from this sample are shown as open triangles, while the SMC star AV~232 is indicated by a star.

We calculate the cumulative integrated power of the power density spectrum $S (\nu)$ as

\begin{equation}
    P_{\rm int} (\nu) = \frac{\int_{\nu_0}^{\nu} S (\nu) {\rm d}\nu}{\int_{\nu_0}^{\nu_{\rm norm}} S (\nu)  {\rm d}\nu},
\end{equation}

where $\nu \in [\nu_0, \nu_{\rm norm}]$. Once again we use $\nu_0 = 1.157\,\mu{\rm Hz}$ as the lower frequency limit. The upper frequency $\nu_{\rm norm}$ defines the frequency at which the cumulatively integrated power density spectrum is normalized to one. Just as in Sect.~\ref{sec:fitting} we exclude data from TESS Cycle 1+2 from this analysis and choose to use the Nyquist frequency of the 10-min cadence data as $\nu_{\rm norm}$.

We use the integrated power to calculate the frequencies below which 20\%, 50\%, and 80\% of the power resides ($\nu_{20\%}$, $\nu_{50\%}$, $\nu_{80\%}$), corresponding to $P_{\rm int} (\nu_{20\%}) = 0.2$, $P_{\rm int} (\nu_{50\%}) = 0.5$, and $P_{\rm int} (\nu_{80\%}) = 0.8$. Additionally, we also calculate the "width"

\begin{equation}
    w = \frac{\nu_{80\%} - \nu_{20\%}}{\nu_{50\%}},
\end{equation}

of the spectrum, which we expect to depend on $\gamma$ and $\nu_{\rm char}$. The bottom right panel of Fig.~\ref{fig:fit_example} provides an example of the normalized cumulative integrated power density for the 10-min cadence FFI data of the star Gaia~EDR3~2059070135632404992 from sector 41. The vertical lines show the value of $\nu_{20\%}$ (orange), $\nu_{50\%}$ (blue), and $\nu_{80\%}$ (green) for this light curve.

As previously mentioned, we choose the same $\nu_{\rm norm}$ in the derivation of $\nu_{20\%}$, $\nu_{50\%}$, and $\nu_{80\%}$ for all sectors and observing cadences. This is done for the sake of consistency, allowing for direct comparisons between the variation of these parameters from sector to sector due to the intrinsic variability of the star. This enforces the use of the Nyquist frequency of the longest cadence FFI data as the highest meaningful value of $\nu_{\rm norm}$. To demonstrate how the $\nu_{20\%}$, $\nu_{50\%}$, $\nu_{80\%}$, and $w$ parameters would change if a different choice of $\nu_{\rm norm}$ was made, we used the Lorentzian-like model in Eq.~(\ref{eq:model_L}) to simulate the SLF variability in PDS for a 2-min cadence data set. The white noise was fixed to $\log C_W = 0.2$, while $\log \alpha_0 \in [1,10]$, $\nu_{\rm char} \in [1,100]$, and $\gamma \in [1,5]$ were varied within the given ranges. The four parameters $\nu_{20\%}$, $\nu_{50\%}$, $\nu_{80\%}$, and $w$ were then calculated for four different values of $\nu_{\rm norm}$, corresponding to the Nyquist frequency of a 30-min, 10-min, 200-sec, and 2-min cadence light curve, respectively. 

We find that the differences between the estimated parameters for the four different cadences are minimal provided that the amplitude of the SLF variability ($\log \alpha_0$) is sufficiently high with little dependence on the observing cadence. The differences are less than 1\,$\mu$Hz for $\nu_{20\%}$, $\nu_{50\%}$, and $\nu_{80\%}$ if $\alpha_0$ is larger than $10^2$, $10^3$, and $10^4$\,ppm$^2\mu$Hz$^{-1}$, respectively, and less than 0.1 for $w$ if $ \alpha_0 > 10^4$\,ppm$^2\mu$Hz$^{-1}$. In comparison, when considering the dependence of the four parameters $\nu_{20\%}$, $\nu_{50\%}$, $\nu_{80\%}$, and $w$ on $\gamma$ and $\nu_{\rm char}$ a much stronger cadence dependence is found, especially for $\nu_{20\%}$, $\nu_{50\%}$, and $\nu_{80\%}$ with the differences being largest for $\nu_{80\%}$. These differences also show a clear correlation between $\gamma$ and $\nu_{\rm char}$ and become larger when both $\gamma$ and $\nu_{\rm char}$ increase in value. The difference between the estimated $\nu_{20\%}$, $\nu_{50\%}$, and $\nu_{80\%}$ values of the 30-min and 2-min cadence data are larger than 5\,$\mu$Hz (20\,$\mu$Hz) for $\gamma$ values larger than 2.5 (1.5), 3 (2.2), and 3.5 (2.8), respectively, for all values of $\nu_{\rm char}$. When comparing the results for the 10-min and 2-min cadence simulated data, the differences between the estimated $\nu_{20\%}$, $\nu_{50\%}$, and $\nu_{80\%}$ values are less than 5\,$\mu$Hz for $\gamma \gtrsim 1.5$, 2, and 2.5, respectively.

Based on the results of these comparisons, we choose to exclude the FFI data from TESS cycle 1 and 2 from our analysis including both the calculations of $\nu_{20\%}$, $\nu_{50\%}$, $\nu_{80\%}$, and $w$ as well as the Lorentzian-like model fit to the PDS described in Sect.~\ref{sec:fitting} and adopt $\nu_{\rm norm} = 833\,\mu$Hz, corresponding to the Nyquist frequency of the 10-min cadence data.

\section{Comparison stars}
\label{sec:comparison_stars}

To compare our results to prior studies of SLF variability we consider the sample of 70 O- and B-type stars studied by \citet{Bowman2020} who also used a Lorentzian-like model to characterise the SLF variability using TESS 2-min cadence data. One important difference is that \citet{Bowman2020} did the fitting in the amplitude spectrum instead of in PDS, meaning that we cannot draw a direct comparison between their resulting parameter estimates and the ones presented here. 

To circumvent this issue, we redo the fitting for the \citet{Bowman2020} sample in the PDS and also derive RMS, $\nu_{50\%}$, and $w$ for the sample. To ensure that we are comparing similar data to those used by \citet{Bowman2020}, we consider only the 2-min cadence TESS Pre-search Data Conditioning Simple Aperture Photometry (PDCSAP) light curves for this sample. These light curves were downloaded from the Mikulski Archive for Space Telescopes (MAST) using \texttt{lightkurve}. We normalised the light curves by dividing by a first order polynomial fit to each sector and change the flux units to ppm. To otherwise keep the analysis method similar to the approach adopted in this work, we estimate the $\alpha_0$, $\nu_{\rm char}$, $\gamma$, $C_W$, RMS, $\nu_{50\%}$, and $w$ parameters separately for each sector. Finally, we exclude stars with a low S/N detection of the SLF variability as well as stars where the dominant variability is due to binarity, rotation, or coherent pulsations. This reduces the sample of comparison stars to 53 O- and B-type stars, see Table~\ref{tab:bowman_sample_overview} in Appendix~\ref{App:bowman_sample} for an overview. For the remainder of this paper, we will refer to this sample as the B20 sample.

\section{Results}
\label{sec:results}

In Table~\ref{tab:results_averages} in Appendix~\ref{App:sample} we provide the average values across all sectors and observing cadences of the estimated parameters $\alpha_0$, $\nu_{\rm char}$, $\gamma$, $C_W$, RMS, $\nu_{50\%}$, and $w$ for the Cyg~OB sample and the SMC star AV~232. The corresponding parameter estimates for the B20 sample are listed in Table~\ref{tab:results_averages_bowman} in Appendix~\ref{App:bowman_sample}. Following previous studies of SLF variability, we now investigate the inter-parameter dependencies as well as the dependence on stellar evolution stage by placing the stars in an HR diagram. Because the B20 sample only has spectroscopic luminosities ($\mathcal{L} = T_{\rm eff}^4 / g$) available, we first convert our bolometric luminosities $L$ to spectroscopic luminosities using 

\begin{equation}
    \frac{\mathcal{L}}{\mathcal{L}_\odot} = \frac{T_{\rm eff}^4}{g} \left(\frac{T_{{\rm eff},\odot}^4}{g_\odot}\right)^{-1} = \frac{L}{L_\odot} \left(\frac{M}{M_\odot}\right)^{-1},
\end{equation}

derived using $L = 4\pi \sigma R^2 T_{\rm eff}^4$ and $g = GM/R^2$. The derived spectroscopic luminosities are likewise listed in Table~\ref{tab:results_averages}. One low-luminosity star in our Cyg~OB sample is a clear outlier from the rest of our sample, and so we choose to exclude it from the rest of this discussion but nevertheless list its SLF parameter estimates in Table~\ref{tab:results_averages}.

\begin{figure*}%[ht!]
\begin{center}
\includegraphics[width=\linewidth]{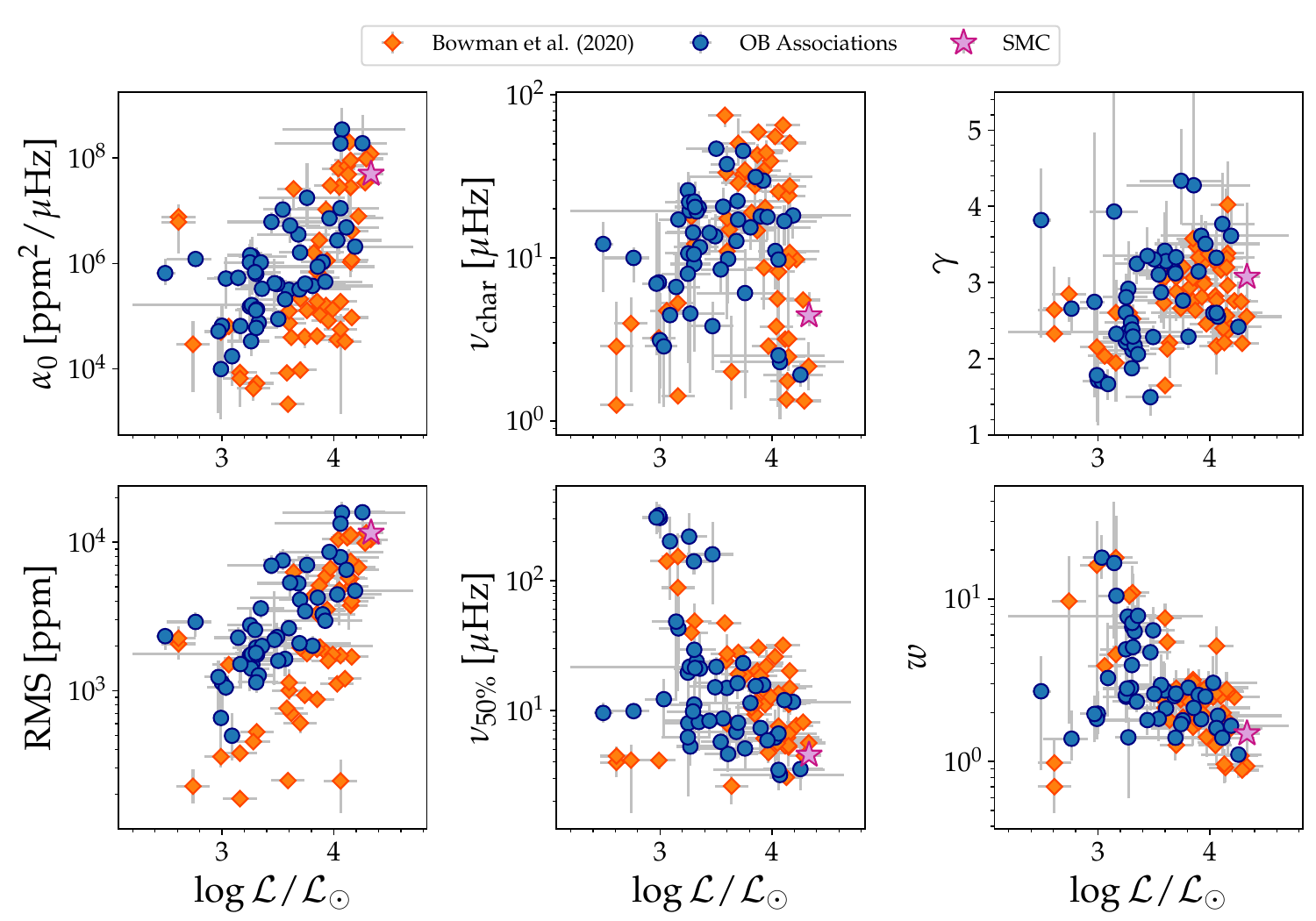}
\caption{Measured parameters from our two methods of determining the characteristics of the SLF variability shown against the spectroscopic luminosities of the stars. The symbols denote the average parameter estimated calculated for a given star across all sectors and observing cadences while the grey bars indicate the range in observed parameters for a given star over all sectors. The Cyg~OB sample is shown as blue circles, the SMC star AV~232 is indicated by a pink star, whereas the orange diamonds represent the B20 sample.}\label{fig:para_vs_lum}
\end{center}
\end{figure*}

Figure~\ref{fig:para_vs_lum} shows the six estimated parameters as a function of $\mathcal{L}$ for the Cyg~OB sample (blue circles), SMC star AV~232 (pink star), and the B20 sample (orange diamonds). The grey bars on $\alpha_0$, $\nu_{\rm char}$, $\gamma$, $C_W$, RMS, $\nu_{50\%}$, and $w$ represent the full retrieved range in the estimated parameter across all sectors and observing cadences for a given star, while the symbols correspond to the average value. Large bars are therefore indicative of large variations in these parameters over the different sectors, which means that the SLF variability changes characteristics over time. 

The trends in the individual panels of Fig.~\ref{fig:para_vs_lum} indicate that there is a correlation between most of these parameters and the spectroscopic luminosity, independently of if all three stellar samples are considered simultaneously or independently. To quantify this, we calculate the Spearman's rank correlation coefficients $r_s$ \citep{Spearman1904}, which are applicable and can be calculated for two variables that share a monotonic relationship. The coefficient takes values between $-1$ and $1$, where $r_s=1$ and $r_s=-1$ corresponds to a perfect positive and negative correlation between the two variables, respectively, and $r_s=0$ means that the parameters are uncorrelated. How close $\left| r_s \right|$ is to one determines the strength of the correlation: $\left| r_s \right| \in \ ]0, 0.19]$ is a very weak correlation, $\left| r_s \right| \in [0.2, 0.39]$ is weak, $\left| r_s \right| \in [0.4, 0.59]$ is moderate, $\left| r_s \right| \in [0.6, 0.79]$ is strong, and $\left| r_s \right| \in [0.8, 1.0]$ is a very strong correlation. Finally, each calculated $r_s$ is associated with a $p$-value which determines the significance of the correlation. For $p > 0.05$ the correlation is considered insignificant, while $0.001 < p \leq 0.05$ corresponds to significant, and $p < 0.001$ is highly significant.

\begin{table}
	\centering
	\caption{The Table elements list the Spearman's rank correlation coefficients $r_s$ between the parameters listed in the first column and the rest of the column header parameters. The colours indicate the $p$-value for a given correlation as explained at the bottom of the table. Greyed out table elements are either duplicates or cases where the correlation is calculated between the same two parameters, e.g., $\alpha_0$ versus $\alpha_0$. The first table segment provides the $r_s$ values when all stars in the Cyg~OB, B20 and AV~232 sample are studied simultaneously. The second segment are the corresponding correlations obtained for the Cyg~OB sample only, while the last table segment are the correlations for the B20 sample.}
	\label{tab:corellations}
	\begin{tabular}{lcccccc} % four columns, alignment for each
    \hline
		  & $\alpha_0$ & $\nu_{\rm char}$  & $\gamma$  & RMS & $\nu_{50\%}$ & $w$\\[0.5ex]
    \hline
		 		 \multicolumn{7}{c}{\textit{Full sample}}\\
\hline
		 $\log \mathcal{L}/\mathcal{L}_\odot$ 	&	\cellcolor{cyan!55}0.524 	&	\cellcolor{red!25}0.061 	&	\cellcolor{cyan!55}0.322 	&	\cellcolor{cyan!55}0.609 	&	\cellcolor{cyan!55}-0.381 	&	\cellcolor{cyan!55}-0.497\\[0.5ex]
		 $\log T_{\rm eff}$ 	&	\cellcolor{red!25}0.001 	&	\cellcolor{cyan!55}0.504 	&	\cellcolor{cyan!55}0.460 	&	\cellcolor{red!25}0.174 	&	\cellcolor{red!25}0.153 	&	\cellcolor{red!25}-0.161\\[0.5ex]
	 $\alpha_0$ 	&	\cellcolor{gray!50} 	&	\cellcolor{cyan!55}-0.509 	&	\cellcolor{red!25}0.133 	&	\cellcolor{cyan!55}0.948 	&	\cellcolor{cyan!55}-0.774 	&	\cellcolor{cyan!55}-0.673\\[0.5ex]
	 $\nu_{\rm char}$ 	&	\cellcolor{gray!50} 	&	\cellcolor{gray!50} 	&	\cellcolor{cyan!55}0.572 	&	\cellcolor{cyan!55}-0.338 	&	\cellcolor{cyan!55}0.516 	&	\cellcolor{cyan!25}0.230\\[0.5ex]
	 $\gamma$ 	&	\cellcolor{gray!50} 	&	\cellcolor{gray!50} 	&	\cellcolor{gray!50} 	&	\cellcolor{cyan!25}0.267 	&	\cellcolor{red!25}-0.031 	&	\cellcolor{cyan!25}-0.304\\[0.5ex]
	 RMS 	&	\cellcolor{gray!50} 	&	\cellcolor{gray!50} 	&	\cellcolor{gray!50} 	&	\cellcolor{gray!50} 	&	\cellcolor{cyan!55}-0.639 	&	\cellcolor{cyan!55}-0.667\\[0.5ex]
	 $\nu_{\rm 50\%}$ 	&	\cellcolor{gray!50} 	&	\cellcolor{gray!50} 	&	\cellcolor{gray!50} 	&	\cellcolor{gray!50} 	&	\cellcolor{gray!50} 	&	\cellcolor{cyan!55}0.471\\[0.5ex]
	 $w$ 	&	\cellcolor{gray!50} 	&	\cellcolor{gray!50} 	&	\cellcolor{gray!50} 	&	\cellcolor{gray!50} 	&	\cellcolor{gray!50} 	&	\cellcolor{gray!50}\\[0.5ex]
		\hline
		 \multicolumn{7}{c}{\textit{Cyg~OB}}\\
\hline
		 $\log \mathcal{L}/\mathcal{L}_\odot$ 	&	\cellcolor{cyan!55}0.607 	&	\cellcolor{red!25}0.166 	&	\cellcolor{cyan!25}0.428 	&	\cellcolor{cyan!55}0.751 	&	\cellcolor{cyan!55}-0.533 	&	\cellcolor{cyan!25}-0.415\\[0.5ex]
		 $\log T_{\rm eff}$ 	&	\cellcolor{cyan!25}0.351 	&	\cellcolor{red!25}0.125 	&	\cellcolor{cyan!25}0.289 	&	\cellcolor{cyan!55}0.550 	&	\cellcolor{red!25}-0.257 	&	\cellcolor{cyan!55}-0.468\\[0.5ex]
	 $\alpha_0$ 	&	\cellcolor{gray!50} 	&	\cellcolor{cyan!25}-0.339 	&	\cellcolor{cyan!25}0.449 	&	\cellcolor{cyan!55}0.891 	&	\cellcolor{cyan!55}-0.894 	&	\cellcolor{cyan!55}-0.506\\[0.5ex]
	 $\nu_{\rm char}$ 	&	\cellcolor{gray!50} 	&	\cellcolor{gray!50} 	&	\cellcolor{cyan!25}0.422 	&	\cellcolor{red!25}-0.078 	&	\cellcolor{red!25}0.262 	&	\cellcolor{red!25}0.098\\[0.5ex]
	 $\gamma$ 	&	\cellcolor{gray!50} 	&	\cellcolor{gray!50} 	&	\cellcolor{gray!50} 	&	\cellcolor{cyan!55}0.577 	&	\cellcolor{cyan!25}-0.336 	&	\cellcolor{cyan!55}-0.478\\[0.5ex]
	 RMS 	&	\cellcolor{gray!50} 	&	\cellcolor{gray!50} 	&	\cellcolor{gray!50} 	&	\cellcolor{gray!50} 	&	\cellcolor{cyan!55}-0.733 	&	\cellcolor{cyan!55}-0.538\\[0.5ex]
	 $\nu_{\rm 50\%}$ 	&	\cellcolor{gray!50} 	&	\cellcolor{gray!50} 	&	\cellcolor{gray!50} 	&	\cellcolor{gray!50} 	&	\cellcolor{gray!50} 	&	\cellcolor{cyan!25}0.398\\[0.5ex]
	 $w$ 	&	\cellcolor{gray!50} 	&	\cellcolor{gray!50} 	&	\cellcolor{gray!50} 	&	\cellcolor{gray!50} 	&	\cellcolor{gray!50} 	&	\cellcolor{gray!50}\\[0.5ex]
		\hline
		 \multicolumn{7}{c}{\textit{B20}}\\
\hline
		 $\log \mathcal{L}/\mathcal{L}_\odot$ 	&	\cellcolor{cyan!55}0.623 	&	\cellcolor{red!25}-0.048 	&	\cellcolor{red!25}0.212 	&	\cellcolor{cyan!55}0.672 	&	\cellcolor{red!25}-0.268 	&	\cellcolor{cyan!55}-0.464\\[0.5ex]
		 $\log T_{\rm eff}$ 	&	\cellcolor{cyan!25}-0.331 	&	\cellcolor{cyan!55}0.835 	&	\cellcolor{cyan!55}0.616 	&	\cellcolor{red!25}-0.171 	&	\cellcolor{cyan!55}0.636 	&	\cellcolor{red!25}0.225\\[0.5ex]
	 $\alpha_0$ 	&	\cellcolor{gray!50} 	&	\cellcolor{cyan!55}-0.553 	&	\cellcolor{red!25}-0.096 	&	\cellcolor{cyan!55}0.961 	&	\cellcolor{cyan!55}-0.714 	&	\cellcolor{cyan!55}-0.824\\[0.5ex]
	 $\nu_{\rm char}$ 	&	\cellcolor{gray!50} 	&	\cellcolor{gray!50} 	&	\cellcolor{cyan!55}0.749 	&	\cellcolor{cyan!25}-0.415 	&	\cellcolor{cyan!55}0.676 	&	\cellcolor{cyan!25}0.350\\[0.5ex]
	 $\gamma$ 	&	\cellcolor{gray!50} 	&	\cellcolor{gray!50} 	&	\cellcolor{gray!50} 	&	\cellcolor{red!25}0.017 	&	\cellcolor{cyan!25}0.322 	&	\cellcolor{red!25}-0.092\\[0.5ex]
	 RMS 	&	\cellcolor{gray!50} 	&	\cellcolor{gray!50} 	&	\cellcolor{gray!50} 	&	\cellcolor{gray!50} 	&	\cellcolor{cyan!55}-0.569 	&	\cellcolor{cyan!55}-0.807\\[0.5ex]
	 $\nu_{\rm 50\%}$ 	&	\cellcolor{gray!50} 	&	\cellcolor{gray!50} 	&	\cellcolor{gray!50} 	&	\cellcolor{gray!50} 	&	\cellcolor{gray!50} 	&	\cellcolor{cyan!55}0.512\\[0.5ex]
	 $w$ 	&	\cellcolor{gray!50} 	&	\cellcolor{gray!50} 	&	\cellcolor{gray!50} 	&	\cellcolor{gray!50} 	&	\cellcolor{gray!50} 	&	\cellcolor{gray!50}\\[0.5ex]
		\hline
\end{tabular}
\begin{tabular}{ccc}
\\[-1.5ex]
		 \multicolumn{3}{c}{\textit{Colour explanation}}\\
		 \cellcolor{red!25}$p > 0.05$ 	&	 \cellcolor{cyan!25}$0.001 < p \leq 0.05$ 	&	 \cellcolor{cyan!55}$p < 0.001$\\ 
 \end{tabular}
\end{table}

The first row of Table~\ref{tab:corellations} lists the Spearman's ranks correlation coefficients between $\log \mathcal{L}/\mathcal{L}_\odot$ and the six estimated parameters used to characterise the SLF variability for the full sample of stars considered in this paper (Cyg~OB + AV~232 + B20). The colour indicates the $p$-value for the considered correlation. As seen in the Table, significant weak to strong positive correlations are found for $\alpha_0$, $\gamma$, and RMS, while significant weak to moderate negative correlations are found for $\nu_{50\%}$ and $w$. For $\nu_{\rm char}$, the correlation is insignificant and we cannot reject the null-hypothesis that $r_s = 0$. Similar generally stronger correlations are found if the Cyg~OB sample is considered on its own, whereas the B20 sample also shows no correlation between $\log \mathcal{L}/\mathcal{L}_\odot$ and the two parameters $\gamma$ and $\nu_{50\%}$.

\begin{figure*}%[ht!]
\begin{center}
\includegraphics[width=0.95\linewidth]{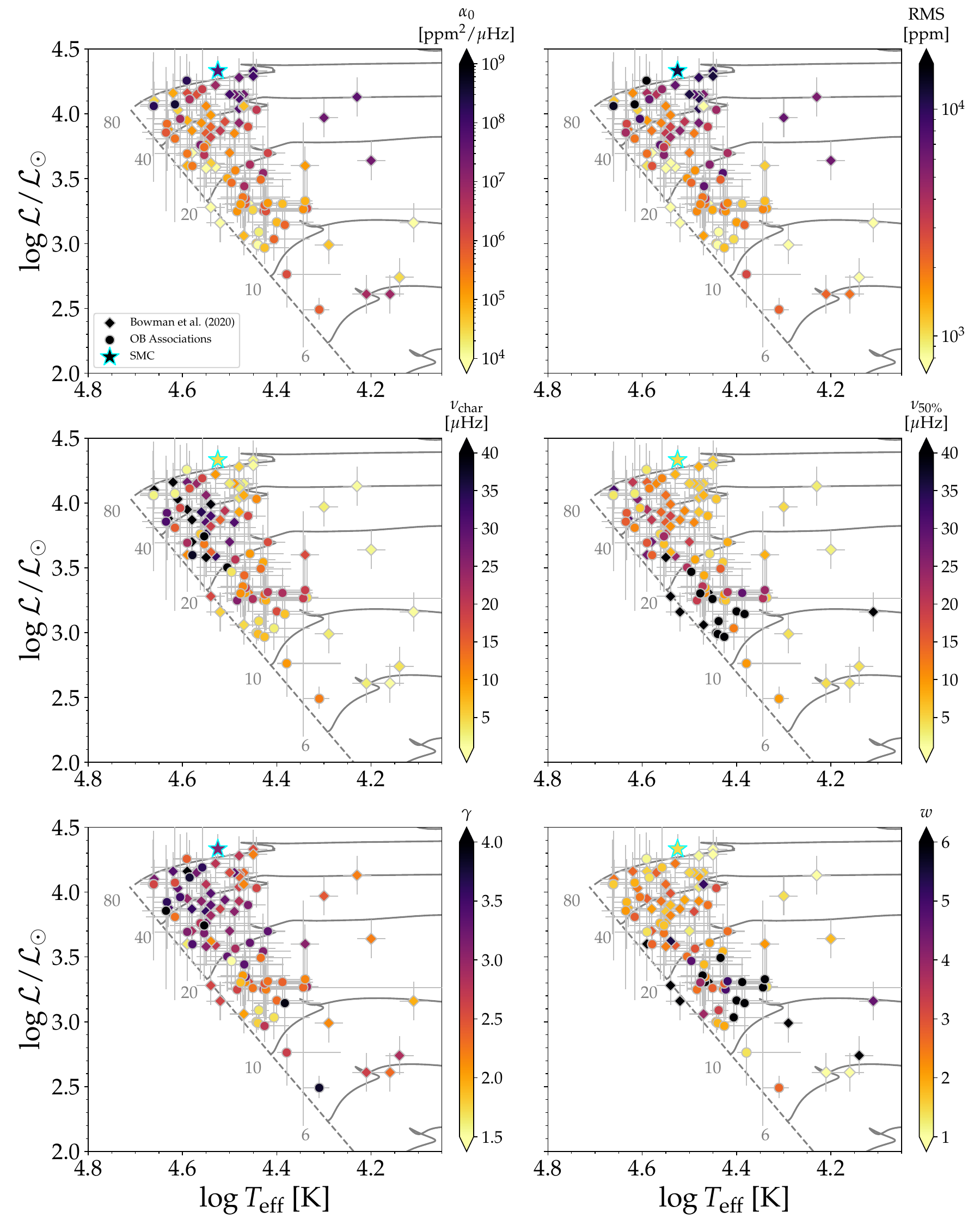}
\caption{Spectroscopic HR diagram showing the full sample of Cyg~OB (circles), AV~232 (star), and B20 (diamonds) stars. The colour of the symbols indicate the value of the average estimates of the six parameters $\alpha_0$, $\nu_{\rm char}$, $\gamma$, RMS, $\nu_{50\%}$ and $w$ as indicated by the colour bars.}\label{fig:para_hrd}
\end{center}
\end{figure*}

The full sample of stars including the Cyg~OB, AV~232, and B20 samples are shown in the spectroscopic HR diagrams in Fig.~\ref{fig:para_hrd}, where the colours of the symbols in each subplot correspond to the derived average parameter estimate for each of the six parameters $\alpha_0$, $\nu_{\rm char}$, $\gamma$, $C_W$, RMS, $\nu_{50\%}$, and $w$ as indicated by the colour bars. The shape of the symbols have the same meaning as in Fig.~\ref{fig:para_vs_lum}. Based on this figure we can see that stars in the same part of the HR diagram generally have, a) high RMS values when $\alpha_0$ is high, b) high $\nu_{50\%}$ values when $\nu_{\rm char}$ is low, and c) high $w$ values when $\gamma$ is low, and vice versa. The trends with evolutionary stage are less clear especially when the full sample is considered, but are indicative of the $\alpha_0$ and RMS both increase when the stars get older, while $\nu_{\rm char}$ and $\nu_{50\%}$ both decrease. No clear correlation with age is seen for either $\gamma$ or $w$. Comparing our results in Fig.~\ref{fig:para_hrd} to the study of the trends in macroturbulence velocities of $\approx 600$ O- and B-type stars from \cite{Serebriakova2024} shows excellent agreement, where stars with high $\alpha_0$ and RMS are situation in the same parts of the HR diagram as stars with high macroturbulence velocities. This is consistent with earlier findings of relations between photometric SLF variability and spectroscopic macroturbulence by \cite{Bowman2020}.

The correlation coefficients between $\log T_{\rm eff}$ and each of the six estimated SLF parameters are shown in the second rows of each of the table segments in Table~\ref{tab:corellations}. When the full sample is considered, we find that neither $\alpha_0$, RMS, $\nu_{50\%}$, or $w$ are significantly correlated with $\log T_{\rm eff}$, but if we consider the Cyg~OB and B20 samples separately a different pattern emerges for these four parameters. When a parameter is significantly positively correlated with $\log T_{\rm eff}$ for Cyg~OB it is negatively correlated with $\log T_{\rm eff}$ for B20, and vise versa. Because of this, when the two samples are combined no correlation with $\log T_{\rm eff}$ is found. For $\nu_{\rm char}$ a very strong correlation with $\log T_{\rm eff}$ is found for B20 while it is insignificant for Cyg~OB, resulting in an overall significant moderate correlation for the full sample. The opposite is seen for the RMS parameter. The $\gamma$ parameter is the only one that is simultaneously found to be both significant and positively correlated with $\log T_{\rm eff}$ for both the Cyg~OB and B20 sample, but the correlation is stronger for the B20 sample. For the Cyg~OB sample, we find a positive correlation between $\log T_{\rm eff}$ and both $\alpha_0$ and RMS, which is consistent with recent findings for a sample of blue supergiants of lower effective temperatures than our sample \citep{Kourniotis2025}.

\begin{figure*}%[ht!]
\begin{center}
\includegraphics[width=\linewidth]{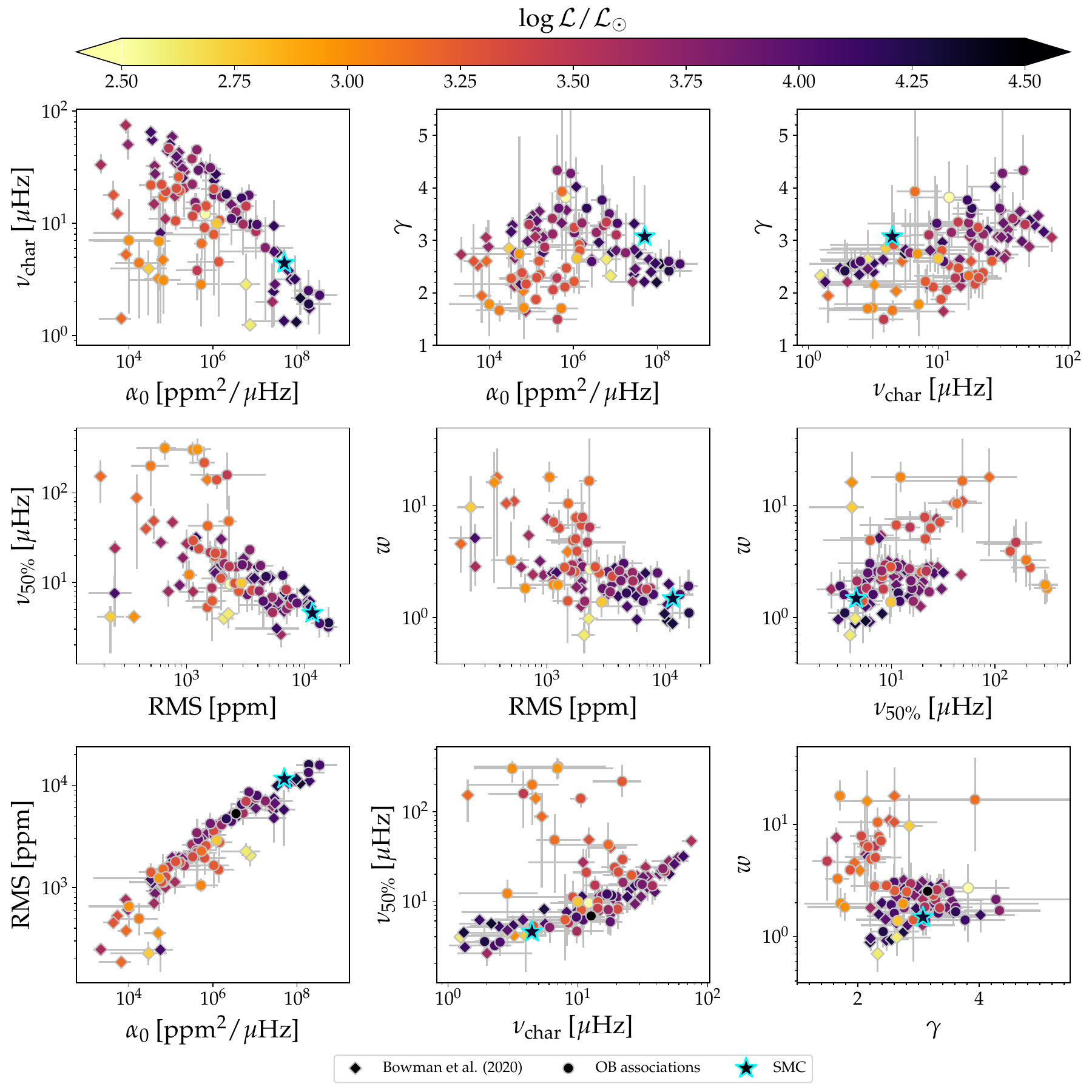}
\caption{Inter-parameter correlations between the six parameters $\alpha_0$, $\nu_{\rm char}$, $\gamma$, RMS, $\nu_{50\%}$ and $w$ used to characterize the SLF variability. The symbols indicate the average parameter estimates and the grey bars the observed range in the estimated parameters. The shapes of the symbols have the same meaning as in Fig.~\ref{fig:para_hrd} while their colours are determined by their spectroscopic luminosities.}\label{fig:para_corr}
\end{center}
\end{figure*}

Figure~\ref{fig:para_corr} plots the six estimated parameters that we use to characterise the SLF variability. The colour of the symbols indicate the spectroscopic luminosity. In combination with the correlation coefficients in Table~\ref{tab:corellations}, we find that $\alpha_0$ is correlated with all other parameters except $\gamma$ for the B20 sample. The $\nu_{\rm char}$ parameter is correlated with $\alpha_0$ and $\gamma$ for Cyg~OB, whereas it is also correlated with $\log T_{\rm eff}$, $\alpha_0$, RMS, $\nu_{50\%}$ and $w$ for the B20 sample. The $\gamma$ parameter is correlated with all other parameters for the Cyg~OB sample, but not with $\log \mathcal{L}$, $\alpha_0$, RMS or $w$ for B20. The RMS is correlated with all other parameters except for $\nu_{\rm char}$ for Cyg~OB, and correlates with $\log \mathcal{L}$, $\alpha_0$, $\nu_{\rm char}$, $\nu_{50\%}$ and $w$ for the B20 sample. The $\nu_{50\%}$ parameter is correlated with all parameters except for $\log T_{\rm eff}$ and $\nu_{\rm char}$ for Cyg~OB and $\log \mathcal{L}$ for B20. Finally, the width $w$ is correlated with all other parameters except for $\nu_{\rm char}$ for the Cyg~OB sample and for $\log T_{\rm eff}$ and $\gamma$ for the B20 sample. When all stars are combined, the only insignificant correlations are found for $\log T_{\rm eff}$ versus $\alpha_0$, $\log T_{\rm eff}$ versus RMS, $\log \mathcal{L}$ versus $\nu_{\rm char}$, $\alpha_0$ versus $\gamma$, $\log T_{\rm eff}$ versus $\nu_{50\%}$, $\log T_{\rm eff}$ versus $w$, and $\gamma$ versus $\nu_{50\%}$, i.e. for seven parameter combinations out of 21. Six of the significant correlations are found to be both strong and highly significant.

For the lower metallicity SMC star AV~232 we find that its SLF parameters are generally at the edges of the parameter estimates for both the Cyg~OB and B20 sample, but are similar to the values and correlations found for stars with similar spectroscopic luminosities.

\section{Discussion}\label{Sec:discussion}

\subsection{Surface granulation}

For solar-like oscillators, the background signal arising from a combination of surface granulation noise and stellar activity is usually modelled and removed by fitting a sum of power laws (i.e. Harvey models) with two to five components to the PDS and including a constant for the white noise component, resulting in

\begin{equation}
    {\rm PDS}_{\rm Bkg} (\nu) = C_W + \sum_i^N \frac{a_i}{1 + \left( \frac{\nu}{b_i}\right)^{c_i}},
    \label{eq:Harvey}
\end{equation}

for the background PDS (${\rm PDS}_{\rm Bkg}$), which simplifies to Eq.~(\ref{eq:model_L}) when only one power law component is included in the background model. Unlike in this work and previous studies of SLF variability in massive stars, the $c_i$ parameter is held fixed when fitting Eq.~(\ref{eq:Harvey}) to the background noise of solar-like oscillators. When initially introducing this model, \cite{Harvey1985} adopted $c_i = 2$ for the Sun, while later it was shown that $c_i = 4$ appears to be a more appropriate value \citep[e.g.][]{Aigrain2004,Michel2009}. As seen in Fig.~\ref{fig:para_vs_lum}, most of our estimated $\gamma$ fall between these two $c_i$ values.

\begin{figure}%[ht!]
\begin{center}
\includegraphics[width=\linewidth]{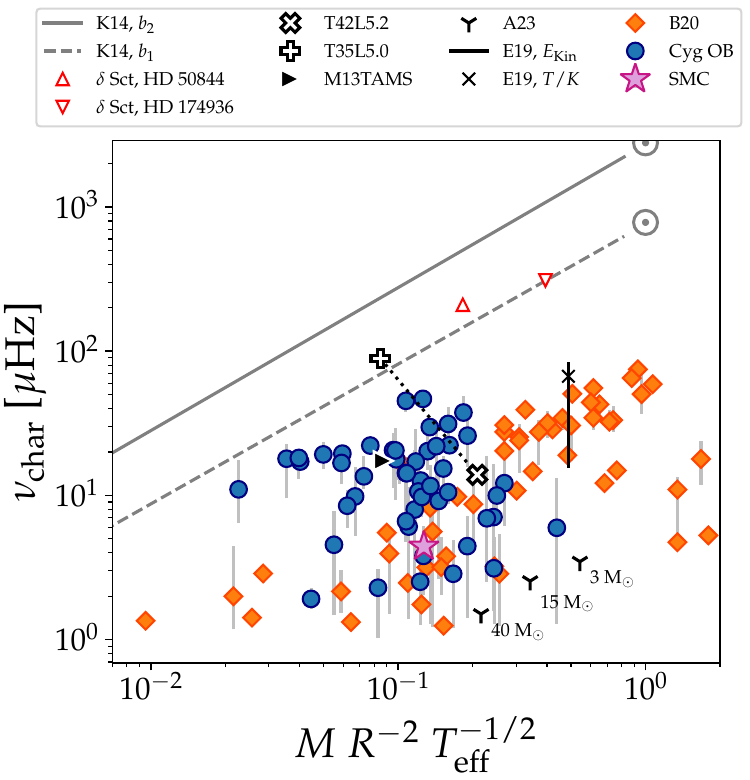}
\caption{Comparison between the characteristic frequency (y-axis) of observed red noise as a function of the expected granulation frequency $\nu_{\rm gran} \propto M R^{-2} T_{\rm eff}^{-1/2}$ in solar units (x-axis) and estimates from hydrodynamical simulations. The full and dashed grey lines show the observationally derived relations for two characteristic granulation frequencies ($b_2$ and $b_1$) for low-mass stars \citep{Kallinger2014}, compared to results for two $\delta$~Sct stars \citep[red triangles;][]{Kallinger2010}. Predictions from hydrodynamical simulations of sub-surface convection are indicated by the open black cross and plus symbol \citep{Schultz2022} and the sideways black triangle \citep{Schultz2023}. Corresponding predictions for internal gravity waves excited by core convection are indicated by the Y symbols \citep{Anders2023} and the cross and vertical black line \citep{Edelmann2019}. Please see text for more details.}\label{fig:result_overview_gran}
\end{center}
\end{figure}

In a study of the connection between granulation noise and oscillation signal in solar-like, subgiant and red giant stars observed by the \emph{Kepler} space telescope, \cite{Kallinger2014} considered eight different models for the granulation background including a two-component Harvey model of the form of Eq.~(\ref{eq:Harvey}). Their derived correlation between the two characteristic granulation frequencies $b_1$ and $b_2$ and the frequency at maximum power $\nu_{\rm max}$ of the solar-like oscillations are shown as dashed and full grey lines in Fig.~\ref{fig:result_overview_gran}, respectively, and compared to our estimates for the Cyg~OB, SMC star and B20 sample. Here we have adopted 

\begin{equation}
    \nu_{\rm gran} \propto \nu_{\rm max} \propto M R^{-2} T_{\rm eff}^{-1/2}, 
    \label{eq:numax}
\end{equation}

following \cite{Kallinger2010} based on \cite{KjeldsenBedding1995} and \cite{Huber2009}. The derived $M R^{-2} T_{\rm eff}^{-1/2}$ values are given with respect to solar values such that $M R^{-2} T_{\rm eff}^{-1/2} = 1$ for the Sun. If a single component is used to fit the background, as done for the stars in this work, the granulation frequency $\nu_{\rm gran}$ should fall between the two grey lines. 

As seen in Fig.~\ref{fig:result_overview_gran}, an offset is seen between the OB stars and the derived $\nu_{\rm char}$ versus $M R^{-2} T_{\rm eff}^{-1/2}$ relations found by \cite{Kallinger2014} for the lower-mass solar-like oscillators, with the massive stars showing consistently lower $\nu_{\rm char}$ values. This is in line with previous findings by \cite{Bowman2019a}, who argued that the observed SLF variability in their sample of (near-) main-sequence OBA stars is unlikely to be caused by surface granulation for this reason. If granulation is the cause, then the properties of the resulting red noise cannot be directly extrapolated from the relations found for cooler stars \citep[see also similar discussion by][]{Blomme2011}. For comparison, we also show the characteristic frequencies derived by \cite{Kallinger2010} for two $\delta$~Sct stars (red triangles), which are oscillating main-sequence A-type stars expected to have very shallow convective envelopes. Contrary to the OB stars, their $\nu_{\rm char}$ values do align with the relation found for the cooler stars, suggesting that their observed red noise could be caused by surface granulation.

For the more evolved massive red supergiants (RSGs), the situation may be different. These stars are known to have large surface convective cells with convective turnover timescales on the order of hundreds of days \citep[e.g.][]{Goldberg2022}, and have also been shown to exhibit SLF variability \citep{Kiss2006,Zhang2024}. Using ground-based photometric light curves of 48 RSGs from the database of the American Association of Variable Star Observers (AAVSO) with a mean time span of 61\,yr, \cite{Kiss2006} showed that the variability followed a $1/\nu$ frequency dependence consistent with expectations for solar granulation background, but could not derive a $\nu_{\rm char}$ due to the much sparser sampling of the data compared to current photometric space missions. On the other hand, \cite{Zhang2024} find granulation time scales on the same order as the predictions from \cite{Goldberg2022} by modelling the SLF variability seen in ground-based OGLE and ASAS-SN light curves of more than 6000 RSGs in the LMC and SMC using Gaussian processes.

\subsection{Sub-surface convection}\label{Sec:subsurface}

\cite{Stothers1993} were some of the first to show that the opacity bumps near the stellar surface of hot massive stars caused by the partial ionization of iron and helium \citep{Iglesias1992} give rise to dynamical instabilities, also known as sub-surface convection zones. \cite{Cantiello2009} showed that while the He opacity bump in the majority of cases does not form a significant sub-surface convection zone, convection is always developed around the iron opacity bump for $L \gtrsim 10^{3.2}\,{\rm L}_\odot$ provided that the metallicity of the star is sufficiently high. In general, this sub-surface convection zone from the iron opacity bump (FeCZ) increases in importance and becomes more prominent for lower surface temperatures and gravities and higher luminosities and metallicities. Later studies by \cite{Jermyn2022} demonstrated that the subsurface convection zones are present over a narrower range in stellar properties than previously estimated and is expected to be absent below $\approx 16\,{\rm M}_\odot$ and $\approx 35\,{\rm M}_\odot$ for stars in the LMC and SMC, respectively, while the FeCZ should be present for all masses above $\approx 8\,{\rm M}_\odot$ for stars of solar metallicity. \cite{Bowman2024} compared the predictions by \cite{Jermyn2022} to the observed positions of SLF variables in the HD diagram, finding SMC stars located within regions predicted to be stable against sub-surface convection in the FeCZ at low metallicities.

The estimated surface velocities arising from gravity waves excited by the FeCZ appear to correlate with the occurrence of significant microturbulence in O- and B-type stars, indicating a possible physical connection between the two \citep{Cantiello2009}. Additionally, it has been shown that stochastic low-frequency photometric variability and macroturbulence could be caused by these sub-surface convection zones \citep{Cantiello2021}. 
With the exception of three stars, the SLF variables in our Cyg~OB sample all fall in the regime where substantial sub-surface FeCZs are expected to be found. This is largely due to our sample selection, in which we focused on stars with $L/L_\odot \geq 4$ and deliberately avoided stars where the photometric variability is dominated by rotational variability, pulsations, or binarity. Of the stars above this threshold, 96\% show SLF variability. The SMC star AV~232, however, is found right at the boundary \citep[$M_{\rm spec} = 35.3 \pm 8.2\,{\rm M}_\odot$,][]{Bouret2021} where prominent sub-surface FeCZ are expected to be developed according to \cite{Jermyn2022} and shows similar SLF amplitudes as the Galactic OB~Cyg sample at similar spectroscopic luminositites. 

The theoretical considerations of sub-surface convection zones mentioned above all relied on the use of 1-D stellar models. In recent years, detailed 3-D numerical simulations of these sub-surface convection zones have been carried out, partially in an attempt to explain the SLF variability. \cite{Schultz2022} performed 3-D radiation hydrodynamical simulations of the outer $\approx 15\%$ of the stellar envelopes of two $35\,{\rm M}_\odot$ stars at two different evolutionary stages, one on the zero-age main-sequence (model T42L5.2) and the other half way through the main-sequence evolution (model T35L5.0). They find that the amplitudes of the tangential and radial velocities from the simulations are comparable to the predictions from 1-D models, however, the velocity profiles from the 3-D simulations are much broader and extend all the way to the surface. While the simulated convection from the Fe opacity bump carries minimal flux, the turbulent motions reach the surface and there is therefore no convectively quiet zone in the outer $\approx 7\%$ of the simulated stars. Through an analysis of the synthetic light curves produced by the two models, \cite{Schultz2022} find similar slopes as comparison SLF variables from \cite{Bowman2020} found around the same position in the HRD, while their derived $\nu_{\rm char}$ are consistent with the thermal time scale in the FeCZ. The same authors later added a $13\,{\rm M}_\odot$ terminal-age main-sequence 3-D envelope model  simulation to their sample of simulated stars \citep[M13TAMS,][]{Schultz2023} and likewise determine the $\nu_{\rm char}$ of the model. All of these simulations were run at solar metallicity. 

\begin{figure}%[ht!]
\begin{center}
\includegraphics[width=\linewidth]{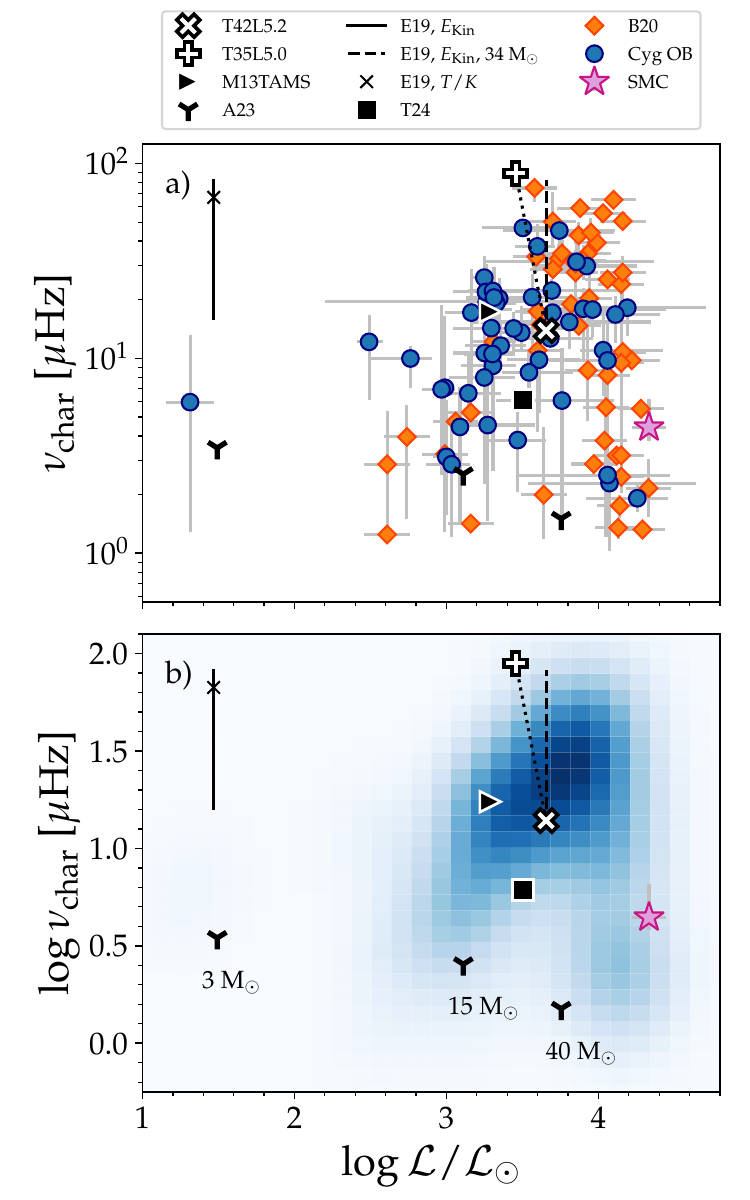}
\caption{Comparison between observed characteristic frequency versus spectroscopic luminosity relations (panel a) for the Cyg~OB, SMC, and B20 samples and theoretical predictions from numerical simulations of sub-surface convection and internal gravity waves excited by core convection. Panel b) shows a density plot of the combined Cyg~OB and B20 samples in panel a). Symbols have the same meaning as in Fig.~\ref{fig:result_overview_gran}. The black square show the additional predictions by \citet{Thompson2024} from IGWs excited by core convection. See text for more details.}\label{fig:nuchar_vs_logL}
\end{center}
\end{figure}

The results for all three simulated stars are included in Fig.~\ref{fig:result_overview_gran}. A dotted line connects the two $35\,{\rm M}_\odot$ models from \cite{Schultz2022} (open black cross and plus symbols) and are indicative of the direction the stars should be moving over time if the SLF variability is caused by sub-surface convection. As seen in the figure, this trend aligns with the parameter range of the majority of the stars in the Cyg~OB sample with the more evolved T35L5.0 model even crossing the observed trend for surface granulations in solar-like oscillators (dashed grey). Similarly, the results for the simulated $13\,{\rm M}_\odot$ star matches the observations of a part of the Cyg~OB sample. 

The same general trends are seen in Fig.~\ref{fig:nuchar_vs_logL} where $\nu_{\rm char}$ is shown as a function of $\log \mathcal{L}/\mathcal{L}_\odot$. As seen in the figure, the observed scatter in $\nu_{\rm char}$ at similar $\log \mathcal{L}/\mathcal{L}_\odot$ values could be the result of a change in characteristics of sub-surface convection zone during the main-sequence evolution. To confirm this, more simulations at similar masses but different evolutionary stages would be required. Additionally, the weakness of the sub-surface convection theory lies in the disappearance of such prominent zones at lower metallicities as predicted from 1-D models \citep{Cantiello2009,Jermyn2022} while SLF variability has previously been identified in stars located in the LMC \citep[e.g.][]{Pedersen2019,Bowman2019b}. With our $M_{\rm spec} = 35.3 \pm 8.2\,{\rm M}_\odot$ SMC star being located right at the boundary where substantial FeCZ should be able to develop at SMC metallicity, having a 3-D hydrodynamical simulation of such a star would be highly valuable, however, to our knowledge no such simulations currently exist. For an extensive review of simulations of massive star envelopes, we refer the reader to \cite{Jiang2023}. 

\subsection{Internal gravity waves excited by core convection}

At interfaces between convective and radiative zones, internal gravity waves (IGWs) are stochastically exited by processes that causes disturbances at the convective boundary such as the overshooting/penetration of convective plumes \citep{Townsend1966,Montalban2000} and Reynolds stresses of turbulent convective eddies \citep{Lighthill1952,Lecoanet2013}%Goldreich1990,Kumar1999}
. Such excited IGWs propagate through the radiative zones and away from the convective regions either towards higher (low-mass stars) or lower (high-mass stars) density regions. As discussed in Sect.~\ref{Sec:subsurface}, such waves are expected to be generated both above and below sub-surface convection zones. For stars with convective cores, IGWs excited by core convection have been demonstrated to be capable of efficient angular momentum transport \citep{Rogers2013}, efficient mixing of chemical elements \citep{Rogers2017,Varghese2023} and potentially cause enhanced mass loss in massive stars towards the end of their evolution \citep{Quataert2012}. For an extensive review on simulations of convective cores please see \cite{Lecoanet2023}.

While the earliest simulations of convective cores focused on the study of convective boundary mixing \citep[e.g.][]{Deupree_2000,Browning2004}, increased attention has been put on the study of the resulting IGW spectrum and its possible connection to SLF variability. \cite{Aerts2015} took the IGW spectrum from the 2-D hydrodynamical simulations of the inner 0.5-98\% in radius of a $3\,{\rm M}_\odot$ star \citep{Rogers2013} and rescaled both the frequency and amplitude of the spectrum to match it to the photometric observations of three young O-type stars showing SLF variability, finding good general agreement between the two except at the very low frequency end. A similar approach was taken by \cite{Ramiaramanantsoa2018} to study the red noise component of the O-type star V973~Scorpii. They showed that doing the rescaling of the simulated IGW spectrum assuming a $45\,{\rm M}_\odot$ terminal-age main-sequence star resulted in a better match to the observations than for a similar evolutionary stage $34\,{\rm M}_\odot$ star.

\cite{Edelmann2019} carried out the first 3-D hydrodynamical simulations covering 1\%-90\% in radius of a rotating $3\,{\rm M}_\odot$ zero-age main-sequence star allowing them to study the properties of the IGWs generated by the convective core throughout the majority of the radiative envelope. They find the generated spectrum to be well represented by a broken power law, where the frequency position of the break point depends on the angular degree $\ell$ of the waves. In Fig.~\ref{fig:result_overview_gran} and \ref{fig:nuchar_vs_logL} we show the range in frequency of these break points found for the Fourier transform of the kinetic energy spectrum of $\ell \in [1,20]$ as a vertical full black line. The overlapping cross is the break point frequency estimated for the combined Fourier spectrum of the corresponding temperature fluctuations over all $\ell$ values. As seen in Fig.~\ref{fig:result_overview_gran}, the range in break-point frequencies matches well with the range in $\nu_{\rm char}$ for the B20 sample at the given $M R^{-2} T_{\rm eff}^{-1/2}$ value. In Fig.~\ref{fig:nuchar_vs_logL}, the results from the 3-D simulations do not overlap with the observations as the stars studied in this work are of much higher masses and therefore higher $\log \mathcal{L}/\mathcal{L}_\odot$. This is different for the 3-D hydrodynamical simulations of a $25\,{\rm M}_\odot$ main-sequence star carried out by \cite{Herwig2023} whose simulations cover the inner 54\% of the star. Based on these simulations, \cite{Thompson2024} derived the $\nu_{\rm char}$ resulting from fitting Eq.~(\ref{eq:model_L}) to the resulting IGW spectrum. Their result is indicated in Fig.~\ref{fig:nuchar_vs_logL} by the black square and is situated at the lower end in the parameter ranges for the observations at similar $\log \mathcal{L}/\mathcal{L}_\odot$. 

The simulations by \cite{Rogers2013}, \cite{Herwig2023} and \cite{Edelmann2019} all required a boosting of the convective core to compensate for the high thermal diffusivity in the simulations. In the case of the 2-D simulations by \cite{Rogers2013}, \cite{Aerts2015} argued that such boosting could mean that the generated IGW spectrum might be more representative of a $\approx 30\,{\rm M}_\odot$ star. Assuming a similar argument can be made for the 3-D simulations by \cite{Edelmann2019}, this would move the full vertical black line in Fig.~\ref{fig:nuchar_vs_logL} to the position of the black dashed line for a $34\,{\rm M}_\odot$ star\footnote{The $\log \mathcal{L}/\mathcal{L}_\odot$ value was taken from a solar metallicity zero-age main-sequence $34\,{\rm M}_\odot$ MIST model.}. In this case the range in break-point frequency approximately spans the observed range in $\nu_{\rm char}$ at the given $\log \mathcal{L}/\mathcal{L}_\odot$, potentially implying that the scatter is caused by a variation in which $\ell$ values are more dominant. 

The required boosting of the convective core is one of the criticisms of the IGW theory as the source of the observed SLF variability as it does not allow for a prediction of the amplitude of the waves. Recently, \cite{Anders2023} carried out 3-D simulations of three zero-age main-sequence stars at $3\,{\rm M}_\odot$, $15\,{\rm M}_\odot$, and $40\,{\rm M}_\odot$ covering the inner 93\% of the star in radius and without boosting the convective core in the simulations. They find the amplitudes of the generated IGW spectrum to be a few orders of magnitudes lower than the observed red noise in O- and B-type stars. The derived $\nu_{\rm char}$ are likewise lower than the observations that they compare their results to \citep{Bowman2020} and decrease with mass. Here we have likewise included the results from \cite{Anders2023} in Fig.~\ref{fig:result_overview_gran} and \ref{fig:nuchar_vs_logL}. As seen in Fig.~\ref{fig:result_overview_gran}, the $\nu_{\rm char}$ values from the simulations fall along the lower boundary of the observed values. Contrary to the findings by \cite{Anders2023} (see their Fig.~3), the $\nu_{\rm char}$ values from the simulations do fall in regions in Fig.~\ref{fig:nuchar_vs_logL} that are covered by the observations, however, the majority of the stars in our combined sample have higher $\nu_{\rm char}$ values at similar $\log \mathcal{L}/\mathcal{L}_\odot$. Furthermore, as seen in both figures the break point frequencies found for the $3\,{\rm M}_\odot$ simulation with core luminosity boosting are much higher than the $\nu_{\rm char}$ value from corresponding $3\,{\rm M}_\odot$ simulation from \cite{Anders2023}. Another point of contingency for the IGW theory is the question if it is possible for gravity waves, that are damped in convective regions, to reach the surface of stars with sub-surface convection zones. Based on 1-D stellar models, \cite{Serebriakova2024} demonstrated that IGWs can indeed tunnel through the FeCZ if the star is less than $30\,{\rm M}_\odot$ and thereby contribute to the observed variability. Full tests of this are currently awaiting 2-D and 3-D simulations as none (to our knowledge) of the available published simulations of core generated IGWs include the subsurface convection zones.

\subsection{Stellar winds}

Massive stars \citep[$M \gtrsim 15\,{\rm M}_\odot$,][]{Abbott1982} have radiatively driven stellar winds which show direct observational features in the stellar spectra when their luminosities are higher than $ 10^4\, {\rm L}_\odot$ \citep{Abbott1979,Kudritzki2000}. When the mass loss rates caused by such stellar winds are variable as due to, e.g., line-drive wind instabilities \citep{Feldmeier1997,Runacres2002} they result in photometric variability due to the effect of wind blanketing
\citep{Abbott1985}. 

Relying on simulations of stellar winds of O-type stars, \cite{Krticka2018} demonstrated that the resulting photometric variability from stellar winds is stochastic and shows similar characteristics to the SLF variability found in four comparison blue supergiants, with amplitudes decreasing for more structured stellar winds. \cite{Krticka2021} expanded upon this study by considering the impact of the choice of inner boundary atmospheric perturbations on the stellar wind. For stochastic 
perturbations expected by, e.g., sub-surface convection zones, \cite{Krticka2021} find that the resulting frequency spectrum 
follows a uniform power law close to the base of the perturbations but develops into a broken power law at larger heights above the boundary. Additionally, they find that the disk integrated photometric variability in mmag shows negative skewness calculated as $\overline{\left(x(t) - \overline{x(t)}\right)^3} / \sigma_{\rm std}^3$, where $\sigma_{\rm std}$ is the standard deviation of the photometric variability $x(t)$ in mmag as a function of time $t$.

\begin{figure}%[ht!]
\begin{center}
\includegraphics[width=\linewidth]{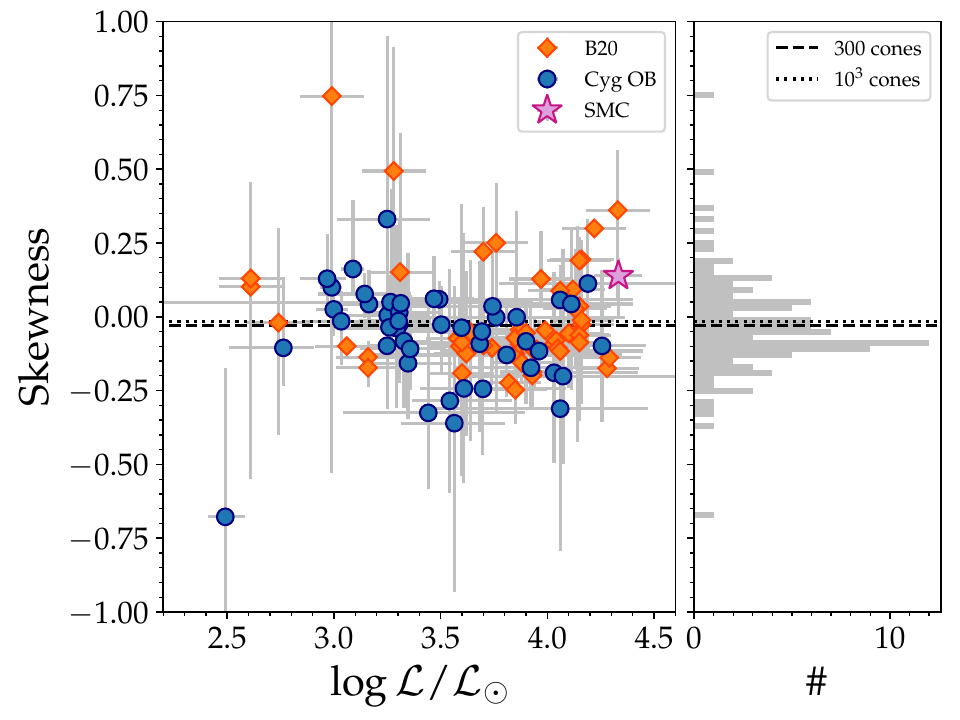}
\caption{Average skewness of the photometric light curves for our three considered stellar samples versus spectroscopic luminosity (\emph{left}). Vertical grey lines represent the range in skewness calculated from individual TESS sectors of a given star. Horizontal dashed and dotted lines show the skewness predictions of radiatively driven stellar winds \citep{Krticka2021} assuming the wind is made up of 300 (dashed) or 1000 (dotted) independent concentric cones. A histogram of the average skewness is likewise provided (\emph{right}).}\label{fig:skewness}
\end{center}
\end{figure}

The studies of the simulated SLF variability caused by stellar winds carried out by \cite{Krticka2018,Krticka2021} do not provide enough information to place the predictions in either Fig.~\ref{fig:result_overview_gran} or \ref{fig:nuchar_vs_logL}. Instead we calculate the skewness of the residual light curves of the Cyg~OB, B20, and SMC samples after converting the flux to mmag and plot the skewness as a function of spectroscopic luminosity in Fig.~\ref{fig:skewness}. While 66\% of the stars show a negative skewness when averaged across all TESS sectors, we find no clear dependence on $\log \mathcal{L}$. The lower metallicity SMC star shows a positive skewness. As \cite{Krticka2018,Krticka2021} predict the signal of the SLF variability caused by stellar winds to be larger for higher mass loss rates $\dot{M}$, this could potentially be consistent with the lower $\dot{M}$ found for lower metallicity stars \citep{Mokiem2007}, however, the sample size is too small to say anything conclusive. In comparison, \cite{Kourniotis2025} also calculated the skewness of their sample of 41 Galactic blue supergiants, finding values in the range of -0.69 to 0.34 with $\approx 54\%$ of the stars having negative skewness values.

The hydrodynamical simulations of sub-surface convection and convective core generated internal gravity waves discussed above do not provide any estimates for the skewness of the predicted SLF variability. It is therefore unclear if negative skewness of SLF light curves can be uniquely linked to stellar winds. However, given how different choices of boundary perturbations impact the predicted photometric variability \citep{Krticka2021} combined with the results from 3-D hydrodynamical simulations showing that the velocities generated by the iron sub-surface convection zone extend all the way to the surface \citep{Schultz2022} and that the wind structure starts to form already at the iron opacity bump \citep{Debnath2024}, it is highly likely that these processes are closely related. 

\section{Conclusions}\label{Sec:conclusions}

In this work we used two different methods to characterize the SLF variability found in a sample of 49 O- and B-type stars located within six Cygnus OB associations and one low-metallicity star in the SMC, AV~232. An additional three SMC stars out of an initial sample of 13 are found to be candidate SLF variables. For comparison, we also included 53 additional O- and B-type stars with SLF variability which was previously studied by \citet{Bowman2020}. For the first method we modelled the SLF variability in the power density spectrum using a Lorentzian-like profile, resulting in estimates of four parameters: $\alpha_0$, $\nu_{\rm char}$, $\gamma$ and $C_W$. The second method uses the photometric time series combined with the cumulative integrated power density spectrum to characterize the SLF variability using three parameters: RMS, $\nu_{50\%}$, and $w$. 

We find that both $\nu_{50\%}$ and $w$ are much less impacted by the observing cadence of the TESS data than $\nu_{\rm char}$ and $\gamma$ and recommend using the RMS, $\nu_{50\%}$ and $w$ parameters in future characterization of SLF variability for this reason. Similarly, \cite{Bowman2022} found the use of Gaussian process regression to be less sensitive to the data quality than fitting models to the variability in the Fourier domain. 
Upon studying the sector-to-sector dependence of the derived SLF parameters we find that the SLF variability changes characteristics over time. Additionally, we find that all derived parameters (except $\nu_{\rm char}$) are significantly correlated with the spectroscopic luminosity of the stars for the combined sample of 103 studied O- and B-type stars, whereas these correlations are no longer significant for $\gamma$ and $\nu_{50\%}$ if the B20 sample is considered on its own. Studying the values of RMS, $\alpha_0$, $\nu_{\rm char}$ and $\nu_{50\%}$ relative to the position of the stars in the spectroscopic HRD indicate that both $\alpha_0$ and RMS increase for more evolved stars, whereas $\nu_{\rm char}$ and $\nu_{50\%}$ decrease. No similar clear correlation is found for either $\gamma$ or $w$. 

In line with previous studies \citep{Bowman2018,Bowman2019a}, we find that the SLF variability is unlikely to be caused by surface granulation. When comparing the position of the full sample of stars in the $\nu_{\rm char}$ versus $\log \mathcal{L}/{\mathcal{L}_\odot}$ diagram to predictions from hydrodynamical simulations of sub-surface convection and internal gravity waves excited by core convection, we find good agreement between the observations and simulations of sub-surface convection zones \citep{Schultz2022,Schultz2023}. Similar agreement can be found for the simulations of IGWs generated by convective cores in some parts of the diagram. 

Having more 2-D and 3-D simulations available covering different masses, ages, and metallicities would be highly valuable to compare against observations and help determine the dominating process causing the SLF variability. Particularly, also publishing and making publicly available the simulated light curves is especially useful for anyone attempting to make comparisons to observations where different methods for characterizing the SLF variability have been applied, as is the case for this work. Here we were limited to making comparisons to $\nu_{\rm char}$. For the same reasons, we could not make similar direct comparisons to simulations of photometric variability caused by variable stellar winds and had to derive a new parameter, the skewness of the photometric variability, to compare to predictions from one simulation. With more than half of the 103 stars studied here showing negative skewness, this might support a connection to stellar winds. 

The low-metallicity SMC star AV~232 shows similar SLF characteristics to the 102 studied Galactic OB stars, and is located at the boundary where \citet{Jermyn2022} predict that a substantial subsurface FeCZ should be able to develop according to 1-D models. Having additional 3-D simulations of the envelope of a low metallicity SMC star near this boundary would therefore be very valuable. \citet{Bowman2024} studied a larger sample of SMC, LMC and Galactic OB stars, demonstrating that the lower metallicity SLF variables are found both within and outside the same predicted FeCZ stability windows. Again, having 3-D simulations of stars in the same position of the HRD but at different metallicities would provide an interesting comparison for the predicted SLF signals. 

Finally, in this work we focused on characterizing the SLF variability present in the TESS light curves of our sample of stars. Previous studies have also linked such derived SLF parameters with spectroscopic observations of macroturbulence \citep{Bowman2020,Shen2024}, which may likewise be caused by core-generated IGWs \citep{Aerts2015} and subsurface convection \citep{Cantiello2009,Cantiello2021,Schultz2023a} with the dominating process likely depending on the position of the star in the HRD \citep[e.g.][]{Serebriakova2024}. \citet{Krticka2018,Krticka2021} suggested that studying the variability in the UV could also be used to distinguish between different effects causing SLF variability \citep{Krticka2016}. Most recently, the SLF variability observed in TESS light curves of the O4 supergiant $\zeta$~Puppis has also been linked to variations in the linear-polarization of the light \citep{Bailey2024}. While carrying out similar studies are beyond the scope of this work, we look forward to following the continued utilization of synergies between different observations and predictions from simulations to study the origin of the SLF variability in massive stars.

\section*{Acknowledgements}

The authors are grateful for discussions with Tim Bedding, Dennis Stello, and Courtney Crawford at earlier stages of the manuscript, and appreciate the comments by Conny Aerts, Dominic Bowman, and Pieterjan Van Daele to the initial submitted version of the manuscript. We thank the anonymous referee for their comments which helped improve the manuscript. This research was supported in part by the NASA ATP grant ATP80NSSC22K0725, and by the National Science Foundation under Grant No. NSF PHY-2309135 at the KITP, as well as through the TESS Guest Investigator program Cycle 4 under Grant No. 80NSSC22K0743 from NASA, and by the Professor Harry Messel Research Fellowship in Physics Endowment, at the University of Sydney. 
This paper includes data collected with the TESS mission, obtained from the MAST data archive at the Space Telescope Science Institute (STScI). Funding for the TESS mission is provided by the NASA Explorer Program. STScI is operated by the Association of Universities for Research in Astronomy, Inc., under NASA contract NAS 5–26555. 
This research has made use of the SIMBAD database, operated at CDS, Strasbourg, France.

We acknowledge the use of the following \texttt{python} packages:
\texttt{astropy} \citep{astropy2013,astropy2018,astropy2022}, 
\texttt{astroquery} \citep{astroquery2019},
\texttt{lightkurve} \citep{Lightkurve2018},
\texttt{tesscut} \citep{tesscut2019},
\texttt{numpy} \citep{harris2020array},  \texttt{matplotlib} \citep{Matplotlib2007},
\texttt{pandas} \citep{reback2020pandas,mckinney-proc-scipy-2010},
\texttt{lmfit} \citep{lmfit},
\texttt{scipy} \citep{scipy},
\texttt{MLFriends} \citep{Buchner2016,Buchner2019}, \texttt{UltraNest} \citep{Buchner2021}.

%%%%%%%%%%%%%%%%%%%%%%%%%%%%%%%%%%%%%%%%%%%%%%%%%%
\section*{Data Availability}

The original TESS FFI data, target pixel files and PDCSAP light curves for the studied SLF variables are publicly available on the Michulski Archive for Space Telescopes (MAST). Residual light curves for each star, data type (FFI or 2-min cadence) and corresponding power density spectra are publicly available on Zenodo (DOI:10.5281/zenodo.15261328). Similarly, tables \ref{tab:sample_overview} to \ref{tab:results_averages_bowman} are publicly available on Zenodo in machine readable format, as are all necessary data for reconstructing figures (\ref{fig:result_overview_gran}-\ref{fig:skewness}) when not already included in one of the tables.

%%%%%%%%%%%%%%%%%%%% REFERENCES %%%%%%%%%%%%%%%%%%

% The best way to enter references is to use BibTeX:

\bibliographystyle{mnras}
\bibliography{references.bib} % if your bibtex file is called example.bib

\begin{thebibliography}{}
\makeatletter
\relax
\def\mn@urlcharsother{\let\do\@makeother \do\$\do\&\do\#\do\^\do\_\do\%\do\~}
\def\mn@doi{\begingroup\mn@urlcharsother \@ifnextchar [ {\mn@doi@} {\mn@doi@[]}}
\def\mn@doi@[#1]#2{\def\@tempa{#1}\ifx\@tempa\@empty \href {http://dx.doi.org/#2} {doi:#2}\else \href {http://dx.doi.org/#2} {#1}\fi \endgroup}
\def\mn@eprint#1#2{\mn@eprint@#1:#2::\@nil}
\def\mn@eprint@arXiv#1{\href {http://arxiv.org/abs/#1} {{\tt arXiv:#1}}}
\def\mn@eprint@dblp#1{\href {http://dblp.uni-trier.de/rec/bibtex/#1.xml} {dblp:#1}}
\def\mn@eprint@#1:#2:#3:#4\@nil{\def\@tempa {#1}\def\@tempb {#2}\def\@tempc {#3}\ifx \@tempc \@empty \let \@tempc \@tempb \let \@tempb \@tempa \fi \ifx \@tempb \@empty \def\@tempb {arXiv}\fi \@ifundefined {mn@eprint@\@tempb}{\@tempb:\@tempc}{\expandafter \expandafter \csname mn@eprint@\@tempb\endcsname \expandafter{\@tempc}}}

\bibitem[\protect\citeauthoryear{{Abbott}}{{Abbott}}{1979}]{Abbott1979}
{Abbott} D.~C.,  1979, in {Conti} P.~S.,  {De Loore} C.~W.~H.,  eds,  IAU Symposium Vol. 83, Mass Loss and Evolution of O-Type Stars. pp 237--239

\bibitem[\protect\citeauthoryear{{Abbott}}{{Abbott}}{1982}]{Abbott1982}
{Abbott} D.~C.,  1982, \mn@doi [\apj] {10.1086/160166}, \href {https://ui.adsabs.harvard.edu/abs/1982ApJ...259..282A} {259, 282}

\bibitem[\protect\citeauthoryear{{Abbott} \& {Hummer}}{{Abbott} \& {Hummer}}{1985}]{Abbott1985}
{Abbott} D.~C.,  {Hummer} D.~G.,  1985, \mn@doi [\apj] {10.1086/163297}, \href {https://ui.adsabs.harvard.edu/abs/1985ApJ...294..286A} {294, 286}

\bibitem[\protect\citeauthoryear{{Aerts} \& {Rogers}}{{Aerts} \& {Rogers}}{2015}]{Aerts2015}
{Aerts} C.,  {Rogers} T.~M.,  2015, \mn@doi [\apjl] {10.1088/2041-8205/806/2/L33}, \href {https://ui.adsabs.harvard.edu/abs/2015ApJ...806L..33A} {806, L33}

\bibitem[\protect\citeauthoryear{{Aerts} et~al.,}{{Aerts} et~al.}{2018}]{Aerts2018}
{Aerts} C.,  et~al., 2018, \mn@doi [\mnras] {10.1093/mnras/sty308}, \href {https://ui.adsabs.harvard.edu/abs/2018MNRAS.476.1234A} {476, 1234}

\bibitem[\protect\citeauthoryear{{Aigrain}, {Favata}  \& {Gilmore}}{{Aigrain} et~al.}{2004}]{Aigrain2004}
{Aigrain} S.,  {Favata} F.,   {Gilmore} G.,  2004, \mn@doi [\aap] {10.1051/0004-6361:20034039}, \href {https://ui.adsabs.harvard.edu/abs/2004A&A...414.1139A} {414, 1139}

\bibitem[\protect\citeauthoryear{{Anders} et~al.,}{{Anders} et~al.}{2023}]{Anders2023}
{Anders} E.~H.,  et~al., 2023, \mn@doi [Nature Astronomy] {10.1038/s41550-023-02040-7}, \href {https://ui.adsabs.harvard.edu/abs/2023NatAs...7.1228A} {7, 1228}

\bibitem[\protect\citeauthoryear{{Astropy Collaboration} et~al.,}{{Astropy Collaboration} et~al.}{2013}]{astropy2013}
{Astropy Collaboration} et~al., 2013, \mn@doi [\aap] {10.1051/0004-6361/201322068}, \href {https://ui.adsabs.harvard.edu/abs/2013A&A...558A..33A} {558, A33}

\bibitem[\protect\citeauthoryear{{Astropy Collaboration} et~al.,}{{Astropy Collaboration} et~al.}{2018}]{astropy2018}
{Astropy Collaboration} et~al., 2018, \mn@doi [\aj] {10.3847/1538-3881/aabc4f}, \href {https://ui.adsabs.harvard.edu/abs/2018AJ....156..123A} {156, 123}

\bibitem[\protect\citeauthoryear{{Astropy Collaboration} et~al.,}{{Astropy Collaboration} et~al.}{2022}]{astropy2022}
{Astropy Collaboration} et~al., 2022, \mn@doi [\apj] {10.3847/1538-4357/ac7c74}, \href {https://ui.adsabs.harvard.edu/abs/2022ApJ...935..167A} {935, 167}

\bibitem[\protect\citeauthoryear{{Bailey}, {Howarth}, {Cotton}, {Kedziora-Chudczer}, {De Horta}, {Martell}, {Eldridge}  \& {Luckas}}{{Bailey} et~al.}{2024}]{Bailey2024}
{Bailey} J.,  {Howarth} I.~D.,  {Cotton} D.~V.,  {Kedziora-Chudczer} L.,  {De Horta} A.,  {Martell} S.~L.,  {Eldridge} C.,   {Luckas} P.,  2024, \mn@doi [\mnras] {10.1093/mnras/stae548}, \href {https://ui.adsabs.harvard.edu/abs/2024MNRAS.529..374B} {529, 374}

\bibitem[\protect\citeauthoryear{{Balona}}{{Balona}}{1992}]{Balona1992}
{Balona} L.~A.,  1992, \mn@doi [\mnras] {10.1093/mnras/254.3.404}, \href {https://ui.adsabs.harvard.edu/abs/1992MNRAS.254..404B} {254, 404}

\bibitem[\protect\citeauthoryear{{Blomme} et~al.,}{{Blomme} et~al.}{2011}]{Blomme2011}
{Blomme} R.,  et~al., 2011, \mn@doi [\aap] {10.1051/0004-6361/201116949}, \href {https://ui.adsabs.harvard.edu/abs/2011A&A...533A...4B} {533, A4}

\bibitem[\protect\citeauthoryear{{Bouma}, {Hartman}, {Bhatti}, {Winn}  \& {Bakos}}{{Bouma} et~al.}{2019}]{Bouma2019}
{Bouma} L.~G.,  {Hartman} J.~D.,  {Bhatti} W.,  {Winn} J.~N.,   {Bakos} G.~{\'A}.,  2019, \mn@doi [\apjs] {10.3847/1538-4365/ab4a7e}, \href {https://ui.adsabs.harvard.edu/abs/2019ApJS..245...13B} {245, 13}

\bibitem[\protect\citeauthoryear{{Bouret}, {Martins}, {Hillier}, {Marcolino}, {Rocha-Pinto}, {Georgy}, {Lanz}  \& {Hubeny}}{{Bouret} et~al.}{2021}]{Bouret2021}
{Bouret} J.~C.,  {Martins} F.,  {Hillier} D.~J.,  {Marcolino} W.~L.~F.,  {Rocha-Pinto} H.~J.,  {Georgy} C.,  {Lanz} T.,   {Hubeny} I.,  2021, \mn@doi [\aap] {10.1051/0004-6361/202039890}, \href {https://ui.adsabs.harvard.edu/abs/2021A&A...647A.134B} {647, A134}

\bibitem[\protect\citeauthoryear{{Bowman} \& {Dorn-Wallenstein}}{{Bowman} \& {Dorn-Wallenstein}}{2022}]{Bowman2022}
{Bowman} D.~M.,  {Dorn-Wallenstein} T.~Z.,  2022, \mn@doi [\aap] {10.1051/0004-6361/202243545}, \href {https://ui.adsabs.harvard.edu/abs/2022A&A...668A.134B} {668, A134}

\bibitem[\protect\citeauthoryear{{Bowman} et~al.,}{{Bowman} et~al.}{2018}]{Bowman2018}
{Bowman} D.~M.,  et~al., 2018, in PHysics of Oscillating STars. p.~31 (\mn@eprint {arXiv} {1811.12930}), \mn@doi{10.5281/zenodo.1745562}

\bibitem[\protect\citeauthoryear{{Bowman} et~al.,}{{Bowman} et~al.}{2019a}]{Bowman2019b}
{Bowman} D.~M.,  et~al., 2019a, \mn@doi [Nature Astronomy] {10.1038/s41550-019-0768-1}, \href {https://ui.adsabs.harvard.edu/abs/2019NatAs...3..760B} {3, 760}

\bibitem[\protect\citeauthoryear{{Bowman} et~al.,}{{Bowman} et~al.}{2019b}]{Bowman2019a}
{Bowman} D.~M.,  et~al., 2019b, \mn@doi [\aap] {10.1051/0004-6361/201833662}, \href {https://ui.adsabs.harvard.edu/abs/2019A&A...621A.135B} {621, A135}

\bibitem[\protect\citeauthoryear{{Bowman}, {Burssens}, {Sim{\'o}n-D{\'\i}az}, {Edelmann}, {Rogers}, {Horst}, {R{\"o}pke}  \& {Aerts}}{{Bowman} et~al.}{2020}]{Bowman2020}
{Bowman} D.~M.,  {Burssens} S.,  {Sim{\'o}n-D{\'\i}az} S.,  {Edelmann} P.~V.~F.,  {Rogers} T.~M.,  {Horst} L.,  {R{\"o}pke} F.~K.,   {Aerts} C.,  2020, \mn@doi [\aap] {10.1051/0004-6361/202038224}, \href {https://ui.adsabs.harvard.edu/abs/2020A&A...640A..36B} {640, A36}

\bibitem[\protect\citeauthoryear{{Bowman}, {Van Daele}, {Michielsen}  \& {Van Reeth}}{{Bowman} et~al.}{2024}]{Bowman2024}
{Bowman} D.~M.,  {Van Daele} P.,  {Michielsen} M.,   {Van Reeth} T.,  2024, \mn@doi [arXiv e-prints] {10.48550/arXiv.2410.12726}, \href {https://ui.adsabs.harvard.edu/abs/2024arXiv241012726B} {p. arXiv:2410.12726}

\bibitem[\protect\citeauthoryear{{Brasseur}, {Phillip}, {Fleming}, {Mullally}  \& {White}}{{Brasseur} et~al.}{2019}]{tesscut2019}
{Brasseur} C.~E.,  {Phillip} C.,  {Fleming} S.~W.,  {Mullally} S.~E.,   {White} R.~L.,  2019, {Astrocut: Tools for creating cutouts of TESS images} (\mn@eprint {ascl} {1905.007})

\bibitem[\protect\citeauthoryear{{Browning}, {Brun}  \& {Toomre}}{{Browning} et~al.}{2004}]{Browning2004}
{Browning} M.~K.,  {Brun} A.~S.,   {Toomre} J.,  2004, \mn@doi [\apj] {10.1086/380198}, \href {https://ui.adsabs.harvard.edu/abs/2004ApJ...601..512B} {601, 512}

\bibitem[\protect\citeauthoryear{{Buchner}}{{Buchner}}{2016}]{Buchner2016}
{Buchner} J.,  2016, \mn@doi [Statistics and Computing] {10.1007/s11222-014-9512-y}, \href {https://ui.adsabs.harvard.edu/abs/2016S&C....26..383B} {26, 383}

\bibitem[\protect\citeauthoryear{{Buchner}}{{Buchner}}{2019}]{Buchner2019}
{Buchner} J.,  2019, \mn@doi [\pasp] {10.1088/1538-3873/aae7fc}, \href {https://ui.adsabs.harvard.edu/abs/2019PASP..131j8005B} {131, 108005}

\bibitem[\protect\citeauthoryear{{Buchner}}{{Buchner}}{2021}]{Buchner2021}
{Buchner} J.,  2021, \mn@doi [The Journal of Open Source Software] {10.21105/joss.03001}, \href {https://ui.adsabs.harvard.edu/abs/2021JOSS....6.3001B} {6, 3001}

\bibitem[\protect\citeauthoryear{{Burssens} et~al.,}{{Burssens} et~al.}{2020}]{Burssens2020}
{Burssens} S.,  et~al., 2020, \mn@doi [\aap] {10.1051/0004-6361/202037700}, \href {https://ui.adsabs.harvard.edu/abs/2020A&A...639A..81B} {639, A81}

\bibitem[\protect\citeauthoryear{{Caldwell} et~al.,}{{Caldwell} et~al.}{2020}]{Caldwell2020}
{Caldwell} D.~A.,  et~al., 2020, \mn@doi [Research Notes of the American Astronomical Society] {10.3847/2515-5172/abc9b3}, \href {https://ui.adsabs.harvard.edu/abs/2020RNAAS...4..201C} {4, 201}

\bibitem[\protect\citeauthoryear{{Cantiello} et~al.,}{{Cantiello} et~al.}{2009}]{Cantiello2009}
{Cantiello} M.,  et~al., 2009, \mn@doi [\aap] {10.1051/0004-6361/200911643}, \href {https://ui.adsabs.harvard.edu/abs/2009A&A...499..279C} {499, 279}

\bibitem[\protect\citeauthoryear{{Cantiello}, {Lecoanet}, {Jermyn}  \& {Grassitelli}}{{Cantiello} et~al.}{2021}]{Cantiello2021}
{Cantiello} M.,  {Lecoanet} D.,  {Jermyn} A.~S.,   {Grassitelli} L.,  2021, \mn@doi [\apj] {10.3847/1538-4357/ac03b0}, \href {https://ui.adsabs.harvard.edu/abs/2021ApJ...915..112C} {915, 112}

\bibitem[\protect\citeauthoryear{{Chaplin} et~al.,}{{Chaplin} et~al.}{2011}]{Chaplin2011}
{Chaplin} W.~J.,  et~al., 2011, \mn@doi [\apj] {10.1088/0004-637X/732/1/54}, \href {https://ui.adsabs.harvard.edu/abs/2011ApJ...732...54C} {732, 54}

\bibitem[\protect\citeauthoryear{{Chaplin}, {Elsworth}, {Davies}, {Campante}, {Handberg}, {Miglio}  \& {Basu}}{{Chaplin} et~al.}{2014}]{Chaplin2014}
{Chaplin} W.~J.,  {Elsworth} Y.,  {Davies} G.~R.,  {Campante} T.~L.,  {Handberg} R.,  {Miglio} A.,   {Basu} S.,  2014, \mn@doi [\mnras] {10.1093/mnras/stu1811}, \href {https://ui.adsabs.harvard.edu/abs/2014MNRAS.445..946C} {445, 946}

\bibitem[\protect\citeauthoryear{{Chen{\'e}} et~al.,}{{Chen{\'e}} et~al.}{2011}]{Chene2011}
{Chen{\'e}} A.~N.,  et~al., 2011, \mn@doi [\apj] {10.1088/0004-637X/735/1/34}, \href {https://ui.adsabs.harvard.edu/abs/2011ApJ...735...34C} {735, 34}

\bibitem[\protect\citeauthoryear{{Choi}, {Dotter}, {Conroy}, {Cantiello}, {Paxton}  \& {Johnson}}{{Choi} et~al.}{2016}]{Choi2016}
{Choi} J.,  {Dotter} A.,  {Conroy} C.,  {Cantiello} M.,  {Paxton} B.,   {Johnson} B.~D.,  2016, \mn@doi [\apj] {10.3847/0004-637X/823/2/102}, \href {https://ui.adsabs.harvard.edu/abs/2016ApJ...823..102C} {823, 102}

\bibitem[\protect\citeauthoryear{{Davies} et~al.,}{{Davies} et~al.}{2016}]{Davies2016}
{Davies} G.~R.,  et~al., 2016, \mn@doi [\mnras] {10.1093/mnras/stv2593}, \href {https://ui.adsabs.harvard.edu/abs/2016MNRAS.456.2183D} {456, 2183}

\bibitem[\protect\citeauthoryear{{Debnath}, {Sundqvist}, {Moens}, {Van der Sijpt}, {Verhamme}  \& {Poniatowski}}{{Debnath} et~al.}{2024}]{Debnath2024}
{Debnath} D.,  {Sundqvist} J.~O.,  {Moens} N.,  {Van der Sijpt} C.,  {Verhamme} O.,   {Poniatowski} L.~G.,  2024, \mn@doi [\aap] {10.1051/0004-6361/202348206}, \href {https://ui.adsabs.harvard.edu/abs/2024A&A...684A.177D} {684, A177}

\bibitem[\protect\citeauthoryear{{Deupree}}{{Deupree}}{2000}]{Deupree_2000}
{Deupree} R.~G.,  2000, \mn@doi [\apj] {10.1086/317068}, 543, 395

\bibitem[\protect\citeauthoryear{{Dorn-Wallenstein}, {Levesque}  \& {Davenport}}{{Dorn-Wallenstein} et~al.}{2019}]{Dorn-Wallenstein2019}
{Dorn-Wallenstein} T.~Z.,  {Levesque} E.~M.,   {Davenport} J. R.~A.,  2019, \mn@doi [\apj] {10.3847/1538-4357/ab223f}, \href {https://ui.adsabs.harvard.edu/abs/2019ApJ...878..155D} {878, 155}

\bibitem[\protect\citeauthoryear{{Dorn-Wallenstein}, {Levesque}, {Neugent}, {Davenport}, {Morris}  \& {Gootkin}}{{Dorn-Wallenstein} et~al.}{2020}]{DornWallenstein2020}
{Dorn-Wallenstein} T.~Z.,  {Levesque} E.~M.,  {Neugent} K.~F.,  {Davenport} J. R.~A.,  {Morris} B.~M.,   {Gootkin} K.,  2020, \mn@doi [\apj] {10.3847/1538-4357/abb318}, \href {https://ui.adsabs.harvard.edu/abs/2020ApJ...902...24D} {902, 24}

\bibitem[\protect\citeauthoryear{{Dotter}}{{Dotter}}{2016}]{Dotter2016}
{Dotter} A.,  2016, \mn@doi [\apjs] {10.3847/0067-0049/222/1/8}, \href {https://ui.adsabs.harvard.edu/abs/2016ApJS..222....8D} {222, 8}

\bibitem[\protect\citeauthoryear{{Duvall} \& {Harvey}}{{Duvall} \& {Harvey}}{1986}]{DuvallHarvey1986}
{Duvall} T.~L. J.,  {Harvey} J.~W.,  1986, in {Gough} D.~O.,  ed.,  NATO Advanced Study Institute (ASI) Series C Vol. 169, Seismology of the Sun and the Distant Stars. pp 105--116

\bibitem[\protect\citeauthoryear{{Edelmann}, {Ratnasingam}, {Pedersen}, {Bowman}, {Prat}  \& {Rogers}}{{Edelmann} et~al.}{2019}]{Edelmann2019}
{Edelmann} P.~V.~F.,  {Ratnasingam} R.~P.,  {Pedersen} M.~G.,  {Bowman} D.~M.,  {Prat} V.,   {Rogers} T.~M.,  2019, \mn@doi [\apj] {10.3847/1538-4357/ab12df}, \href {https://ui.adsabs.harvard.edu/abs/2019ApJ...876....4E} {876, 4}

\bibitem[\protect\citeauthoryear{{Elliott} et~al.,}{{Elliott} et~al.}{2022}]{Elliott2022}
{Elliott} A.,  et~al., 2022, \mn@doi [\mnras] {10.1093/mnras/stab3112}, \href {https://ui.adsabs.harvard.edu/abs/2022MNRAS.509.4246E} {509, 4246}

\bibitem[\protect\citeauthoryear{{Feldmeier}, {Puls}  \& {Pauldrach}}{{Feldmeier} et~al.}{1997}]{Feldmeier1997}
{Feldmeier} A.,  {Puls} J.,   {Pauldrach} A.~W.~A.,  1997, \aap, \href {https://ui.adsabs.harvard.edu/abs/1997A&A...322..878F} {322, 878}

\bibitem[\protect\citeauthoryear{{Ginsburg} et~al.,}{{Ginsburg} et~al.}{2019}]{astroquery2019}
{Ginsburg} A.,  et~al., 2019, \mn@doi [\aj] {10.3847/1538-3881/aafc33}, \href {https://ui.adsabs.harvard.edu/abs/2019AJ....157...98G} {157, 98}

\bibitem[\protect\citeauthoryear{{Goldberg}, {Jiang}  \& {Bildsten}}{{Goldberg} et~al.}{2022}]{Goldberg2022}
{Goldberg} J.~A.,  {Jiang} Y.-F.,   {Bildsten} L.,  2022, \mn@doi [\apj] {10.3847/1538-4357/ac5ab3}, \href {https://ui.adsabs.harvard.edu/abs/2022ApJ...929..156G} {929, 156}

\bibitem[\protect\citeauthoryear{{Handberg} et~al.,}{{Handberg} et~al.}{2021}]{Handberg2021}
{Handberg} R.,  et~al., 2021, \mn@doi [\aj] {10.3847/1538-3881/ac09f1}, \href {https://ui.adsabs.harvard.edu/abs/2021AJ....162..170H} {162, 170}

\bibitem[\protect\citeauthoryear{Harris et~al.,}{Harris et~al.}{2020}]{harris2020array}
Harris C.~R.,  et~al., 2020, \mn@doi [Nature] {10.1038/s41586-020-2649-2}, 585, 357

\bibitem[\protect\citeauthoryear{{Harvey}}{{Harvey}}{1985}]{Harvey1985}
{Harvey} J.,  1985, in {Rolfe} E.,  {Battrick} B.,  eds,  ESA Special Publication Vol. 235, Future Missions in Solar, Heliospheric \& Space Plasma Physics. p.~199

\bibitem[\protect\citeauthoryear{{Herwig} et~al.,}{{Herwig} et~al.}{2023}]{Herwig2023}
{Herwig} F.,  et~al., 2023, \mn@doi [\mnras] {10.1093/mnras/stad2157}, \href {https://ui.adsabs.harvard.edu/abs/2023MNRAS.525.1601H} {525, 1601}

\bibitem[\protect\citeauthoryear{{Higgins} \& {Bell}}{{Higgins} \& {Bell}}{2023}]{Higgins2023}
{Higgins} M.~E.,  {Bell} K.~J.,  2023, \mn@doi [\aj] {10.3847/1538-3881/acb20c}, \href {https://ui.adsabs.harvard.edu/abs/2023AJ....165..141H} {165, 141}

\bibitem[\protect\citeauthoryear{{Hon} et~al.,}{{Hon} et~al.}{2021}]{Hon2021}
{Hon} M.,  et~al., 2021, \mn@doi [\apj] {10.3847/1538-4357/ac14b1}, \href {https://ui.adsabs.harvard.edu/abs/2021ApJ...919..131H} {919, 131}

\bibitem[\protect\citeauthoryear{{Huber}, {Stello}, {Bedding}, {Chaplin}, {Arentoft}, {Quirion}  \& {Kjeldsen}}{{Huber} et~al.}{2009}]{Huber2009}
{Huber} D.,  {Stello} D.,  {Bedding} T.~R.,  {Chaplin} W.~J.,  {Arentoft} T.,  {Quirion} P.~O.,   {Kjeldsen} H.,  2009, \mn@doi [Communications in Asteroseismology] {10.48550/arXiv.0910.2764}, \href {https://ui.adsabs.harvard.edu/abs/2009CoAst.160...74H} {160, 74}

\bibitem[\protect\citeauthoryear{{Huber} et~al.,}{{Huber} et~al.}{2022}]{Huber2022}
{Huber} D.,  et~al., 2022, \mn@doi [\aj] {10.3847/1538-3881/ac3000}, \href {https://ui.adsabs.harvard.edu/abs/2022AJ....163...79H} {163, 79}

\bibitem[\protect\citeauthoryear{Hunter}{Hunter}{2007}]{Matplotlib2007}
Hunter J.~D.,  2007, \mn@doi [Computing in Science \& Engineering] {10.1109/MCSE.2007.55}, 9, 90

\bibitem[\protect\citeauthoryear{{Iglesias}, {Rogers}  \& {Wilson}}{{Iglesias} et~al.}{1992}]{Iglesias1992}
{Iglesias} C.~A.,  {Rogers} F.~J.,   {Wilson} B.~G.,  1992, \mn@doi [\apj] {10.1086/171827}, \href {https://ui.adsabs.harvard.edu/abs/1992ApJ...397..717I} {397, 717}

\bibitem[\protect\citeauthoryear{{Jenkins} et~al.,}{{Jenkins} et~al.}{2016}]{Jenkins2016}
{Jenkins} J.~M.,  et~al., 2016, in {Chiozzi} G.,  {Guzman} J.~C.,  eds,  Society of Photo-Optical Instrumentation Engineers (SPIE) Conference Series Vol. 9913, Software and Cyberinfrastructure for Astronomy IV. p. 99133E, \mn@doi{10.1117/12.2233418}

\bibitem[\protect\citeauthoryear{{Jermyn}, {Anders}  \& {Cantiello}}{{Jermyn} et~al.}{2022}]{Jermyn2022}
{Jermyn} A.~S.,  {Anders} E.~H.,   {Cantiello} M.,  2022, \mn@doi [\apj] {10.3847/1538-4357/ac4e89}, \href {https://ui.adsabs.harvard.edu/abs/2022ApJ...926..221J} {926, 221}

\bibitem[\protect\citeauthoryear{{Jiang}}{{Jiang}}{2023}]{Jiang2023}
{Jiang} Y.-F.,  2023, \mn@doi [Galaxies] {10.3390/galaxies11050105}, \href {https://ui.adsabs.harvard.edu/abs/2023Galax..11..105J} {11, 105}

\bibitem[\protect\citeauthoryear{{Kallinger} \& {Matthews}}{{Kallinger} \& {Matthews}}{2010}]{Kallinger2010}
{Kallinger} T.,  {Matthews} J.~M.,  2010, \mn@doi [\apjl] {10.1088/2041-8205/711/1/L35}, \href {https://ui.adsabs.harvard.edu/abs/2010ApJ...711L..35K} {711, L35}

\bibitem[\protect\citeauthoryear{{Kallinger} et~al.,}{{Kallinger} et~al.}{2014}]{Kallinger2014}
{Kallinger} T.,  et~al., 2014, \mn@doi [\aap] {10.1051/0004-6361/201424313}, \href {https://ui.adsabs.harvard.edu/abs/2014A&A...570A..41K} {570, A41}

\bibitem[\protect\citeauthoryear{{Kiss}, {Szab{\'o}}  \& {Bedding}}{{Kiss} et~al.}{2006}]{Kiss2006}
{Kiss} L.~L.,  {Szab{\'o}} G.~M.,   {Bedding} T.~R.,  2006, \mn@doi [\mnras] {10.1111/j.1365-2966.2006.10973.x}, \href {https://ui.adsabs.harvard.edu/abs/2006MNRAS.372.1721K} {372, 1721}

\bibitem[\protect\citeauthoryear{{Kjeldsen} \& {Bedding}}{{Kjeldsen} \& {Bedding}}{1995}]{KjeldsenBedding1995}
{Kjeldsen} H.,  {Bedding} T.~R.,  1995, \mn@doi [\aap] {10.48550/arXiv.astro-ph/9403015}, \href {https://ui.adsabs.harvard.edu/abs/1995A&A...293...87K} {293, 87}

\bibitem[\protect\citeauthoryear{{Kjeldsen} et~al.,}{{Kjeldsen} et~al.}{2005}]{Kjeldsen2005}
{Kjeldsen} H.,  et~al., 2005, \mn@doi [\apj] {10.1086/497530}, \href {https://ui.adsabs.harvard.edu/abs/2005ApJ...635.1281K} {635, 1281}

\bibitem[\protect\citeauthoryear{{Kourniotis}, {Cidale}, {Kraus}, {Ruiz Diaz}  \& {Alberici Adam}}{{Kourniotis} et~al.}{2025}]{Kourniotis2025}
{Kourniotis} M.,  {Cidale} L.~S.,  {Kraus} M.,  {Ruiz Diaz} M.~A.,   {Alberici Adam} A.,  2025, \mn@doi [arXiv e-prints] {10.48550/arXiv.2503.20860}, \href {https://ui.adsabs.harvard.edu/abs/2025arXiv250320860K} {p. arXiv:2503.20860}

\bibitem[\protect\citeauthoryear{{Krti{\v{c}}ka}}{{Krti{\v{c}}ka}}{2016}]{Krticka2016}
{Krti{\v{c}}ka} J.,  2016, \mn@doi [\aap] {10.1051/0004-6361/201629222}, \href {https://ui.adsabs.harvard.edu/abs/2016A&A...594A..75K} {594, A75}

\bibitem[\protect\citeauthoryear{{Krti{\v{c}}ka} \& {Feldmeier}}{{Krti{\v{c}}ka} \& {Feldmeier}}{2018}]{Krticka2018}
{Krti{\v{c}}ka} J.,  {Feldmeier} A.,  2018, \mn@doi [\aap] {10.1051/0004-6361/201731614}, \href {https://ui.adsabs.harvard.edu/abs/2018A&A...617A.121K} {617, A121}

\bibitem[\protect\citeauthoryear{{Krti{\v{c}}ka} \& {Feldmeier}}{{Krti{\v{c}}ka} \& {Feldmeier}}{2021}]{Krticka2021}
{Krti{\v{c}}ka} J.,  {Feldmeier} A.,  2021, \mn@doi [\aap] {10.1051/0004-6361/202040148}, \href {https://ui.adsabs.harvard.edu/abs/2021A&A...648A..79K} {648, A79}

\bibitem[\protect\citeauthoryear{{Kudritzki} \& {Puls}}{{Kudritzki} \& {Puls}}{2000}]{Kudritzki2000}
{Kudritzki} R.-P.,  {Puls} J.,  2000, \mn@doi [\araa] {10.1146/annurev.astro.38.1.613}, \href {https://ui.adsabs.harvard.edu/abs/2000ARA&A..38..613K} {38, 613}

\bibitem[\protect\citeauthoryear{{Lamontagne} \& {Moffat}}{{Lamontagne} \& {Moffat}}{1987}]{Lamontagne1987}
{Lamontagne} R.,  {Moffat} A. F.~J.,  1987, \mn@doi [\aj] {10.1086/114535}, \href {https://ui.adsabs.harvard.edu/abs/1987AJ.....94.1008L} {94, 1008}

\bibitem[\protect\citeauthoryear{{Lecoanet} \& {Edelmann}}{{Lecoanet} \& {Edelmann}}{2023}]{Lecoanet2023}
{Lecoanet} D.,  {Edelmann} P. V.~F.,  2023, \mn@doi [Galaxies] {10.3390/galaxies11040089}, \href {https://ui.adsabs.harvard.edu/abs/2023Galax..11...89L} {11, 89}

\bibitem[\protect\citeauthoryear{{Lecoanet} \& {Quataert}}{{Lecoanet} \& {Quataert}}{2013}]{Lecoanet2013}
{Lecoanet} D.,  {Quataert} E.,  2013, \mn@doi [\mnras] {10.1093/mnras/stt055}, \href {https://ui.adsabs.harvard.edu/abs/2013MNRAS.430.2363L} {430, 2363}

\bibitem[\protect\citeauthoryear{{Lenoir-Craig} et~al.,}{{Lenoir-Craig} et~al.}{2022}]{Lenoir-Craig2022}
{Lenoir-Craig} G.,  et~al., 2022, \mn@doi [\apj] {10.3847/1538-4357/ac397d}, \href {https://ui.adsabs.harvard.edu/abs/2022ApJ...925...79L} {925, 79}

\bibitem[\protect\citeauthoryear{{L{\'e}pine} \& {Moffat}}{{L{\'e}pine} \& {Moffat}}{1999}]{Lepine1999}
{L{\'e}pine} S.,  {Moffat} A. F.~J.,  1999, \mn@doi [\apj] {10.1086/306958}, \href {https://ui.adsabs.harvard.edu/abs/1999ApJ...514..909L} {514, 909}

\bibitem[\protect\citeauthoryear{{Li}, {Bedding}, {Li}, {Bi}, {Stello}, {Zhou}  \& {White}}{{Li} et~al.}{2020}]{LiY2020}
{Li} Y.,  {Bedding} T.~R.,  {Li} T.,  {Bi} S.,  {Stello} D.,  {Zhou} Y.,   {White} T.~R.,  2020, \mn@doi [\mnras] {10.1093/mnras/staa1335}, \href {https://ui.adsabs.harvard.edu/abs/2020MNRAS.495.2363L} {495, 2363}

\bibitem[\protect\citeauthoryear{{Lighthill}}{{Lighthill}}{1952}]{Lighthill1952}
{Lighthill} M.~J.,  1952, \mn@doi [Proceedings of the Royal Society of London Series A] {10.1098/rspa.1952.0060}, \href {https://ui.adsabs.harvard.edu/abs/1952RSPSA.211..564L} {211, 564}

\bibitem[\protect\citeauthoryear{{Lightkurve Collaboration} et~al.,}{{Lightkurve Collaboration} et~al.}{2018}]{Lightkurve2018}
{Lightkurve Collaboration} et~al., 2018, {Lightkurve: Kepler and TESS time series analysis in Python}, Astrophysics Source Code Library (\mn@eprint {ascl} {1812.013})

\bibitem[\protect\citeauthoryear{{Lomb}}{{Lomb}}{1976}]{Lomb1976}
{Lomb} N.~R.,  1976, \mn@doi [\apss] {10.1007/BF00648343}, \href {https://ui.adsabs.harvard.edu/abs/1976Ap&SS..39..447L} {39, 447}

\bibitem[\protect\citeauthoryear{{Lund} et~al.,}{{Lund} et~al.}{2017}]{Lund2017}
{Lund} M.~N.,  et~al., 2017, \mn@doi [\apj] {10.3847/1538-4357/835/2/172}, \href {https://ui.adsabs.harvard.edu/abs/2017ApJ...835..172L} {835, 172}

\bibitem[\protect\citeauthoryear{{Ma}, {Johnston}, {Bellinger}  \& {de Mink}}{{Ma} et~al.}{2024}]{Ma2024}
{Ma} L.,  {Johnston} C.,  {Bellinger} E.~P.,   {de Mink} S.~E.,  2024, \mn@doi [\apj] {10.3847/1538-4357/ad38bc}, \href {https://ui.adsabs.harvard.edu/abs/2024ApJ...966..196M} {966, 196}

\bibitem[\protect\citeauthoryear{{Michel}, {Samadi}, {Baudin}, {Barban}, {Appourchaux}  \& {Auvergne}}{{Michel} et~al.}{2009}]{Michel2009}
{Michel} E.,  {Samadi} R.,  {Baudin} F.,  {Barban} C.,  {Appourchaux} T.,   {Auvergne} M.,  2009, \mn@doi [\aap] {10.1051/0004-6361:200810353}, \href {https://ui.adsabs.harvard.edu/abs/2009A&A...495..979M} {495, 979}

\bibitem[\protect\citeauthoryear{{Moffat} et~al.,}{{Moffat} et~al.}{2008}]{Moffat2008}
{Moffat} A.~F.~J.,  et~al., 2008, in {de Koter} A.,  {Smith} L.~J.,   {Waters} L. B.~F.~M.,  eds,  Astronomical Society of the Pacific Conference Series Vol. 388, Mass Loss from Stars and the Evolution of Stellar Clusters. p.~29

\bibitem[\protect\citeauthoryear{{Mokiem} et~al.,}{{Mokiem} et~al.}{2007}]{Mokiem2007}
{Mokiem} M.~R.,  et~al., 2007, \mn@doi [\aap] {10.1051/0004-6361:20077545}, \href {https://ui.adsabs.harvard.edu/abs/2007A&A...473..603M} {473, 603}

\bibitem[\protect\citeauthoryear{{Montalb{\'a}n} \& {Schatzman}}{{Montalb{\'a}n} \& {Schatzman}}{2000}]{Montalban2000}
{Montalb{\'a}n} J.,  {Schatzman} E.,  2000, \aap, \href {https://ui.adsabs.harvard.edu/abs/2000A&A...354..943M} {354, 943}

\bibitem[\protect\citeauthoryear{{Nardiello} et~al.,}{{Nardiello} et~al.}{2019}]{Nardiello2019}
{Nardiello} D.,  et~al., 2019, \mn@doi [\mnras] {10.1093/mnras/stz2878}, \href {https://ui.adsabs.harvard.edu/abs/2019MNRAS.490.3806N} {490, 3806}

\bibitem[\protect\citeauthoryear{{Naz{\'e}}, {Rauw}  \& {Gosset}}{{Naz{\'e}} et~al.}{2021}]{Naze2021}
{Naz{\'e}} Y.,  {Rauw} G.,   {Gosset} E.,  2021, \mn@doi [\mnras] {10.1093/mnras/stab133}, \href {https://ui.adsabs.harvard.edu/abs/2021MNRAS.502.5038N} {502, 5038}

\bibitem[\protect\citeauthoryear{{Newville} et~al.,}{{Newville} et~al.}{2024}]{lmfit}
{Newville} M.,  et~al., 2024, lmfit/lmfit-py: 1.3.2, Zenodo, \mn@doi{10.5281/zenodo.12785036}

\bibitem[\protect\citeauthoryear{{Nielsen} et~al.,}{{Nielsen} et~al.}{2021}]{Nielsen2021}
{Nielsen} M.~B.,  et~al., 2021, \mn@doi [\aj] {10.3847/1538-3881/abcd39}, \href {https://ui.adsabs.harvard.edu/abs/2021AJ....161...62N} {161, 62}

\bibitem[\protect\citeauthoryear{{P{\'a}pics} et~al.,}{{P{\'a}pics} et~al.}{2017}]{Papics2017}
{P{\'a}pics} P.~I.,  et~al., 2017, \mn@doi [\aap] {10.1051/0004-6361/201629814}, \href {https://ui.adsabs.harvard.edu/abs/2017A&A...598A..74P} {598, A74}

\bibitem[\protect\citeauthoryear{{Paxton}, {Bildsten}, {Dotter}, {Herwig}, {Lesaffre}  \& {Timmes}}{{Paxton} et~al.}{2011}]{Paxton2011}
{Paxton} B.,  {Bildsten} L.,  {Dotter} A.,  {Herwig} F.,  {Lesaffre} P.,   {Timmes} F.,  2011, \mn@doi [\apjs] {10.1088/0067-0049/192/1/3}, \href {https://ui.adsabs.harvard.edu/abs/2011ApJS..192....3P} {192, 3}

\bibitem[\protect\citeauthoryear{{Paxton} et~al.,}{{Paxton} et~al.}{2013}]{Paxton2013}
{Paxton} B.,  et~al., 2013, \mn@doi [\apjs] {10.1088/0067-0049/208/1/4}, \href {https://ui.adsabs.harvard.edu/abs/2013ApJS..208....4P} {208, 4}

\bibitem[\protect\citeauthoryear{{Paxton} et~al.,}{{Paxton} et~al.}{2015}]{Paxton2015}
{Paxton} B.,  et~al., 2015, \mn@doi [\apjs] {10.1088/0067-0049/220/1/15}, \href {https://ui.adsabs.harvard.edu/abs/2015ApJS..220...15P} {220, 15}

\bibitem[\protect\citeauthoryear{{Pedersen} \& {Bell}}{{Pedersen} \& {Bell}}{2023}]{Pedersen2023}
{Pedersen} M.~G.,  {Bell} K.~J.,  2023, \mn@doi [\aj] {10.3847/1538-3881/accc31}, \href {https://ui.adsabs.harvard.edu/abs/2023AJ....165..239P} {165, 239}

\bibitem[\protect\citeauthoryear{{Pedersen} et~al.,}{{Pedersen} et~al.}{2019}]{Pedersen2019}
{Pedersen} M.~G.,  et~al., 2019, \mn@doi [\apjl] {10.3847/2041-8213/ab01e1}, \href {https://ui.adsabs.harvard.edu/abs/2019ApJ...872L...9P} {872, L9}

\bibitem[\protect\citeauthoryear{{Quataert} \& {Shiode}}{{Quataert} \& {Shiode}}{2012}]{Quataert2012}
{Quataert} E.,  {Shiode} J.,  2012, \mn@doi [\mnras] {10.1111/j.1745-3933.2012.01264.x}, \href {https://ui.adsabs.harvard.edu/abs/2012MNRAS.423L..92Q} {423, L92}

\bibitem[\protect\citeauthoryear{{Quintana} \& {Wright}}{{Quintana} \& {Wright}}{2021}]{Quintana2021}
{Quintana} A.~L.,  {Wright} N.~J.,  2021, \mn@doi [\mnras] {10.1093/mnras/stab2663}, \href {https://ui.adsabs.harvard.edu/abs/2021MNRAS.508.2370Q} {508, 2370}

\bibitem[\protect\citeauthoryear{{Quintana} \& {Wright}}{{Quintana} \& {Wright}}{2022}]{Quintana2022}
{Quintana} A.~L.,  {Wright} N.~J.,  2022, \mn@doi [\mnras] {10.1093/mnras/stac232}, \href {https://ui.adsabs.harvard.edu/abs/2022MNRAS.511.1224Q} {511, 1224}

\bibitem[\protect\citeauthoryear{{Ramiaramanantsoa} et~al.,}{{Ramiaramanantsoa} et~al.}{2018}]{Ramiaramanantsoa2018}
{Ramiaramanantsoa} T.,  et~al., 2018, \mn@doi [\mnras] {10.1093/mnras/sty1897}, \href {https://ui.adsabs.harvard.edu/abs/2018MNRAS.480..972R} {480, 972}

\bibitem[\protect\citeauthoryear{{Ramiaramanantsoa} et~al.,}{{Ramiaramanantsoa} et~al.}{2019}]{Ramiaramanantsoa2019}
{Ramiaramanantsoa} T.,  et~al., 2019, \mn@doi [\mnras] {10.1093/mnras/stz2895}, \href {https://ui.adsabs.harvard.edu/abs/2019MNRAS.490.5921R} {490, 5921}

\bibitem[\protect\citeauthoryear{{Ricker} et~al.,}{{Ricker} et~al.}{2014}]{TESS}
{Ricker} G.~R.,  et~al., 2014, in {Oschmann} Jacobus~M. J.,  {Clampin} M.,  {Fazio} G.~G.,   {MacEwen} H.~A.,  eds,  Society of Photo-Optical Instrumentation Engineers (SPIE) Conference Series Vol. 9143, Space Telescopes and Instrumentation 2014: Optical, Infrared, and Millimeter Wave. p. 914320 (\mn@eprint {arXiv} {1406.0151}), \mn@doi{10.1117/12.2063489}

\bibitem[\protect\citeauthoryear{{Rogers} \& {McElwaine}}{{Rogers} \& {McElwaine}}{2017}]{Rogers2017}
{Rogers} T.~M.,  {McElwaine} J.~N.,  2017, \mn@doi [\apjl] {10.3847/2041-8213/aa8d13}, \href {https://ui.adsabs.harvard.edu/abs/2017ApJ...848L...1R} {848, L1}

\bibitem[\protect\citeauthoryear{{Rogers}, {Lin}, {McElwaine}  \& {Lau}}{{Rogers} et~al.}{2013}]{Rogers2013}
{Rogers} T.~M.,  {Lin} D.~N.~C.,  {McElwaine} J.~N.,   {Lau} H.~H.~B.,  2013, \mn@doi [\apj] {10.1088/0004-637X/772/1/21}, \href {https://ui.adsabs.harvard.edu/abs/2013ApJ...772...21R} {772, 21}

\bibitem[\protect\citeauthoryear{{Runacres} \& {Owocki}}{{Runacres} \& {Owocki}}{2002}]{Runacres2002}
{Runacres} M.~C.,  {Owocki} S.~P.,  2002, \mn@doi [\aap] {10.1051/0004-6361:20011526}, \href {https://ui.adsabs.harvard.edu/abs/2002A&A...381.1015R} {381, 1015}

\bibitem[\protect\citeauthoryear{{Scargle}}{{Scargle}}{1982}]{Scargle1982}
{Scargle} J.~D.,  1982, \mn@doi [\apj] {10.1086/160554}, \href {https://ui.adsabs.harvard.edu/abs/1982ApJ...263..835S} {263, 835}

\bibitem[\protect\citeauthoryear{{Schofield} et~al.,}{{Schofield} et~al.}{2019}]{Schofield2019}
{Schofield} M.,  et~al., 2019, \mn@doi [\apjs] {10.3847/1538-4365/ab04f5}, \href {https://ui.adsabs.harvard.edu/abs/2019ApJS..241...12S} {241, 12}

\bibitem[\protect\citeauthoryear{{Schultz}, {Bildsten}  \& {Jiang}}{{Schultz} et~al.}{2022}]{Schultz2022}
{Schultz} W.~C.,  {Bildsten} L.,   {Jiang} Y.-F.,  2022, \mn@doi [\apjl] {10.3847/2041-8213/ac441f}, \href {https://ui.adsabs.harvard.edu/abs/2022ApJ...924L..11S} {924, L11}

\bibitem[\protect\citeauthoryear{{Schultz}, {Tsang}, {Bildsten}  \& {Jiang}}{{Schultz} et~al.}{2023a}]{Schultz2023a}
{Schultz} W.~C.,  {Tsang} B. T.~H.,  {Bildsten} L.,   {Jiang} Y.-F.,  2023a, \mn@doi [\apj] {10.3847/1538-4357/acb701}, \href {https://ui.adsabs.harvard.edu/abs/2023ApJ...945...58S} {945, 58}

\bibitem[\protect\citeauthoryear{{Schultz}, {Bildsten}  \& {Jiang}}{{Schultz} et~al.}{2023b}]{Schultz2023}
{Schultz} W.~C.,  {Bildsten} L.,   {Jiang} Y.-F.,  2023b, \mn@doi [\apjl] {10.3847/2041-8213/acdf50}, \href {https://ui.adsabs.harvard.edu/abs/2023ApJ...951L..42S} {951, L42}

\bibitem[\protect\citeauthoryear{{Serebriakova}, {Tkachenko}  \& {Aerts}}{{Serebriakova} et~al.}{2024}]{Serebriakova2024}
{Serebriakova} N.,  {Tkachenko} A.,   {Aerts} C.,  2024, \mn@doi [arXiv e-prints] {10.48550/arXiv.2408.15888}, \href {https://ui.adsabs.harvard.edu/abs/2024arXiv240815888S} {p. arXiv:2408.15888}

\bibitem[\protect\citeauthoryear{{Shen}, {Zhu}, {L{\"u}}, {Lu}  \& {He}}{{Shen} et~al.}{2024}]{Shen2024}
{Shen} D.-X.,  {Zhu} C.-H.,  {L{\"u}} G.-L.,  {Lu} X.-z.,   {He} X.-l.,  2024, \mn@doi [\apjs] {10.3847/1538-4365/ad71d3}, \href {https://ui.adsabs.harvard.edu/abs/2024ApJS..275....2S} {275, 2}

\bibitem[\protect\citeauthoryear{Spearman}{Spearman}{1904}]{Spearman1904}
Spearman C.,  1904, The American Journal of Psychology, 15, 72

\bibitem[\protect\citeauthoryear{{Stothers} \& {Chin}}{{Stothers} \& {Chin}}{1993}]{Stothers1993}
{Stothers} R.~B.,  {Chin} C.-W.,  1993, \mn@doi [\apjl] {10.1086/186837}, \href {https://ui.adsabs.harvard.edu/abs/1993ApJ...408L..85S} {408, L85}

\bibitem[\protect\citeauthoryear{{Sullivan} et~al.,}{{Sullivan} et~al.}{2015}]{Sullivan2015}
{Sullivan} P.~W.,  et~al., 2015, \mn@doi [\apj] {10.1088/0004-637X/809/1/77}, \href {https://ui.adsabs.harvard.edu/abs/2015ApJ...809...77S} {809, 77}

\bibitem[\protect\citeauthoryear{{Thompson}, {Herwig}, {Woodward}, {Mao}, {Denissenkov}, {Bowman}  \& {Blouin}}{{Thompson} et~al.}{2024}]{Thompson2024}
{Thompson} W.,  {Herwig} F.,  {Woodward} P.~R.,  {Mao} H.,  {Denissenkov} P.,  {Bowman} D.~M.,   {Blouin} S.,  2024, \mn@doi [\mnras] {10.1093/mnras/stae1162}, \href {https://ui.adsabs.harvard.edu/abs/2024MNRAS.531.1316T} {531, 1316}

\bibitem[\protect\citeauthoryear{{Tkachenko} et~al.,}{{Tkachenko} et~al.}{2014}]{Tkachenko2014}
{Tkachenko} A.,  et~al., 2014, \mn@doi [\mnras] {10.1093/mnras/stt2421}, \href {https://ui.adsabs.harvard.edu/abs/2014MNRAS.438.3093T} {438, 3093}

\bibitem[\protect\citeauthoryear{{Toutain} \& {Appourchaux}}{{Toutain} \& {Appourchaux}}{1994}]{Toutain1994}
{Toutain} T.,  {Appourchaux} T.,  1994, \aap, \href {https://ui.adsabs.harvard.edu/abs/1994A&A...289..649T} {289, 649}

\bibitem[\protect\citeauthoryear{{Townsend}}{{Townsend}}{1966}]{Townsend1966}
{Townsend} A.~A.,  1966, \mn@doi [Journal of Fluid Mechanics] {10.1017/S0022112066000661}, \href {https://ui.adsabs.harvard.edu/abs/1966JFM....24..307T} {24, 307}

\bibitem[\protect\citeauthoryear{{Varghese}, {Ratnasingam}, {Vanon}, {Edelmann}  \& {Rogers}}{{Varghese} et~al.}{2023}]{Varghese2023}
{Varghese} A.,  {Ratnasingam} R.~P.,  {Vanon} R.,  {Edelmann} P.~V.~F.,   {Rogers} T.~M.,  2023, \mn@doi [\apj] {10.3847/1538-4357/aca092}, \href {https://ui.adsabs.harvard.edu/abs/2023ApJ...942...53V} {942, 53}

\bibitem[\protect\citeauthoryear{Virtanen et~al.,}{Virtanen et~al.}{2020}]{scipy}
Virtanen P.,  et~al., 2020, \mn@doi [Nature Methods] {10.1038/s41592-019-0686-2}, \href {https://rdcu.be/b08Wh} {17, 261}

\bibitem[\protect\citeauthoryear{{W}es {M}c{K}inney}{{W}es {M}c{K}inney}{2010}]{mckinney-proc-scipy-2010}
{W}es {M}c{K}inney 2010, in {S}t\'efan van~der {W}alt {J}arrod {M}illman eds, {P}roceedings of the 9th {P}ython in {S}cience {C}onference. pp 56 -- 61, \mn@doi{10.25080/Majora-92bf1922-00a}

\bibitem[\protect\citeauthoryear{{Zhang}, {Ren}, {Jiang}, {Soszynski}  \& {Jayasinghe}}{{Zhang} et~al.}{2024}]{Zhang2024}
{Zhang} Z.,  {Ren} Y.,  {Jiang} B.,  {Soszynski} I.,   {Jayasinghe} T.,  2024, \mn@doi [arXiv e-prints] {10.48550/arXiv.2405.01405}, \href {https://ui.adsabs.harvard.edu/abs/2024arXiv240501405Z} {p. arXiv:2405.01405}

\bibitem[\protect\citeauthoryear{pandas~development team}{pandas~development team}{2020}]{reback2020pandas}
pandas~development team T.,  2020, pandas-dev/pandas: Pandas, \mn@doi{10.5281/zenodo.3509134}, \url {https://doi.org/10.5281/zenodo.3509134}

\makeatother
\end{thebibliography}

%%%%%%%%%%%%%%%%%%%%%%%%%%%%%%%%%%%%%%%%%%%%%%%%%%

%%%%%%%%%%%%%%%%% APPENDICES %%%%%%%%%%%%%%%%%%%%%

\appendix

\section{Overview of and fitting results for the Cyg~OB and SMC sample}\label{App:sample}

An overview of the sample fo 49 SLF variables found across the six Cygnus OB associations including their name (Gaia EDR3 and TIC ID), association group, TESS (T) and \emph{Gaia} (G) magnitudes, effective temperatures, luminosities and masses, as well as the available TESS sectors with FFI and 2-min cadence data is provided in Table~\ref{tab:sample_overview}. A similar overview of the six SMC stars from \citet{Bouret2021} that are found to either an SLF variable, SLF candidate or not variable is provided in Table~\ref{tab:sample_overview_smc}. The corresponding derived SLF parameters $\alpha_0$, $\nu_{\rm char}$, $\gamma$, $C_W$, RMS, $\nu_{50\%}$, and $w$ are provided in Table~\ref{tab:results_averages} along with the derived spectroscopic luminosity $\log \mathcal{L}/\mathcal{L}_\odot$.

\begin{table*}
    \scriptsize
	\centering
	\caption{
 The sample of 49 O- and B-type stars showing SLF variability and the available TESS data listed in the order of their Gaia EDR3 IDs. The first two columns list the Gaia EDR3 and TIC IDs. Group denotes which of the six Cygnus OB associations identified by \citet{Quintana2021} the star belongs to, while $T$ and $G$ are the TESS and Gaia G-band magnitudes. The effective temperature $\log T_{\rm eff}$, luminosity $\log L$, and mass $M$ are from \citet{Quintana2021}. The last two columns list the number of sectors with FFI data (FFI sectors) and 2-min cadence data (2-min sectors) available.}
	\label{tab:sample_overview}
	\begin{tabular}{cccccccccc} % four columns, alignment for each
		\hline
		Gaia EDR3 & TIC & Group & $T$ & $G$ & $\log T_{\rm eff}$ & $\log L$ & $M$ & FFI sectors & 2-min sectors\\
      &  &  & [mag] & [mag] & [K] & [L$_\odot$]  & [M$_\odot$] &  & \\
		\hline
		 2057949905056008960 	&	 13980753 	&	 D 	&	 8.91 	&	 9.66 	&	 $4.55^{0.06}_{0.13}$ 	&	 $5.01^{0.23}_{0.20}$ 	&	 $21.43^{11.08}_{10.64}$ 	&	 14,15,41,55,75,82 	&	 \\[0.75ex]
		 2057965916694269312 	&	 13252071 	&	 C 	&	 10.24 	&	 10.94 	&	 $4.44^{0.02}_{0.03}$ 	&	 $4.07^{0.05}_{0.06}$ 	&	 $11.89^{1.09}_{1.65}$ 	&	 14,15,41,55,75,82 	&	 41,55,75,82\\[0.75ex]
		 2057970692697712896 	&	 13980437 	&	 D 	&	 10.07 	&	 10.72 	&	 $4.41^{0.04}_{0.08}$ 	&	 $4.04^{0.11}_{0.08}$ 	&	 $10.05^{2.00}_{2.77}$ 	&	 14,15,41,55,75,82 	&	 \\[0.75ex]
		 2058985262750809088 	&	 43417017 	&	 F 	&	 7.71 	&	 7.88 	&	 $4.62^{0.06}_{0.17}$ 	&	 $5.33^{0.50}_{0.43}$ 	&	 $33.19^{29.18}_{21.25}$ 	&	 14,15,41,54,55,75,81,82 	&	 \\[0.75ex]
		 2059014782091284608 	&	 89751848 	&	 A 	&	 10.82 	&	 10.89 	&	 $4.31^{0.03}_{0.02}$ 	&	 $3.32^{0.06}_{0.08}$ 	&	 $6.68^{0.97}_{0.52}$ 	&	 14,15,41,54,55,74,75,81,82 	&	 \\[0.75ex]
		 2059070135632404992 	&	 89753650 	&	 A 	&	 7.61 	&	 7.68 	&	 $4.58^{0.02}_{0.04}$ 	&	 $5.00^{0.07}_{0.12}$ 	&	 $25.18^{4.13}_{5.50}$ 	&	 14,15,41,54,55,74,75,81,82 	&	 41,54,55,74,75,81,82\\[0.75ex]
		 2059072197216667776 	&	 89758873 	&	 A 	&	 7.05 	&	 7.19 	&	 $4.56^{0.05}_{0.14}$ 	&	 $5.11^{0.28}_{0.32}$ 	&	 $22.59^{10.75}_{11.68}$ 	&	 14,15,41,54,55,74,75,81,82 	&	 41,54,55,74,75,81,82\\[0.75ex]
		 2059076148587289728 	&	 89757859 	&	 A 	&	 6.67 	&	 6.73 	&	 $4.44^{0.15}_{0.03}$ 	&	 $5.11^{0.22}_{0.05}$ 	&	 $11.89^{15.40}_{1.34}$ 	&	 14,15,41,54,55,74,75,81,82 	&	 \\[0.75ex]
		 2059076251666505728 	&	 89757779 	&	 A 	&	 7.15 	&	 7.28 	&	 $4.46^{0.09}_{0.07}$ 	&	 $4.71^{0.18}_{0.12}$ 	&	 $12.74^{8.45}_{3.55}$ 	&	 14,15,41,54,55,74,75,81,82 	&	 41,54,55\\[0.75ex]
		 2059129303105142144 	&	 41192804 	&	 F 	&	 9.36 	&	 9.59 	&	 $4.34^{0.14}_{0.02}$ 	&	 $4.14^{0.17}_{0.05}$ 	&	 $7.43^{6.83}_{0.70}$ 	&	 14,15,41,54,55,74,75,81 	&	 \\[0.75ex]
		 2059130368252069888 	&	 378273410 	&	 F 	&	 7.51 	&	 7.70 	&	 $4.49^{0.07}_{0.09}$ 	&	 $4.74^{0.16}_{0.19}$ 	&	 $14.89^{6.88}_{5.01}$ 	&	 14,15,41,54,55,75,81,82 	&	 14,15,41,54,55,75,81,82\\[0.75ex]
		 2059150640497117952 	&	 42256981 	&	 F 	&	 9.24 	&	 9.38 	&	 $4.44^{0.01}_{0.02}$ 	&	 $4.06^{0.04}_{0.04}$ 	&	 $11.69^{0.84}_{0.91}$ 	&	 14,15,41,54,55,75,81,82 	&	 41,54,55,75,81,82\\[0.75ex]
		 2059223002113939840 	&	 41315581 	&	 F 	&	 7.79 	&	 7.90 	&	 $4.46^{0.10}_{0.04}$ 	&	 $5.01^{0.22}_{0.13}$ 	&	 $12.94^{9.29}_{2.08}$ 	&	 14,15,41,54,55,75,81 	&	 41,54,55,75,81\\[0.75ex]
		 2060508777869902592 	&	 10989276 	&	 F 	&	 7.52 	&	 7.75 	&	 $4.59^{0.07}_{0.05}$ 	&	 $5.12^{0.50}_{0.21}$ 	&	 $27.04^{23.66}_{6.76}$ 	&	 14,15,41,54,55,75,81,82 	&	 \\[0.75ex]
		 2060669306564757888 	&	 12017514 	&	 D 	&	 7.34 	&	 7.53 	&	 $4.43^{0.11}_{0.06}$ 	&	 $4.59^{0.17}_{0.10}$ 	&	 $11.12^{8.84}_{2.55}$ 	&	 14,15,41,55,75,81 	&	 41,55,75,81,82\\[0.75ex]
		 2060968682953119872 	&	 13252343 	&	 C 	&	 9.66 	&	 10.41 	&	 $4.47^{0.04}_{0.09}$ 	&	 $4.43^{0.10}_{0.12}$ 	&	 $13.34^{3.38}_{4.52}$ 	&	 14,15,41,55,75,81,82 	&	 55,75,81,82\\[0.75ex]
		 2061008780768493440 	&	 13778939 	&	 C 	&	 9.76 	&	 10.60 	&	 $4.43^{0.06}_{0.08}$ 	&	 $4.30^{0.14}_{0.14}$ 	&	 $11.09^{4.08}_{3.36}$ 	&	 14,15,41,55,81,82 	&	 \\[0.75ex]
		 2061009742841196160 	&	 13554172 	&	 C 	&	 9.59 	&	 10.25 	&	 $4.48^{0.02}_{0.03}$ 	&	 $4.41^{0.05}_{0.06}$ 	&	 $14.55^{1.63}_{2.05}$ 	&	 14,15,41,55,75,81,82 	&	 \\[0.75ex]
		 2061014342741333504 	&	 13327889 	&	 C 	&	 9.15 	&	 9.63 	&	 $4.42^{0.07}_{0.10}$ 	&	 $4.28^{0.13}_{0.17}$ 	&	 $10.84^{4.02}_{3.66}$ 	&	 14,15,41,55,75,81,82 	&	 \\[0.75ex]
		 2061020119482237056 	&	 13328752 	&	 C 	&	 8.80 	&	 9.48 	&	 $4.47^{0.06}_{0.08}$ 	&	 $4.47^{0.17}_{0.13}$ 	&	 $13.24^{4.95}_{3.87}$ 	&	 14,15,41,55,75,81,82 	&	 \\[0.75ex]
		 2061129727045269248 	&	 14206948 	&	 C 	&	 8.58 	&	 9.09 	&	 $4.63^{0.03}_{0.05}$ 	&	 $5.52^{0.22}_{0.39}$ 	&	 $39.26^{12.50}_{13.80}$ 	&	 14,15,41,55,81 	&	 41,55,81,82\\[0.75ex]
		 2062343725324664064 	&	 274081363 	&	 B 	&	 10.12 	&	 11.17 	&	 $4.43^{0.07}_{0.09}$ 	&	 $4.33^{0.16}_{0.16}$ 	&	 $10.96^{4.85}_{3.62}$ 	&	 14,15,41,55,81,82 	&	 \\[0.75ex]
		 2062546275984196224 	&	 11798617 	&	 B 	&	 8.07 	&	 8.49 	&	 $4.42^{0.15}_{0.04}$ 	&	 $4.72^{0.28}_{0.07}$ 	&	 $10.64^{12.43}_{1.58}$ 	&	 14,15,41,55,81,82 	&	 14,15,41,55,81,82\\[0.75ex]
		 2062549471439722624 	&	 11345781 	&	 B 	&	 8.37 	&	 9.06 	&	 $4.50^{0.07}_{0.10}$ 	&	 $4.71^{0.18}_{0.16}$ 	&	 $16.22^{7.50}_{6.17}$ 	&	 14,15,41,55,81,82 	&	 \\[0.75ex]
		 2064739041458261120 	&	 63824831 	&	 E 	&	 9.56 	&	 10.73 	&	 $4.66^{0.03}_{0.05}$ 	&	 $5.78^{0.26}_{0.37}$ 	&	 $52.72^{19.39}_{21.03}$ 	&	 14,15,41,55,75 	&	 \\[0.75ex]
		 2064871124590004096 	&	 64418011 	&	 E 	&	 15.96 	&	 17.15 	&	 $4.03^{0.03}_{0.03}$ 	&	 $1.71^{0.19}_{0.11}$ 	&	 $2.47^{0.27}_{0.24}$ 	&	 14,15,41,55,56,75,76,82 	&	 \\[0.75ex]
		 2067625637438421248 	&	 15893610 	&	 E 	&	 8.94 	&	 10.30 	&	 $4.62^{0.07}_{0.09}$ 	&	 $5.60^{0.43}_{0.50}$ 	&	 $33.34^{37.94}_{15.27}$ 	&	 14,15,41,55,75,82 	&	 \\[0.75ex]
		 2067745076187793024 	&	 17031739 	&	 E 	&	 9.42 	&	 10.63 	&	 $4.61^{0.07}_{0.09}$ 	&	 $5.44^{0.48}_{0.26}$ 	&	 $30.13^{30.54}_{13.38}$ 	&	 14,15,41,55,75 	&	 \\[0.75ex]
		 2067780947755025920 	&	 17449258 	&	 E 	&	 8.37 	&	 9.67 	&	 $4.59^{0.03}_{0.11}$ 	&	 $5.69^{0.14}_{0.21}$ 	&	 $27.23^{7.45}_{13.07}$ 	&	 14,15,41,55,75 	&	 41,55,75\\[0.75ex]
		 2067783623515353728 	&	 17449204 	&	 E 	&	 8.32 	&	 9.61 	&	 $4.56^{0.11}_{0.11}$ 	&	 $5.53^{0.37}_{0.29}$ 	&	 $21.98^{32.72}_{9.56}$ 	&	 14,15,41,55,75 	&	 41,55,75\\[0.75ex]
		 2067784246289931776 	&	 17450158 	&	 E 	&	 8.55 	&	 9.48 	&	 $4.63^{0.05}_{0.14}$ 	&	 $5.46^{0.42}_{0.48}$ 	&	 $39.90^{28.80}_{24.23}$ 	&	 14,15,41,55,56,75,76 	&	 41,55,75\\[0.75ex]
		 2067784516868550016 	&	 17126660 	&	 E 	&	 10.49 	&	 11.77 	&	 $4.34^{0.11}_{0.56}$ 	&	 $4.15^{0.28}_{0.96}$ 	&	 $7.71^{5.06}_{6.55}$ 	&	 14,15,41,55,75 	&	 \\[0.75ex]
		 2067785070923663104 	&	 17450609 	&	 E 	&	 8.75 	&	 9.74 	&	 $4.55^{0.07}_{0.11}$ 	&	 $5.08^{0.34}_{0.20}$ 	&	 $21.48^{15.59}_{9.34}$ 	&	 14,15,56,75,76 	&	 41,55,75\\[0.75ex]
		 2067793351620401024 	&	 16398592 	&	 E 	&	 10.83 	&	 12.23 	&	 $4.43^{0.08}_{0.07}$ 	&	 $4.55^{0.09}_{0.09}$ 	&	 $11.43^{5.44}_{2.88}$ 	&	 14,15,41,55,75,82 	&	 \\[0.75ex]
		 2067807885789820160 	&	 15987627 	&	 E 	&	 10.74 	&	 12.12 	&	 $4.50^{0.05}_{0.09}$ 	&	 $4.66^{0.12}_{0.14}$ 	&	 $15.45^{5.49}_{5.34}$ 	&	 14,15,41,55,75,82 	&	 \\[0.75ex]
		 2067813005390784896 	&	 16726220 	&	 E 	&	 8.77 	&	 10.09 	&	 $4.66^{0.03}_{0.13}$ 	&	 $5.77^{0.24}_{0.52}$ 	&	 $51.40^{20.71}_{32.35}$ 	&	 14,15,41,55,75,82 	&	 \\[0.75ex]
		 2067827466541470080 	&	 17034121 	&	 E 	&	 10.82 	&	 12.05 	&	 $4.45^{0.05}_{0.10}$ 	&	 $4.35^{0.11}_{0.12}$ 	&	 $12.33^{3.81}_{4.26}$ 	&	 14,15,41,55,75 	&	 \\[0.75ex]
		 2067832624801783040 	&	 17125844 	&	 E 	&	 10.32 	&	 11.24 	&	 $4.40^{0.06}_{0.08}$ 	&	 $4.15^{0.09}_{0.10}$ 	&	 $9.77^{3.41}_{2.73}$ 	&	 14,15,41,55,56,75,76 	&	 41,55,56,75,76\\[0.75ex]
		 2067834926904094848 	&	 17125591 	&	 E 	&	 9.77 	&	 10.57 	&	 $4.34^{0.06}_{0.02}$ 	&	 $4.21^{0.09}_{0.04}$ 	&	 $7.59^{2.12}_{0.49}$ 	&	 14,15,41,55,56,75 	&	 \\[0.75ex]
		 2067835614098871040 	&	 17035378 	&	 E 	&	 10.07 	&	 10.91 	&	 $4.39^{0.08}_{0.06}$ 	&	 $4.27^{0.12}_{0.11}$ 	&	 $9.27^{4.04}_{2.21}$ 	&	 14,15,41,55,75 	&	 \\[0.75ex]
		 2067837263366481536 	&	 17125275 	&	 E 	&	 10.31 	&	 10.98 	&	 $4.43^{0.02}_{0.05}$ 	&	 $4.01^{0.07}_{0.08}$ 	&	 $10.96^{1.20}_{2.24}$ 	&	 14,15,41,75,76 	&	 41,75\\[0.75ex]
		 2067840149584105344 	&	 16733810 	&	 E 	&	 10.46 	&	 11.76 	&	 $4.47^{0.05}_{0.10}$ 	&	 $4.49^{0.12}_{0.14}$ 	&	 $13.68^{3.86}_{5.01}$ 	&	 14,15,41,55,75 	&	 \\[0.75ex]
		 2067887840900094848 	&	 63457037 	&	 E 	&	 10.13 	&	 11.13 	&	 $4.42^{0.07}_{0.08}$ 	&	 $4.34^{0.12}_{0.11}$ 	&	 $10.64^{4.08}_{2.97}$ 	&	 14,15,41,55,56,75,76,82 	&	 14,15,41,55,56,75,76,82\\[0.75ex]
		 2067888218857234304 	&	 63456866 	&	 E 	&	 8.24 	&	 9.24 	&	 $4.58^{0.04}_{0.10}$ 	&	 $5.53^{0.19}_{0.32}$ 	&	 $26.12^{9.94}_{11.60}$ 	&	 14,15,41,55,56,75,76,82 	&	 41,55,56,75,76,82\\[0.75ex]
		 2068008164405551104 	&	 16315781 	&	 E 	&	 11.41 	&	 12.89 	&	 $4.48^{0.04}_{0.07}$ 	&	 $4.45^{0.07}_{0.16}$ 	&	 $14.00^{3.22}_{3.81}$ 	&	 14,15,41,55,75 	&	 \\[0.75ex]
		 2068074620437883520 	&	 15556616 	&	 E 	&	 9.95 	&	 11.16 	&	 $4.47^{0.11}_{0.08}$ 	&	 $4.57^{0.49}_{0.23}$ 	&	 $13.46^{11.26}_{4.23}$ 	&	 14,15,41,55,75,82 	&	 \\[0.75ex]
		 2068619802109825280 	&	 12360661 	&	 B 	&	 11.46 	&	 12.73 	&	 $4.38^{0.07}_{0.08}$ 	&	 $4.10^{0.16}_{0.18}$ 	&	 $9.06^{3.19}_{2.44}$ 	&	 14,15,41,55,75,81,82 	&	 \\[0.75ex]
		 2068698765090280704 	&	 405944953 	&	 B 	&	 10.86 	&	 11.29 	&	 $4.38^{0.02}_{0.11}$ 	&	 $3.71^{0.05}_{0.22}$ 	&	 $8.93^{0.82}_{3.37}$ 	&	 14,15,41,75,81,82 	&	 \\[0.75ex]
		 2074699040563479936 	&	 43073133 	&	 B 	&	 8.61 	&	 8.69 	&	 $4.44^{0.02}_{0.02}$ 	&	 $4.15^{0.05}_{0.03}$ 	&	 $11.59^{0.97}_{1.02}$ 	&	 14,15,41,54,55,75,81,82 	&	 75,81,82\\[0.75ex]
		\hline
	\end{tabular}
\end{table*}

\begin{table*}
\scriptsize
	\centering
	\caption{
 Sample of O-type stars in the SMC from \citet{Bouret2021} identified as either an SLF variable, SLF candidate, or showing only white noise (None). The columns show the name of the star, TIC ID, TESS magnitude $T$, Gaia G-band magnitude $G$, effective temperature $\log T_{\rm eff}$, bolometric luminosity $\log L$, spectroscopic mass $M_{\rm spec}$, the number of sectors with TESS FFI data and 2-min cadence data, and variability type. Stellar parameters are taken from \citet{Bouret2021}.}
	\label{tab:sample_overview_smc}
	\begin{tabular}{cccccccccc} % four columns, alignment for each
		\hline
		Star name & TIC & $T$ & $G$ & $\log T_{\rm eff}$ & $\log L$ & $M_{\rm spec}$ & FFI sectors & 2-min sectors & Variability\\
      &  & [mag] & [mag] & [K] & [L$_\odot$]  & [M$_\odot$] &  & \\
		\hline
		 AV 232 	&	182294089   &	11.23  	&	12.26  	&	$4.53\pm0.01$  	&	$5.89\pm 0.1$  	&	 $35.3\pm 8.2$ 	&	 1,28,67,68 	&	 67,68   & SLF\\[0.5ex]
		 AV 43 	&	180610340   &	14.29 &	13.94  	&	$4.45\pm0.02$  	&	$5.13\pm 0.1$  	&	 $22.4\pm 11.0$ 	&	 1,27,28,67,68 	&	    & SLF candidate\\[0.5ex]
		 AV 83 	&	181043369   &	13.83 &	13.50  	&	$4.52\pm0.01$  	&	$5.54\pm 0.1$  	&	 $22.1\pm 8.3$ 	&	 1,27,67,68 	&	    & SLF candidate\\[0.5ex]
		 AV 327 	&	182730274   &	12.45 &	13.28  	&	$4.48\pm0.01$  	&	$5.54\pm 0.1$  	&	 $22.8\pm 5.8$ 	&	1,2,28,67,68  	&	    & SLF candidate\\[0.5ex]
		 AV 77 	&	181051294   &	13.10 &	13.91  	&	$4.57\pm0.01$  	&	$5.40\pm 0.1$  	&	 $28.0\pm 12.2$ 	&	 1,27,28,67,68 	&	    & None\\[0.5ex]
		 AV 307 	&	182517696   &	14.32 &	14.00  	&	$4.48\pm0.01$  	&	$5.15\pm 0.1$  	&	 $22.5\pm 13.2$ 	&	 1,28,67,68 	&	    & None\\[0.5ex]
		\hline
	\end{tabular}
\end{table*}

\begin{table*}
\scriptsize
	\centering
	\caption{Average parameter estimates for the sample of 49 O- and B-type stars in the six Cygnus OB associations group A-F and the SMC star AV~232. The averages are calculated across all sectors and observing cadences (excluding FFI data from cycle 1) and listed in this table. The errors cover the range in the estimated parameters across all sectors for a given star. We also list the spectroscopic luminosities $ \frac{\mathcal{L}}{\mathcal{L}_\odot} = \frac{T_{\rm eff}^4}{g} \left(\frac{T_{{\rm eff},\odot}^4}{g_\odot}\right)^{-1} = \frac{L}{L_\odot} \left(\frac{M}{M_\odot}\right)^{-1}$ required for comparison to the sample from \citet{Bowman2020}.}
	\label{tab:results_averages}
	\begin{tabular}{ccccccccc} % four columns, alignment for each
		\hline
		Gaia EDR3 & $\log \mathcal{L}$ & $\log \alpha_0$ & $\nu_{\rm char}$  & $\gamma$ & $\log C_W$ & $\log {\rm RMS}$ & $\nu_{50\%}$ & $w$\\
       & [$\mathcal{L}_\odot$] & [ppm$^2 \mu$Hz$^{-1}$] & [$\mu{\rm Hz}$] &  & [ppm$^2 \mu$Hz$^{-1}$] & [ppm] & [$\mu{\rm Hz}$] & \\
		\hline
		 2057949905056008960 	&	 $3.68^{0.28}_{0.37}$ 	&	 $6.52^{0.26}_{0.16}$ 	&	 $12.67^{1.51}_{1.40}$ 	&	 $3.15^{0.14}_{0.30}$ 	&	 $3.19^{0.73}_{0.43}$ 	&	 $3.72^{0.09}_{0.06}$ 	&	 $6.82^{2.20}_{2.61}$ 	&	 $2.53^{0.74}_{0.75}$\\[1.5ex]
		 2057965916694269312 	&	 $3.00^{0.07}_{0.09}$ 	&	 $4.65^{0.58}_{0.65}$ 	&	 $3.12^{2.98}_{1.55}$ 	&	 $1.72^{0.36}_{0.59}$ 	&	 $2.95^{0.07}_{0.05}$ 	&	 $3.04^{0.12}_{0.19}$ 	&	 $304.15^{69.54}_{93.24}$ 	&	 $1.97^{0.72}_{0.49}$\\[1.5ex]
		 2057970692697712896 	&	 $3.03^{0.12}_{0.17}$ 	&	 $5.56^{0.57}_{0.41}$ 	&	 $2.86^{2.04}_{1.65}$ 	&	 $1.70^{0.09}_{0.09}$ 	&	 $2.70^{0.05}_{0.05}$ 	&	 $3.02^{0.07}_{0.06}$ 	&	 $12.19^{5.24}_{5.86}$ 	&	 $17.95^{6.72}_{4.65}$\\[1.5ex]
		 2058985262750809088 	&	 $3.81^{0.54}_{0.64}$ 	&	 $5.56^{0.14}_{0.21}$ 	&	 $15.37^{3.29}_{4.18}$ 	&	 $2.29^{0.12}_{0.15}$ 	&	 $2.12^{0.14}_{0.21}$ 	&	 $3.30^{0.07}_{0.06}$ 	&	 $11.41^{3.15}_{2.70}$ 	&	 $2.85^{0.91}_{0.47}$\\[1.5ex]
		 2059014782091284608 	&	 $2.49^{0.09}_{0.08}$ 	&	 $5.78^{0.31}_{0.19}$ 	&	 $12.15^{4.46}_{6.01}$ 	&	 $3.82^{0.68}_{1.54}$ 	&	 $3.22^{0.06}_{0.05}$ 	&	 $3.36^{0.06}_{0.09}$ 	&	 $9.56^{1.86}_{1.50}$ 	&	 $2.71^{1.74}_{1.82}$\\[1.5ex]
		 2059070135632404992 	&	 $3.60^{0.12}_{0.14}$ 	&	 $5.49^{0.12}_{0.22}$ 	&	 $37.56^{11.27}_{5.32}$ 	&	 $3.42^{0.50}_{0.34}$ 	&	 $2.12^{0.37}_{0.25}$ 	&	 $3.42^{0.05}_{0.06}$ 	&	 $14.89^{7.08}_{5.79}$ 	&	 $2.73^{1.29}_{0.68}$\\[1.5ex]
		 2059072197216667776 	&	 $3.76^{0.34}_{0.43}$ 	&	 $7.12^{0.78}_{0.43}$ 	&	 $6.07^{5.28}_{4.69}$ 	&	 $2.76^{0.76}_{0.56}$ 	&	 $2.13^{0.98}_{0.31}$ 	&	 $3.84^{0.07}_{0.10}$ 	&	 $5.09^{1.44}_{1.68}$ 	&	 $1.80^{0.65}_{0.42}$\\[1.5ex]
		 2059076148587289728 	&	 $4.03^{0.39}_{0.15}$ 	&	 $6.39^{0.23}_{0.34}$ 	&	 $11.01^{6.58}_{4.58}$ 	&	 $2.60^{0.31}_{0.50}$ 	&	 $2.87^{0.18}_{0.32}$ 	&	 $3.64^{0.17}_{0.07}$ 	&	 $6.17^{3.92}_{3.15}$ 	&	 $3.05^{1.39}_{0.67}$\\[1.5ex]
		 2059076251666505728 	&	 $3.61^{0.27}_{0.21}$ 	&	 $6.69^{0.27}_{0.30}$ 	&	 $9.83^{6.03}_{4.61}$ 	&	 $3.28^{0.79}_{0.70}$ 	&	 $2.95^{0.26}_{0.29}$ 	&	 $3.73^{0.04}_{0.12}$ 	&	 $4.62^{1.81}_{1.27}$ 	&	 $2.13^{0.64}_{0.29}$\\[1.5ex]
		 2059129303105142144 	&	 $3.27^{0.31}_{0.12}$ 	&	 $5.97^{0.54}_{0.57}$ 	&	 $4.55^{3.27}_{3.08}$ 	&	 $2.92^{0.62}_{0.94}$ 	&	 $2.59^{0.04}_{0.06}$ 	&	 $3.17^{0.13}_{0.12}$ 	&	 $5.27^{2.93}_{1.54}$ 	&	 $1.41^{0.46}_{0.82}$\\[1.5ex]
		 2059130368252069888 	&	 $3.56^{0.24}_{0.25}$ 	&	 $5.26^{0.44}_{0.35}$ 	&	 $20.60^{6.35}_{9.27}$ 	&	 $2.87^{0.36}_{0.40}$ 	&	 $2.09^{0.37}_{0.23}$ 	&	 $3.21^{0.14}_{0.11}$ 	&	 $8.66^{5.33}_{2.56}$ 	&	 $2.96^{1.59}_{0.58}$\\[1.5ex]
		 2059150640497117952 	&	 $2.99^{0.05}_{0.05}$ 	&	 $3.82^{0.72}_{0.77}$ 	&	 $7.07^{9.45}_{5.49}$ 	&	 $1.78^{0.77}_{0.61}$ 	&	 $2.47^{0.09}_{0.06}$ 	&	 $2.80^{0.11}_{0.23}$ 	&	 $318.36^{69.37}_{82.65}$ 	&	 $1.83^{0.62}_{0.46}$\\[1.5ex]
		 2059223002113939840 	&	 $3.90^{0.30}_{0.19}$ 	&	 $5.99^{0.33}_{0.16}$ 	&	 $17.96^{4.92}_{8.41}$ 	&	 $3.14^{0.37}_{0.57}$ 	&	 $2.23^{0.39}_{0.23}$ 	&	 $3.51^{0.07}_{0.06}$ 	&	 $7.31^{1.07}_{2.67}$ 	&	 $2.57^{1.11}_{0.37}$\\[1.5ex]
		 2060508777869902592 	&	 $3.69^{0.46}_{0.38}$ 	&	 $5.51^{0.02}_{0.03}$ 	&	 $22.29^{3.34}_{2.04}$ 	&	 $3.12^{0.27}_{0.18}$ 	&	 $2.19^{0.17}_{0.11}$ 	&	 $3.32^{0.02}_{0.03}$ 	&	 $16.21^{2.60}_{3.94}$ 	&	 $1.40^{0.54}_{0.38}$\\[1.5ex]
		 2060669306564757888 	&	 $3.54^{0.29}_{0.18}$ 	&	 $6.99^{0.26}_{0.23}$ 	&	 $8.48^{2.44}_{2.51}$ 	&	 $3.10^{0.47}_{0.19}$ 	&	 $2.26^{0.24}_{0.18}$ 	&	 $3.88^{0.08}_{0.06}$ 	&	 $5.72^{1.25}_{1.52}$ 	&	 $1.83^{1.10}_{0.49}$\\[1.5ex]
		 2060968682953119872 	&	 $3.31^{0.15}_{0.21}$ 	&	 $5.61^{0.69}_{0.40}$ 	&	 $9.16^{6.20}_{6.51}$ 	&	 $2.11^{0.25}_{0.19}$ 	&	 $2.92^{0.30}_{0.24}$ 	&	 $3.28^{0.13}_{0.15}$ 	&	 $11.09^{5.73}_{6.19}$ 	&	 $6.71^{9.23}_{3.85}$\\[1.5ex]
		 2061008780768493440 	&	 $3.25^{0.19}_{0.21}$ 	&	 $6.03^{0.43}_{0.44}$ 	&	 $10.65^{5.56}_{3.66}$ 	&	 $2.81^{0.31}_{0.17}$ 	&	 $2.74^{0.16}_{0.06}$ 	&	 $3.43^{0.13}_{0.13}$ 	&	 $7.97^{4.62}_{3.44}$ 	&	 $2.50^{0.98}_{0.76}$\\[1.5ex]
		 2061009742841196160 	&	 $3.25^{0.07}_{0.09}$ 	&	 $5.15^{0.28}_{0.20}$ 	&	 $25.97^{7.64}_{10.24}$ 	&	 $2.61^{0.31}_{0.31}$ 	&	 $2.63^{0.21}_{0.14}$ 	&	 $3.24^{0.07}_{0.04}$ 	&	 $19.54^{2.44}_{3.63}$ 	&	 $2.56^{0.47}_{0.48}$\\[1.5ex]
		 2061014342741333504 	&	 $3.25^{0.20}_{0.23}$ 	&	 $5.77^{0.60}_{0.96}$ 	&	 $7.98^{12.51}_{5.71}$ 	&	 $2.22^{0.24}_{0.23}$ 	&	 $2.39^{0.14}_{0.11}$ 	&	 $3.20^{0.13}_{0.21}$ 	&	 $6.20^{13.96}_{4.02}$ 	&	 $4.90^{3.05}_{2.64}$\\[1.5ex]
		 2061020119482237056 	&	 $3.35^{0.21}_{0.22}$ 	&	 $5.98^{0.24}_{0.27}$ 	&	 $20.27^{5.51}_{5.31}$ 	&	 $3.25^{0.15}_{0.27}$ 	&	 $2.36^{0.19}_{0.14}$ 	&	 $3.55^{0.05}_{0.06}$ 	&	 $8.14^{4.46}_{3.27}$ 	&	 $2.34^{0.51}_{0.33}$\\[1.5ex]
		 2061129727045269248 	&	 $3.92^{0.33}_{0.36}$ 	&	 $5.65^{0.10}_{0.11}$ 	&	 $29.77^{1.65}_{2.23}$ 	&	 $3.61^{0.27}_{0.27}$ 	&	 $2.73^{0.39}_{0.26}$ 	&	 $3.47^{0.06}_{0.04}$ 	&	 $15.76^{0.77}_{1.58}$ 	&	 $1.82^{0.32}_{0.16}$\\[1.5ex]
		 2062343725324664064 	&	 $3.29^{0.23}_{0.24}$ 	&	 $5.81^{0.21}_{0.19}$ 	&	 $14.30^{4.19}_{3.74}$ 	&	 $2.47^{0.14}_{0.09}$ 	&	 $2.82^{0.03}_{0.05}$ 	&	 $3.41^{0.03}_{0.03}$ 	&	 $9.81^{2.97}_{1.89}$ 	&	 $2.84^{0.36}_{0.28}$\\[1.5ex]
		 2062546275984196224 	&	 $3.70^{0.39}_{0.19}$ 	&	 $6.19^{0.20}_{0.14}$ 	&	 $17.16^{3.12}_{2.12}$ 	&	 $3.33^{0.18}_{0.22}$ 	&	 $2.24^{0.18}_{0.22}$ 	&	 $3.61^{0.07}_{0.04}$ 	&	 $8.03^{3.53}_{3.74}$ 	&	 $2.62^{1.03}_{0.85}$\\[1.5ex]
		 2062549471439722624 	&	 $3.50^{0.24}_{0.27}$ 	&	 $4.94^{0.07}_{0.08}$ 	&	 $46.72^{1.10}_{1.80}$ 	&	 $3.30^{0.11}_{0.08}$ 	&	 $2.09^{0.11}_{0.08}$ 	&	 $3.20^{0.02}_{0.04}$ 	&	 $21.64^{4.03}_{5.16}$ 	&	 $2.61^{0.73}_{0.42}$\\[1.5ex]
		 2064739041458261120 	&	 $4.06^{0.34}_{0.39}$ 	&	 $7.04^{0.06}_{0.12}$ 	&	 $9.76^{2.10}_{2.64}$ 	&	 $3.32^{0.36}_{0.57}$ 	&	 $2.81^{0.18}_{0.24}$ 	&	 $3.90^{0.04}_{0.02}$ 	&	 $6.63^{0.98}_{0.97}$ 	&	 $1.62^{0.18}_{0.17}$\\[1.5ex]
		 2064871124590004096 	&	 $1.31^{0.16}_{0.16}$ 	&	 $9.00^{1.27}_{1.00}$ 	&	 $5.96^{7.21}_{4.67}$ 	&	 $2.88^{0.48}_{0.54}$ 	&	 $6.05^{0.53}_{0.29}$ 	&	 $4.92^{0.19}_{0.33}$ 	&	 $2.91^{2.07}_{1.14}$ 	&	 $9.86^{16.31}_{8.89}$\\[1.5ex]
		 2067625637438421248 	&	 $4.07^{0.57}_{0.54}$ 	&	 $8.38^{0.57}_{0.40}$ 	&	 $2.29^{0.80}_{1.27}$ 	&	 $2.55^{0.25}_{0.25}$ 	&	 $2.68^{0.09}_{0.08}$ 	&	 $4.19^{0.08}_{0.06}$ 	&	 $3.18^{0.49}_{0.22}$ 	&	 $1.92^{0.75}_{0.30}$\\[1.5ex]
		 2067745076187793024 	&	 $3.96^{0.49}_{0.46}$ 	&	 $6.83^{0.21}_{0.19}$ 	&	 $17.80^{4.38}_{4.97}$ 	&	 $3.51^{0.30}_{0.29}$ 	&	 $2.58^{0.07}_{0.06}$ 	&	 $3.93^{0.02}_{0.02}$ 	&	 $5.88^{2.75}_{1.77}$ 	&	 $2.51^{0.70}_{0.76}$\\[1.5ex]
		 2067780947755025920 	&	 $4.26^{0.20}_{0.33}$ 	&	 $8.12^{0.70}_{0.14}$ 	&	 $1.91^{0.36}_{0.30}$ 	&	 $2.42^{0.59}_{0.20}$ 	&	 $2.49^{0.17}_{0.15}$ 	&	 $4.20^{0.06}_{0.13}$ 	&	 $3.53^{1.71}_{1.09}$ 	&	 $1.10^{0.11}_{0.31}$\\[1.5ex]
		 2067783623515353728 	&	 $4.19^{0.52}_{0.41}$ 	&	 $6.31^{0.13}_{0.08}$ 	&	 $18.19^{2.31}_{5.12}$ 	&	 $3.61^{0.39}_{0.63}$ 	&	 $2.34^{0.17}_{0.11}$ 	&	 $3.67^{0.01}_{0.01}$ 	&	 $11.62^{1.27}_{0.78}$ 	&	 $1.66^{0.10}_{0.15}$\\[1.5ex]
		 2067784246289931776 	&	 $3.86^{0.51}_{0.60}$ 	&	 $5.92^{0.20}_{0.20}$ 	&	 $31.31^{10.03}_{6.09}$ 	&	 $4.28^{2.09}_{0.69}$ 	&	 $3.23^{0.66}_{0.33}$ 	&	 $3.62^{0.11}_{0.08}$ 	&	 $15.40^{2.20}_{1.52}$ 	&	 $2.14^{1.09}_{0.40}$\\[1.5ex]
		 2067784516868550016 	&	 $3.27^{0.67}_{1.07}$ 	&	 $5.21^{0.06}_{0.06}$ 	&	 $19.49^{2.31}_{2.70}$ 	&	 $2.35^{0.09}_{0.13}$ 	&	 $3.17^{0.05}_{0.09}$ 	&	 $3.25^{0.04}_{0.02}$ 	&	 $21.71^{0.59}_{0.85}$ 	&	 $7.80^{1.22}_{2.06}$\\[1.5ex]
		 2067785070923663104 	&	 $3.74^{0.36}_{0.37}$ 	&	 $5.62^{0.03}_{0.04}$ 	&	 $45.29^{5.66}_{7.26}$ 	&	 $4.33^{0.68}_{0.57}$ 	&	 $2.87^{0.54}_{0.47}$ 	&	 $3.53^{0.04}_{0.02}$ 	&	 $23.13^{2.43}_{4.18}$ 	&	 $1.70^{0.38}_{0.16}$\\[1.5ex]
		 2067793351620401024 	&	 $3.49^{0.19}_{0.15}$ 	&	 $5.57^{0.22}_{0.29}$ 	&	 $13.56^{5.06}_{2.48}$ 	&	 $2.28^{0.15}_{0.09}$ 	&	 $3.25^{0.20}_{0.17}$ 	&	 $3.35^{0.11}_{0.12}$ 	&	 $15.07^{3.88}_{3.93}$ 	&	 $6.41^{2.55}_{1.55}$\\[1.5ex]
		 2067807885789820160 	&	 $3.47^{0.19}_{0.23}$ 	&	 $5.41^{0.67}_{0.29}$ 	&	 $3.81^{1.58}_{1.76}$ 	&	 $1.49^{0.14}_{0.25}$ 	&	 $3.48^{0.72}_{0.38}$ 	&	 $3.26^{0.41}_{0.23}$ 	&	 $159.40^{124.35}_{94.03}$ 	&	 $4.70^{2.85}_{2.63}$\\[1.5ex]
		 2067813005390784896 	&	 $4.06^{0.41}_{0.59}$ 	&	 $8.16^{0.50}_{0.24}$ 	&	 $2.52^{1.34}_{1.27}$ 	&	 $2.60^{0.23}_{0.15}$ 	&	 $2.78^{0.48}_{0.32}$ 	&	 $4.12^{0.12}_{0.07}$ 	&	 $3.48^{0.74}_{1.07}$ 	&	 $1.61^{0.73}_{0.63}$\\[1.5ex]
		 2067827466541470080 	&	 $3.26^{0.16}_{0.22}$ 	&	 $4.49^{0.19}_{0.25}$ 	&	 $21.94^{8.73}_{8.86}$ 	&	 $2.27^{0.35}_{0.40}$ 	&	 $3.37^{0.41}_{0.22}$ 	&	 $3.14^{0.10}_{0.12}$ 	&	 $218.36^{113.75}_{73.95}$ 	&	 $2.81^{0.82}_{1.05}$\\[1.5ex]
		 2067832624801783040 	&	 $3.16^{0.16}_{0.17}$ 	&	 $4.77^{0.34}_{0.25}$ 	&	 $17.18^{11.65}_{10.61}$ 	&	 $2.33^{0.70}_{0.65}$ 	&	 $3.03^{0.20}_{0.12}$ 	&	 $3.17^{0.15}_{0.16}$ 	&	 $42.82^{32.90}_{20.21}$ 	&	 $10.46^{3.63}_{4.06}$\\[1.5ex]
		\hline
	\end{tabular}
\end{table*}

\begin{table*}
\scriptsize
	\centering
	\contcaption{}
	\begin{tabular}{cccccccccc} % four columns, alignment for each
		\hline
		Gaia EDR3 & $\log \mathcal{L}$ & $\log \alpha_0$ & $\nu_{\rm char}$  & $\gamma$ & $\log C_W$ & $\log {\rm RMS}$ & $\nu_{50\%}$ & $w$\\
       & [$\mathcal{L}_\odot$] & [ppm$^2 \mu$Hz$^{-1}$] & [$\mu{\rm Hz}$] &  & [ppm$^2 \mu$Hz$^{-1}$] & [ppm] & [$\mu{\rm Hz}$] & \\
		\hline
		 2067834926904094848 	&	 $3.33^{0.13}_{0.07}$ 	&	 $4.87^{0.14}_{0.14}$ 	&	 $19.24^{4.23}_{4.02}$ 	&	 $2.17^{0.06}_{0.18}$ 	&	 $2.76^{0.16}_{0.11}$ 	&	 $3.10^{0.09}_{0.05}$ 	&	 $23.80^{4.06}_{3.15}$ 	&	 $6.31^{2.81}_{1.87}$\\[1.5ex]
		 2067835614098871040 	&	 $3.31^{0.19}_{0.16}$ 	&	 $4.76^{0.12}_{0.18}$ 	&	 $22.16^{3.47}_{2.29}$ 	&	 $2.38^{0.04}_{0.02}$ 	&	 $2.82^{0.01}_{0.02}$ 	&	 $3.05^{0.07}_{0.07}$ 	&	 $29.31^{6.88}_{5.47}$ 	&	 $7.12^{1.33}_{1.40}$\\[1.5ex]
		 2067837263366481536 	&	 $2.97^{0.09}_{0.13}$ 	&	 $4.39^{0.59}_{1.23}$ 	&	 $6.94^{11.88}_{4.42}$ 	&	 $2.75^{2.23}_{1.26}$ 	&	 $3.01^{0.25}_{0.17}$ 	&	 $3.07^{0.14}_{0.30}$ 	&	 $306.51^{98.35}_{78.22}$ 	&	 $1.96^{0.58}_{0.65}$\\[1.5ex]
		 2067840149584105344 	&	 $3.36^{0.17}_{0.24}$ 	&	 $5.51^{0.05}_{0.09}$ 	&	 $11.62^{3.82}_{2.53}$ 	&	 $2.06^{0.21}_{0.19}$ 	&	 $3.23^{0.39}_{0.20}$ 	&	 $3.30^{0.03}_{0.04}$ 	&	 $20.95^{17.89}_{9.87}$ 	&	 $7.88^{3.01}_{2.26}$\\[1.5ex]
		 2067887840900094848 	&	 $3.31^{0.18}_{0.18}$ 	&	 $5.09^{0.45}_{0.24}$ 	&	 $20.52^{8.81}_{11.23}$ 	&	 $2.29^{0.43}_{0.39}$ 	&	 $2.92^{0.21}_{0.14}$ 	&	 $3.24^{0.11}_{0.13}$ 	&	 $21.22^{8.68}_{5.99}$ 	&	 $5.06^{2.43}_{1.11}$\\[1.5ex]
		 2067888218857234304 	&	 $4.11^{0.29}_{0.36}$ 	&	 $6.66^{0.24}_{0.24}$ 	&	 $16.80^{4.16}_{4.87}$ 	&	 $3.77^{0.67}_{0.51}$ 	&	 $2.29^{0.54}_{0.16}$ 	&	 $3.81^{0.06}_{0.08}$ 	&	 $11.99^{2.99}_{2.26}$ 	&	 $1.40^{0.35}_{0.27}$\\[1.5ex]
		 2068008164405551104 	&	 $3.30^{0.15}_{0.18}$ 	&	 $5.10^{0.12}_{0.14}$ 	&	 $10.53^{1.43}_{1.75}$ 	&	 $1.87^{0.04}_{0.06}$ 	&	 $3.52^{0.05}_{0.05}$ 	&	 $3.25^{0.11}_{0.09}$ 	&	 $140.62^{15.50}_{29.81}$ 	&	 $3.92^{0.75}_{0.38}$\\[1.5ex]
		 2068074620437883520 	&	 $3.44^{0.45}_{0.40}$ 	&	 $6.76^{0.21}_{0.19}$ 	&	 $14.24^{3.03}_{3.72}$ 	&	 $3.35^{0.38}_{0.32}$ 	&	 $2.81^{0.04}_{0.06}$ 	&	 $3.84^{0.07}_{0.06}$ 	&	 $8.32^{2.85}_{2.02}$ 	&	 $1.80^{0.44}_{0.34}$\\[1.5ex]
		 2068619802109825280 	&	 $3.14^{0.22}_{0.22}$ 	&	 $5.71^{0.17}_{0.16}$ 	&	 $6.62^{3.98}_{1.85}$ 	&	 $3.93^{2.97}_{1.30}$ 	&	 $3.61^{0.16}_{0.10}$ 	&	 $3.35^{0.07}_{0.09}$ 	&	 $48.29^{45.79}_{35.98}$ 	&	 $16.63^{22.94}_{10.89}$\\[1.5ex]
		 2068698765090280704 	&	 $2.76^{0.14}_{0.26}$ 	&	 $6.08^{0.06}_{0.07}$ 	&	 $9.98^{1.57}_{2.91}$ 	&	 $2.66^{0.22}_{0.30}$ 	&	 $3.13^{0.15}_{0.08}$ 	&	 $3.46^{0.07}_{0.10}$ 	&	 $9.87^{0.54}_{0.56}$ 	&	 $1.38^{0.68}_{0.37}$\\[1.5ex]
		 2074699040563479936 	&	 $3.09^{0.05}_{0.06}$ 	&	 $4.11^{0.57}_{0.43}$ 	&	 $4.45^{2.69}_{3.03}$ 	&	 $1.67^{0.20}_{0.21}$ 	&	 $2.25^{0.21}_{0.14}$ 	&	 $2.68^{0.16}_{0.15}$ 	&	 $200.59^{108.92}_{129.43}$ 	&	 $3.27^{3.90}_{1.39}$\\[1.5ex]
   \hline
   \textit{SMC - Star name}\\[1.5ex]
   %\hline
   %Star name\\[1.5ex]
   AV~232 	&	 $4.33^{0.11}_{0.11}$ 	&	 $7.69^{0.09}_{0.13}$ 	&	 $4.43^{1.73}_{0.65}$ 	&	 $3.07^{0.98}_{0.25}$ 	&	 $4.28^{0.03}_{0.04}$ 	&	 $4.06^{0.03}_{0.03}$ 	&	 $4.54^{0.09}_{0.13}$ 	&	 $1.49^{0.12}_{0.47}$\\[1.5ex]
		\hline
	\end{tabular}
\end{table*}

\section{Overview of and fitting results for the B20}\label{App:bowman_sample}

Similar to Appendix~\ref{App:sample}, an overview of the sample of 53 comparison OB stars from \cite{Bowman2020} known to be SLF variables is provided in Table~\ref{tab:bowman_sample_overview} and their corresponding derived SLF parameters are given in Table~\ref{tab:results_averages_bowman}.

\begin{table*}
\scriptsize
	\centering
    \caption{The included sample of 53 Galactic O- and B-type stars from \citet{Bowman2020} as described in Sect.~\ref{sec:comparison_stars}. The stars are listed according to their TIC IDs. The table columns provide the star name, TIC ID, TESS $T$ and Gaia $G$ magnitudes, effective temperatures $T_{\rm eff}$, spectroscopic luminosity $\mathcal{L}$, and the TESS sectors with 2-min cadence data. The $T_{\rm eff}$ and $\mathcal{L}$ values are taken from \citet{Bowman2020}.}
	\label{tab:bowman_sample_overview}
	\begin{tabular}{ccccccc} % four columns, alignment for each
		\hline
		Star name & TIC & $T$ & $G$ & $\log T_{\rm eff}$ & $\log \mathcal{L}$ &   2-min sectors\\
      &  &  [mag] & [mag] & [K] & [$\mathcal{L}_\odot$]   & \\
		\hline
		 HD 37742    	&	 11360636 	&	 4.21 	&	 3.98 	&	 $4.47\pm 0.03$ 	&	 $4.15\pm 0.15$ 	&	 6\\[0.5ex]
		 CPD-47 2963 	&	 30653985 	&	 7.32 	&	 8.00 	&	 $4.57\pm 0.03$ 	&	 $4.16\pm 0.15$ 	&	 8,9\\[0.5ex]
		 HD 27563    	&	 37777866 	&	 5.98 	&	 5.83 	&	 $4.16\pm 0.03$ 	&	 $2.61\pm 0.15$ 	&	 5,32\\[0.5ex]
		 HD 154368   	&	 41792209 	&	 5.54 	&	 5.93 	&	 $4.48\pm 0.03$ 	&	 $4.28\pm 0.15$ 	&	 12,66\\[0.5ex]
		 HD 154643   	&	 43284243 	&	 6.88 	&	 7.06 	&	 $4.49\pm 0.03$ 	&	 $3.85\pm 0.15$ 	&	 12\\[0.5ex]
		 HD 37711    	&	 59215060 	&	 4.89 	&	 4.74 	&	 $4.21\pm 0.03$ 	&	 $2.61\pm 0.15$ 	&	 6,43,44,45,71,72\\[0.5ex]
		 HD 52089    	&	 63198307 	&	 2.48 	&	 $\dots$ 	&	 $4.34\pm 0.03$ 	&	 $3.60\pm 0.15$ 	&	 6,7,33,34,87\\[0.5ex]
		 HD 38771    	&	 66651575 	&	 8.28 	&	 8.71 	&	 $4.47\pm 0.03$ 	&	 $4.06\pm 0.15$ 	&	 6,33\\[0.5ex]
		 HD 53138    	&	 80466973 	&	 1.70 	&	 3.14 	&	 $4.23\pm 0.03$ 	&	 $4.13\pm 0.15$ 	&	 7,87\\[0.5ex]
		 HD 54764    	&	 95513457 	&	 5.99 	&	 6.00 	&	 $4.30\pm 0.03$ 	&	 $3.97\pm 0.15$ 	&	 7,33,87\\[0.5ex]
		 HD 38666    	&	 100589904 	&	 5.47 	&	 5.14 	&	 $4.53\pm 0.03$ 	&	 $3.59\pm 0.15$ 	&	 5,6,32,33,87\\[0.5ex]
		 HD 51309    	&	 146908355 	&	 4.46 	&	 4.39 	&	 $4.20\pm 0.03$ 	&	 $3.64\pm 0.15$ 	&	 6,7,33,87\\[0.5ex]
		 HD 53244    	&	 148109427 	&	 4.24 	&	 4.12 	&	 $4.14\pm 0.03$ 	&	 $2.74\pm 0.15$ 	&	 7,33,87\\[0.5ex]
		 HD 53975    	&	 148506724 	&	 6.59 	&	 6.43 	&	 $4.56\pm 0.03$ 	&	 $3.74\pm 0.15$ 	&	 7,33,87\\[0.5ex]
		 HD 156154   	&	 152659955 	&	 7.30 	&	 7.79 	&	 $4.53\pm 0.03$ 	&	 $4.22\pm 0.15$ 	&	 12\\[0.5ex]
		 HD 54879    	&	 177860391 	&	 7.68 	&	 7.58 	&	 $4.52\pm 0.03$ 	&	 $3.16\pm 0.15$ 	&	 7,33\\[0.5ex]
		 HD 55879    	&	 178489528 	&	 6.19 	&	 5.97 	&	 $4.49\pm 0.03$ 	&	 $3.85\pm 0.15$ 	&	 7,33,87\\[0.5ex]
		 HD 57682    	&	 187458882 	&	 6.60 	&	 6.37 	&	 $4.54\pm 0.03$ 	&	 $3.62\pm 0.15$ 	&	 7,34\\[0.5ex]
		 HD 156738   	&	 195288472 	&	 8.23 	&	 8.92 	&	 $4.58\pm 0.03$ 	&	 $3.87\pm 0.15$ 	&	 12,39,66\\[0.5ex]
		 HD 155913   	&	 216662610 	&	 7.80 	&	 8.16 	&	 $4.63\pm 0.03$ 	&	 $3.88\pm 0.15$ 	&	 12,39\\[0.5ex]
		 HD 47839    	&	 220322383 	&	 4.86 	&	 4.57 	&	 $4.58\pm 0.03$ 	&	 $3.70\pm 0.15$ 	&	 6,33,87\\[0.5ex]
		 HD 48279    	&	 234009943 	&	 7.84 	&	 7.88 	&	 $4.55\pm 0.03$ 	&	 $3.76\pm 0.15$ 	&	 6,33,87\\[0.5ex]
		 HD 48434    	&	 234052684 	&	 5.91 	&	 5.83 	&	 $4.48\pm 0.03$ 	&	 $3.93\pm 0.15$ 	&	 6,33,87\\[0.5ex]
		 HD 150574   	&	 234648113 	&	 7.17 	&	 8.41 	&	 $4.52\pm 0.03$ 	&	 $3.87\pm 0.15$ 	&	 12\\[0.5ex]
		 HD 46056    	&	 234834992 	&	 8.10 	&	 8.18 	&	 $4.55\pm 0.03$ 	&	 $3.58\pm 0.15$ 	&	 6,33,87\\[0.5ex]
		 HD 46150    	&	 234840662 	&	 6.71 	&	 6.70 	&	 $4.61\pm 0.03$ 	&	 $4.03\pm 0.15$ 	&	 6,87\\[0.5ex]
		 HD 46223    	&	 234881667 	&	 7.06 	&	 7.19 	&	 $4.62\pm 0.03$ 	&	 $4.16\pm 0.15$ 	&	 6,33,87\\[0.5ex]
		 HD 46573    	&	 234947719 	&	 7.63 	&	 7.82 	&	 $4.56\pm 0.03$ 	&	 $3.93\pm 0.15$ 	&	 6,33,87\\[0.5ex]
		 HD 152147   	&	 246953610 	&	 5.72 	&	 7.10 	&	 $4.48\pm 0.03$ 	&	 $4.04\pm 0.15$ 	&	 12,39\\[0.5ex]
		 HD 152424   	&	 247267245 	&	 5.92 	&	 6.13 	&	 $4.48\pm 0.03$ 	&	 $4.14\pm 0.15$ 	&	 12,39\\[0.5ex]
		 HD 46769    	&	 281148636 	&	 5.80 	&	 5.76 	&	 $4.11\pm 0.03$ 	&	 $3.16\pm 0.15$ 	&	 6,87\\[0.5ex]
		 HD 41997    	&	 294114621 	&	 8.07 	&	 8.29 	&	 $4.55\pm 0.03$ 	&	 $3.85\pm 0.15$ 	&	 6,33,43,44,45,71,72,87\\[0.5ex]
		 HD 96715    	&	 306491594 	&	 8.11 	&	 8.18 	&	 $4.66\pm 0.03$ 	&	 $4.10\pm 0.15$ 	&	 10,11,37,63,64\\[0.5ex]
		 HD 123056   	&	 330281456 	&	 8.13 	&	 8.11 	&	 $4.50\pm 0.03$ 	&	 $3.70\pm 0.15$ 	&	 11\\[0.5ex]
		 HD 151804   	&	 337793038 	&	 5.24 	&	 5.16 	&	 $4.45\pm 0.03$ 	&	 $4.33\pm 0.15$ 	&	 12,39,66\\[0.5ex]
		 HD 152003   	&	 338640317 	&	 6.76 	&	 6.83 	&	 $4.48\pm 0.03$ 	&	 $4.12\pm 0.15$ 	&	 12\\[0.5ex]
		 HD 152249   	&	 339567904 	&	 5.04 	&	 6.36 	&	 $4.49\pm 0.03$ 	&	 $4.15\pm 0.15$ 	&	 12,39\\[0.5ex]
		 HD 326331   	&	 339568114 	&	 6.27 	&	 7.43 	&	 $4.54\pm 0.03$ 	&	 $3.82\pm 0.15$ 	&	 12\\[0.5ex]
		 HD 152247   	&	 339570292 	&	 5.76 	&	 7.08 	&	 $4.51\pm 0.03$ 	&	 $3.94\pm 0.15$ 	&	 12,39,66\\[0.5ex]
		 HD 35468    	&	 365572007 	&	 2.65 	&	 $\dots$ 	&	 $4.29\pm 0.03$ 	&	 $2.99\pm 0.15$ 	&	 6,32\\[0.5ex]
		 HD 112244   	&	 406050497 	&	 5.37 	&	 5.32 	&	 $4.50\pm 0.03$ 	&	 $4.15\pm 0.15$ 	&	 11,37,38,64\\[0.5ex]
		 HD 36960    	&	 427373484 	&	 5.19 	&	 4.74 	&	 $4.46\pm 0.03$ 	&	 $3.31\pm 0.15$ 	&	 6,32\\[0.5ex]
		 HD 37041    	&	 427395049 	&	 5.02 	&	 4.97 	&	 $4.54\pm 0.03$ 	&	 $3.28\pm 0.15$ 	&	 6,32\\[0.5ex]
		 HD 37042    	&	 427395058 	&	 6.37 	&	 6.29 	&	 $4.47\pm 0.03$ 	&	 $3.06\pm 0.15$ 	&	 6\\[0.5ex]
		 HD 37128    	&	 427451176 	&	 2.69 	&	 $\dots$ 	&	 $4.47\pm 0.03$ 	&	 $4.05\pm 0.15$ 	&	 6,32\\[0.5ex]
		 HD 74920    	&	 430625455 	&	 7.62 	&	 7.48 	&	 $4.54\pm 0.03$ 	&	 $3.90\pm 0.15$ 	&	 8,9\\[0.5ex]
		 HD 110360   	&	 433738620 	&	 9.19 	&	 9.24 	&	 $4.59\pm 0.03$ 	&	 $3.60\pm 0.15$ 	&	 11,37,38,64,65\\[0.5ex]
		 HD 36861    	&	 436103278 	&	 3.83 	&	 3.53 	&	 $4.55\pm 0.03$ 	&	 $4.06\pm 0.15$ 	&	 6\\[0.5ex]
		 HD 135591   	&	 455675248 	&	 5.56 	&	 5.39 	&	 $4.54\pm 0.03$ 	&	 $3.99\pm 0.15$ 	&	 12,38,39,65\\[0.5ex]
		 HD 303492   	&	 459532732 	&	 8.13 	&	 8.61 	&	 $4.45\pm 0.03$ 	&	 $4.29\pm 0.15$ 	&	 10\\[0.5ex]
		 HD 90273    	&	 464295672 	&	 8.86 	&	 9.00 	&	 $4.59\pm 0.03$ 	&	 $3.95\pm 0.15$ 	&	 9,10,36,37,63,64\\[0.5ex]
		 HD 93843    	&	 465012898 	&	 6.23 	&	 7.26 	&	 $4.57\pm 0.03$ 	&	 $4.15\pm 0.15$ 	&	 10,11,63,64\\[0.5ex]
		 HD 97253    	&	 467065657 	&	 6.89 	&	 7.02 	&	 $4.59\pm 0.03$ 	&	 $4.16\pm 0.15$ 	&	 10,37,64\\[0.5ex]
		\hline
	\end{tabular}
\end{table*}

\begin{table*}
\scriptsize
	\centering
	\caption{Average parameter estimates for the 53 O- and B-type stars from the \citet{Bowman2020} sample. he averages are calculated across all sectors and observing cadences (excluding FFI data from cycle 1) and listed in this table. The errors cover the range in the estimated parameters across all sectors for a given star.
 }
	\label{tab:results_averages_bowman}
	\begin{tabular}{ccccccccc} % four columns, alignment for each
		\hline
		TIC ID &  $\log \alpha_0$ & $\nu_{\rm char}$  & $\gamma$ & $\log C_W$ & $\log {\rm RMS}$ & $\nu_{50\%}$ & $w$\\
       & [ppm$^2 \mu$Hz$^{-1}$] & [$\mu{\rm Hz}$] &  & [ppm$^2 \mu$Hz$^{-1}$] & [ppm] & [$\mu{\rm Hz}$] & \\
		\hline
		 11360636 	&	 $7.44^{0.00}_{0.00}$ 	&	 $2.47^{0.00}_{0.00}$ 	&	 $2.35^{0.00}_{0.00}$ 	&	 $2.24^{0.00}_{0.00}$ 	&	 $3.75^{0.00}_{0.00}$ 	&	 $5.21^{0.00}_{0.00}$ 	&	 $1.57^{0.00}_{0.00}$\\[1.5ex]
		 30653985 	&	 $6.61^{0.05}_{0.05}$ 	&	 $10.79^{0.06}_{0.06}$ 	&	 $2.96^{0.00}_{0.00}$ 	&	 $1.94^{0.01}_{0.01}$ 	&	 $3.70^{0.02}_{0.02}$ 	&	 $11.26^{2.06}_{2.06}$ 	&	 $1.78^{0.25}_{0.25}$\\[1.5ex]
		 37777866 	&	 $6.85^{0.17}_{0.17}$ 	&	 $1.25^{0.01}_{0.01}$ 	&	 $2.33^{0.04}_{0.04}$ 	&	 $1.34^{0.04}_{0.04}$ 	&	 $3.30^{0.10}_{0.10}$ 	&	 $3.97^{0.29}_{0.29}$ 	&	 $0.70^{0.22}_{0.22}$\\[1.5ex]
		 41792209 	&	 $7.52^{0.10}_{0.10}$ 	&	 $5.51^{0.50}_{0.50}$ 	&	 $2.75^{0.10}_{0.10}$ 	&	 $2.67^{0.19}_{0.19}$ 	&	 $3.99^{0.02}_{0.02}$ 	&	 $8.11^{1.07}_{1.07}$ 	&	 $1.09^{0.00}_{0.00}$\\[1.5ex]
		 43284243 	&	 $5.28^{0.00}_{0.00}$ 	&	 $30.74^{0.00}_{0.00}$ 	&	 $3.06^{0.00}_{0.00}$ 	&	 $1.81^{0.00}_{0.00}$ 	&	 $3.28^{0.00}_{0.00}$ 	&	 $11.22^{0.00}_{0.00}$ 	&	 $3.07^{0.00}_{0.00}$\\[1.5ex]
		 59215060 	&	 $6.67^{0.46}_{0.48}$ 	&	 $2.86^{2.51}_{1.62}$ 	&	 $2.64^{0.56}_{0.38}$ 	&	 $1.27^{0.37}_{0.27}$ 	&	 $3.35^{0.08}_{0.08}$ 	&	 $4.45^{0.82}_{1.39}$ 	&	 $0.98^{0.39}_{0.32}$\\[1.5ex]
		 63198307 	&	 $5.08^{0.17}_{0.20}$ 	&	 $17.42^{3.35}_{5.37}$ 	&	 $3.07^{0.32}_{0.52}$ 	&	 $1.42^{0.34}_{0.25}$ 	&	 $3.05^{0.10}_{0.11}$ 	&	 $7.89^{2.32}_{1.88}$ 	&	 $2.14^{0.61}_{0.49}$\\[1.5ex]
		 66651575 	&	 $4.10^{0.95}_{0.95}$ 	&	 $8.21^{6.93}_{6.93}$ 	&	 $2.17^{0.38}_{0.38}$ 	&	 $1.00^{0.00}_{0.00}$ 	&	 $2.35^{0.18}_{0.18}$ 	&	 $7.56^{4.38}_{4.38}$ 	&	 $5.14^{1.69}_{1.69}$\\[1.5ex]
		 80466973 	&	 $7.42^{0.54}_{0.54}$ 	&	 $1.35^{0.16}_{0.16}$ 	&	 $2.21^{0.02}_{0.02}$ 	&	 $1.72^{0.72}_{0.72}$ 	&	 $3.68^{0.27}_{0.27}$ 	&	 $3.05^{0.37}_{0.37}$ 	&	 $0.96^{0.22}_{0.22}$\\[1.5ex]
		 95513457 	&	 $7.47^{0.04}_{0.06}$ 	&	 $2.87^{0.25}_{0.39}$ 	&	 $2.46^{0.15}_{0.08}$ 	&	 $1.83^{0.37}_{0.34}$ 	&	 $3.82^{0.03}_{0.02}$ 	&	 $4.72^{0.90}_{0.78}$ 	&	 $1.42^{0.51}_{0.35}$\\[1.5ex]
		 100589904 	&	 $3.32^{0.16}_{0.12}$ 	&	 $33.27^{8.22}_{5.55}$ 	&	 $2.73^{0.33}_{0.34}$ 	&	 $1.01^{0.04}_{0.01}$ 	&	 $2.39^{0.02}_{0.03}$ 	&	 $24.04^{5.50}_{12.45}$ 	&	 $2.86^{1.25}_{0.95}$\\[1.5ex]
		 146908355 	&	 $7.37^{0.20}_{0.36}$ 	&	 $2.00^{2.43}_{0.82}$ 	&	 $2.21^{0.75}_{0.46}$ 	&	 $1.81^{0.83}_{0.69}$ 	&	 $3.79^{0.09}_{0.11}$ 	&	 $2.61^{0.56}_{0.73}$ 	&	 $1.80^{0.76}_{0.59}$\\[1.5ex]
		 148109427 	&	 $4.05^{0.86}_{0.49}$ 	&	 $3.95^{1.80}_{2.45}$ 	&	 $2.85^{0.22}_{0.37}$ 	&	 $1.00^{0.00}_{0.00}$ 	&	 $2.35^{0.13}_{0.10}$ 	&	 $4.13^{1.29}_{2.52}$ 	&	 $9.69^{8.66}_{7.99}$\\[1.5ex]
		 148506724 	&	 $4.59^{0.17}_{0.19}$ 	&	 $32.45^{6.39}_{3.31}$ 	&	 $2.66^{0.07}_{0.06}$ 	&	 $1.52^{0.03}_{0.03}$ 	&	 $2.97^{0.05}_{0.04}$ 	&	 $18.84^{3.26}_{4.61}$ 	&	 $2.40^{0.43}_{0.80}$\\[1.5ex]
		 152659955 	&	 $6.89^{0.00}_{0.00}$ 	&	 $9.77^{0.00}_{0.00}$ 	&	 $2.76^{0.00}_{0.00}$ 	&	 $2.67^{0.00}_{0.00}$ 	&	 $3.83^{0.00}_{0.00}$ 	&	 $7.64^{0.00}_{0.00}$ 	&	 $2.49^{0.00}_{0.00}$\\[1.5ex]
		 177860391 	&	 $3.90^{0.17}_{0.17}$ 	&	 $5.28^{0.70}_{0.70}$ 	&	 $2.60^{0.02}_{0.02}$ 	&	 $1.76^{0.06}_{0.06}$ 	&	 $2.58^{0.02}_{0.02}$ 	&	 $88.26^{74.11}_{74.11}$ 	&	 $17.93^{14.40}_{14.40}$\\[1.5ex]
		 178489528 	&	 $4.62^{0.05}_{0.03}$ 	&	 $27.59^{2.19}_{1.41}$ 	&	 $2.75^{0.06}_{0.11}$ 	&	 $1.31^{0.07}_{0.09}$ 	&	 $2.94^{0.01}_{0.01}$ 	&	 $9.30^{1.23}_{1.06}$ 	&	 $3.19^{0.26}_{0.32}$\\[1.5ex]
		 187458882 	&	 $4.58^{0.12}_{0.12}$ 	&	 $14.90^{1.71}_{1.71}$ 	&	 $2.13^{0.03}_{0.03}$ 	&	 $1.54^{0.05}_{0.05}$ 	&	 $2.84^{0.03}_{0.03}$ 	&	 $7.92^{2.56}_{2.56}$ 	&	 $5.43^{1.21}_{1.21}$\\[1.5ex]
		 195288472 	&	 $5.17^{0.17}_{0.15}$ 	&	 $42.80^{7.12}_{8.92}$ 	&	 $3.55^{0.16}_{0.31}$ 	&	 $2.13^{0.10}_{0.06}$ 	&	 $3.29^{0.04}_{0.04}$ 	&	 $20.11^{2.69}_{1.45}$ 	&	 $2.09^{0.40}_{0.27}$\\[1.5ex]
		 216662610 	&	 $5.03^{0.01}_{0.01}$ 	&	 $59.05^{1.26}_{1.26}$ 	&	 $3.47^{0.03}_{0.03}$ 	&	 $2.06^{0.01}_{0.01}$ 	&	 $3.29^{0.01}_{0.01}$ 	&	 $30.45^{0.19}_{0.19}$ 	&	 $1.81^{0.03}_{0.03}$\\[1.5ex]
		 220322383 	&	 $3.96^{0.18}_{0.13}$ 	&	 $50.37^{21.31}_{13.56}$ 	&	 $2.88^{0.51}_{0.39}$ 	&	 $1.51^{0.41}_{0.51}$ 	&	 $2.78^{0.07}_{0.07}$ 	&	 $27.96^{1.37}_{2.64}$ 	&	 $2.83^{0.62}_{0.51}$\\[1.5ex]
		 234009943 	&	 $5.10^{0.09}_{0.13}$ 	&	 $34.56^{5.89}_{6.09}$ 	&	 $3.20^{0.32}_{0.24}$ 	&	 $2.43^{0.20}_{0.31}$ 	&	 $3.25^{0.03}_{0.02}$ 	&	 $18.73^{3.56}_{3.90}$ 	&	 $1.88^{0.31}_{0.17}$\\[1.5ex]
		 234052684 	&	 $6.92^{0.28}_{0.47}$ 	&	 $8.68^{7.48}_{3.89}$ 	&	 $2.81^{0.65}_{0.41}$ 	&	 $1.97^{0.47}_{0.24}$ 	&	 $3.77^{0.06}_{0.07}$ 	&	 $6.07^{1.43}_{1.01}$ 	&	 $2.25^{0.35}_{0.24}$\\[1.5ex]
		 234648113 	&	 $6.44^{0.00}_{0.00}$ 	&	 $14.69^{0.00}_{0.00}$ 	&	 $2.64^{0.00}_{0.00}$ 	&	 $2.77^{0.00}_{0.00}$ 	&	 $3.71^{0.00}_{0.00}$ 	&	 $11.54^{0.00}_{0.00}$ 	&	 $2.27^{0.00}_{0.00}$\\[1.5ex]
		 234834992 	&	 $3.92^{0.10}_{0.10}$ 	&	 $74.89^{8.55}_{11.51}$ 	&	 $3.06^{0.20}_{0.27}$ 	&	 $2.01^{0.04}_{0.05}$ 	&	 $2.88^{0.02}_{0.02}$ 	&	 $47.09^{0.89}_{0.96}$ 	&	 $2.42^{0.07}_{0.10}$\\[1.5ex]
		 234840662 	&	 $4.55^{0.07}_{0.07}$ 	&	 $55.54^{5.16}_{5.16}$ 	&	 $3.29^{0.16}_{0.16}$ 	&	 $1.64^{0.09}_{0.09}$ 	&	 $3.04^{0.02}_{0.02}$ 	&	 $25.91^{2.41}_{2.41}$ 	&	 $2.49^{0.27}_{0.27}$\\[1.5ex]
		 234881667 	&	 $4.97^{0.04}_{0.05}$ 	&	 $50.63^{3.75}_{5.19}$ 	&	 $3.38^{0.13}_{0.21}$ 	&	 $1.76^{0.17}_{0.16}$ 	&	 $3.23^{0.01}_{0.01}$ 	&	 $20.16^{1.40}_{0.73}$ 	&	 $2.68^{0.07}_{0.05}$\\[1.5ex]
		 234947719 	&	 $5.19^{0.11}_{0.12}$ 	&	 $34.47^{4.88}_{5.90}$ 	&	 $3.04^{0.10}_{0.17}$ 	&	 $1.80^{0.06}_{0.08}$ 	&	 $3.26^{0.02}_{0.03}$ 	&	 $14.24^{4.63}_{4.13}$ 	&	 $2.55^{0.45}_{0.40}$\\[1.5ex]
		 246953610 	&	 $7.75^{0.21}_{0.21}$ 	&	 $3.78^{0.22}_{0.22}$ 	&	 $2.63^{0.13}_{0.13}$ 	&	 $2.50^{0.40}_{0.40}$ 	&	 $4.01^{0.08}_{0.08}$ 	&	 $6.17^{1.56}_{1.56}$ 	&	 $1.31^{0.01}_{0.01}$\\[1.5ex]
		 247267245 	&	 $8.30^{0.01}_{0.01}$ 	&	 $1.75^{0.24}_{0.24}$ 	&	 $2.47^{0.02}_{0.02}$ 	&	 $2.57^{0.37}_{0.37}$ 	&	 $4.04^{0.06}_{0.06}$ 	&	 $6.15^{0.21}_{0.21}$ 	&	 $0.92^{0.17}_{0.17}$\\[1.5ex]
		 281148636 	&	 $3.67^{0.39}_{0.39}$ 	&	 $1.42^{0.05}_{0.05}$ 	&	 $1.94^{0.52}_{0.52}$ 	&	 $1.16^{0.07}_{0.07}$ 	&	 $2.27^{0.04}_{0.04}$ 	&	 $153.42^{76.34}_{76.34}$ 	&	 $4.56^{2.04}_{2.04}$\\[1.5ex]
		 294114621 	&	 $5.80^{0.38}_{0.17}$ 	&	 $30.57^{5.24}_{13.31}$ 	&	 $3.58^{0.21}_{0.62}$ 	&	 $1.95^{0.21}_{0.09}$ 	&	 $3.52^{0.08}_{0.05}$ 	&	 $15.42^{4.68}_{2.39}$ 	&	 $1.63^{0.56}_{0.38}$\\[1.5ex]
		 306491594 	&	 $4.52^{0.06}_{0.09}$ 	&	 $65.10^{4.78}_{3.22}$ 	&	 $3.16^{0.09}_{0.10}$ 	&	 $1.95^{0.02}_{0.02}$ 	&	 $3.08^{0.02}_{0.02}$ 	&	 $31.72^{4.06}_{3.41}$ 	&	 $2.52^{0.22}_{0.33}$\\[1.5ex]
		 330281456 	&	 $5.33^{0.00}_{0.00}$ 	&	 $28.70^{0.00}_{0.00}$ 	&	 $3.07^{0.00}_{0.00}$ 	&	 $2.06^{0.00}_{0.00}$ 	&	 $3.29^{0.00}_{0.00}$ 	&	 $19.05^{0.00}_{0.00}$ 	&	 $1.26^{0.00}_{0.00}$\\[1.5ex]
		 337793038 	&	 $8.04^{0.25}_{0.19}$ 	&	 $2.16^{0.89}_{0.63}$ 	&	 $2.55^{0.28}_{0.35}$ 	&	 $2.42^{0.74}_{0.54}$ 	&	 $4.02^{0.03}_{0.03}$ 	&	 $5.61^{0.66}_{0.50}$ 	&	 $0.94^{0.07}_{0.07}$\\[1.5ex]
		 338640317 	&	 $7.85^{0.00}_{0.00}$ 	&	 $3.17^{0.00}_{0.00}$ 	&	 $2.40^{0.00}_{0.00}$ 	&	 $2.93^{0.00}_{0.00}$ 	&	 $4.03^{0.00}_{0.00}$ 	&	 $5.32^{0.00}_{0.00}$ 	&	 $2.00^{0.00}_{0.00}$\\[1.5ex]
		 339567904 	&	 $7.15^{0.57}_{0.57}$ 	&	 $9.49^{6.64}_{6.64}$ 	&	 $3.32^{0.91}_{0.91}$ 	&	 $3.05^{0.49}_{0.49}$ 	&	 $3.87^{0.07}_{0.07}$ 	&	 $6.62^{1.43}_{1.43}$ 	&	 $1.67^{0.21}_{0.21}$\\[1.5ex]
		 339568114 	&	 $6.21^{0.00}_{0.00}$ 	&	 $18.98^{0.00}_{0.00}$ 	&	 $2.93^{0.00}_{0.00}$ 	&	 $2.63^{0.00}_{0.00}$ 	&	 $3.64^{0.00}_{0.00}$ 	&	 $10.49^{0.00}_{0.00}$ 	&	 $2.33^{0.00}_{0.00}$\\[1.5ex]
		 339570292 	&	 $6.03^{0.13}_{0.23}$ 	&	 $20.36^{6.88}_{5.26}$ 	&	 $3.25^{0.37}_{0.31}$ 	&	 $2.00^{0.11}_{0.19}$ 	&	 $3.55^{0.04}_{0.05}$ 	&	 $11.10^{2.05}_{1.85}$ 	&	 $1.96^{0.10}_{0.07}$\\[1.5ex]
		\hline
	\end{tabular}
\end{table*}

%\contcaption{}
\begin{table*}
\scriptsize
	\centering
	\contcaption{}
	\begin{tabular}{ccccccccc} % four columns, alignment for each
		\hline
		TIC ID &  $\log \alpha_0$ & $\nu_{\rm char}$  & $\gamma$ & $\log C_W$ & $\log {\rm RMS}$ & $\nu_{50\%}$ & $w$\\
       & [ppm$^2 \mu$Hz$^{-1}$] & [$\mu{\rm Hz}$] &  & [ppm$^2 \mu$Hz$^{-1}$] & [ppm] & [$\mu{\rm Hz}$] & \\
		\hline
		 365572007 	&	 $4.62^{0.26}_{0.26}$ 	&	 $3.23^{1.95}_{1.95}$ 	&	 $2.15^{0.49}_{0.49}$ 	&	 $1.39^{0.26}_{0.26}$ 	&	 $2.55^{0.07}_{0.07}$ 	&	 $4.12^{0.61}_{0.61}$ 	&	 $16.14^{14.09}_{14.09}$\\[1.5ex]
		 406050497 	&	 $7.91^{0.20}_{0.38}$ 	&	 $3.18^{1.98}_{1.15}$ 	&	 $2.56^{0.19}_{0.41}$ 	&	 $1.97^{0.82}_{0.50}$ 	&	 $4.05^{0.03}_{0.06}$ 	&	 $5.34^{1.52}_{1.63}$ 	&	 $1.50^{0.26}_{0.15}$\\[1.5ex]
		 427373484 	&	 $3.73^{0.04}_{0.04}$ 	&	 $12.17^{0.60}_{0.60}$ 	&	 $2.52^{0.09}_{0.09}$ 	&	 $2.05^{0.11}_{0.11}$ 	&	 $2.72^{0.05}_{0.05}$ 	&	 $48.68^{18.30}_{18.30}$ 	&	 $10.93^{3.38}_{3.38}$\\[1.5ex]
		 427395049 	&	 $3.59^{0.19}_{0.19}$ 	&	 $17.88^{6.04}_{6.04}$ 	&	 $2.60^{0.71}_{0.71}$ 	&	 $1.89^{0.01}_{0.01}$ 	&	 $2.66^{0.02}_{0.02}$ 	&	 $39.77^{2.44}_{2.44}$ 	&	 $10.48^{1.34}_{1.34}$\\[1.5ex]
		 427395058 	&	 $4.81^{0.00}_{0.00}$ 	&	 $4.74^{0.00}_{0.00}$ 	&	 $2.03^{0.00}_{0.00}$ 	&	 $2.99^{0.00}_{0.00}$ 	&	 $3.17^{0.00}_{0.00}$ 	&	 $141.29^{0.00}_{0.00}$ 	&	 $3.88^{0.00}_{0.00}$\\[1.5ex]
		 427451176 	&	 $6.95^{0.79}_{0.79}$ 	&	 $5.61^{4.39}_{4.39}$ 	&	 $2.82^{0.78}_{0.78}$ 	&	 $2.21^{0.62}_{0.62}$ 	&	 $3.64^{0.19}_{0.19}$ 	&	 $5.58^{1.45}_{1.45}$ 	&	 $1.26^{0.08}_{0.08}$\\[1.5ex]
		 430625455 	&	 $5.70^{0.14}_{0.14}$ 	&	 $31.77^{6.59}_{6.59}$ 	&	 $3.33^{0.27}_{0.27}$ 	&	 $1.88^{0.03}_{0.03}$ 	&	 $3.48^{0.03}_{0.03}$ 	&	 $18.44^{0.20}_{0.20}$ 	&	 $1.75^{0.39}_{0.39}$\\[1.5ex]
		 433738620 	&	 $4.71^{0.65}_{0.24}$ 	&	 $10.98^{2.47}_{6.81}$ 	&	 $1.65^{0.03}_{0.07}$ 	&	 $2.39^{0.04}_{0.04}$ 	&	 $3.00^{0.07}_{0.03}$ 	&	 $27.22^{11.60}_{18.65}$ 	&	 $7.63^{1.77}_{1.03}$\\[1.5ex]
		 436103278 	&	 $5.28^{0.00}_{0.00}$ 	&	 $25.42^{0.00}_{0.00}$ 	&	 $2.76^{0.00}_{0.00}$ 	&	 $1.34^{0.00}_{0.00}$ 	&	 $3.24^{0.00}_{0.00}$ 	&	 $13.48^{0.00}_{0.00}$ 	&	 $2.65^{0.00}_{0.00}$\\[1.5ex]
		 455675248 	&	 $5.13^{0.05}_{0.07}$ 	&	 $39.27^{4.26}_{2.54}$ 	&	 $3.37^{0.14}_{0.10}$ 	&	 $1.31^{0.26}_{0.17}$ 	&	 $3.24^{0.01}_{0.02}$ 	&	 $12.72^{3.34}_{2.50}$ 	&	 $2.94^{0.67}_{0.79}$\\[1.5ex]
		 459532732 	&	 $7.99^{0.00}_{0.00}$ 	&	 $1.32^{0.00}_{0.00}$ 	&	 $2.20^{0.00}_{0.00}$ 	&	 $2.30^{0.00}_{0.00}$ 	&	 $4.06^{0.00}_{0.00}$ 	&	 $4.46^{0.00}_{0.00}$ 	&	 $0.89^{0.00}_{0.00}$\\[1.5ex]
		 464295672 	&	 $4.91^{0.09}_{0.10}$ 	&	 $44.21^{8.07}_{4.25}$ 	&	 $2.98^{0.25}_{0.12}$ 	&	 $2.28^{0.03}_{0.03}$ 	&	 $3.20^{0.03}_{0.03}$ 	&	 $21.88^{7.13}_{3.72}$ 	&	 $2.79^{0.36}_{0.59}$\\[1.5ex]
		 465012898 	&	 $5.99^{0.20}_{0.25}$ 	&	 $24.06^{7.30}_{9.69}$ 	&	 $3.20^{0.25}_{0.60}$ 	&	 $2.04^{0.16}_{0.12}$ 	&	 $3.57^{0.05}_{0.07}$ 	&	 $11.14^{2.45}_{3.81}$ 	&	 $2.78^{0.79}_{0.60}$\\[1.5ex]
		 467065657 	&	 $6.03^{0.26}_{0.18}$ 	&	 $27.56^{5.83}_{11.27}$ 	&	 $4.02^{0.57}_{0.89}$ 	&	 $1.95^{0.11}_{0.10}$ 	&	 $3.60^{0.04}_{0.05}$ 	&	 $14.90^{3.16}_{5.34}$ 	&	 $1.55^{0.52}_{0.47}$\\[1.5ex]
		\hline
	\end{tabular}
\end{table*}

\section{Cadence dependence of estimated parameters}
\label{App:cadence_dependence}

With a few exceptions \citep{Naze2021,Lenoir-Craig2022}, the analysis of SLF variability based on TESS observations has generally relied on the use of the 2-min cadence data. There are a couple of reasons for this. One is that the TESS 2-min cadence light curves are easily accessible as they are automatically made available on MAST by the TESS Science Processing Operations Center \citep[SPOC;][]{Jenkins2016}, whose TESS science pipeline produces both Simple Aperture Photometry (SAP) and PDCSAP light curves. The same is not the case for light curves constructed from FFI data, which therefore require additional steps to be prepared and ready for analysis such as those described in Sect.~\ref{sec:tess_data}. The TESS SPOC team has recently started running their pipeline also on FFI data \citep{Caldwell2020}, while other authors have also used their own pipelines to produce TESS FFI light curves that are made available publicly on MAST \citep[e.g.,][]{Bouma2019,Nardiello2019,Hon2021,Handberg2021}, none of which are complete. 

Another reason for using the 2-min cadence TESS data instead of FFI data is related to amplitude attenuation \citep[e.g.][]{Chaplin2014} also sometimes called amplitude suppression \citep[e.g.][]{Bowman2020}, where the time averaging of high-frequency signals due to increased exposure times results in an attenuation of their amplitudes. In the case of TESS data, the exposure time equals the observing cadence and the fractional attenuation in power can be written as Eq.~\ref{eq:P_att}.

As most of the stars our OB~Cyg sample only have FFI data available, we quantify the impact of the sampling rate on the estimated SLF parameters by taking all of the 2-min cadence residual light curves for both the OB~Cyg and B20 sample and resampling them to match an observing cadence of 240\,sec, 10\,min, and 30\,min. We use 240\,sec instead of 200\,sec as it corresponds to a multiple integer of the 2-min cadence sampling. The resampling is done by binning the normalised residual flux to the new cadence and dividing by the number of data points in each bin to get the averaged flux, while the time stamps are changed to the center values of each bin. Once the light curves have been resampled, we derive the $\alpha_0$, $\nu_{\rm char}$, $\gamma$, $C_W$, $RMS$, $\nu_{50\%}$, and $w$ parameters again and compare them to the estimated values of the 2-min cadence data.

\begin{figure}%[ht!]
\begin{center}
\includegraphics[width=\linewidth]{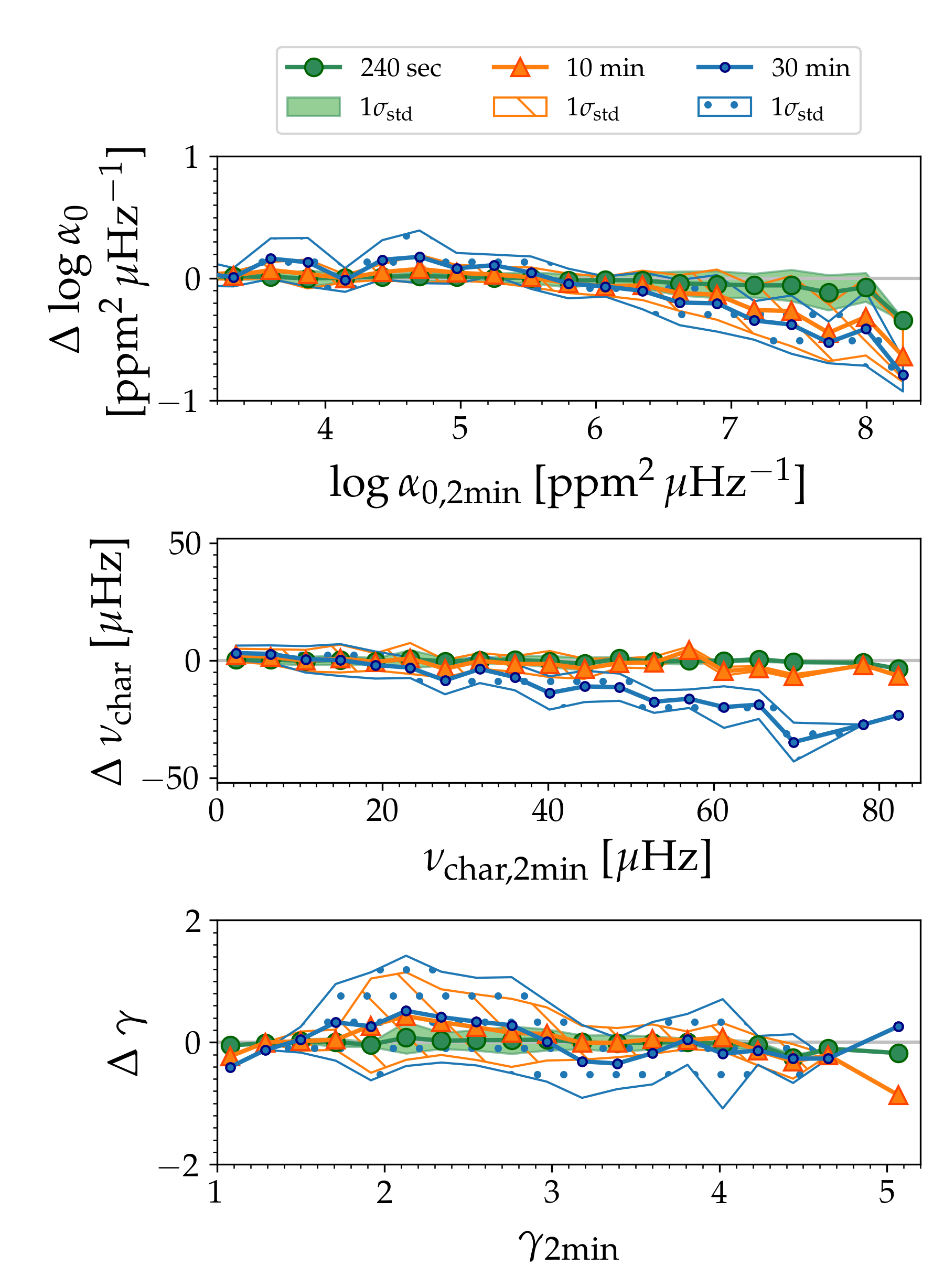}
\caption{Differences in the parameter estimates of $\alpha_0$ (top), $\nu_{\rm char}$ (middle) and $\gamma$ (bottom) obtained when 2-min cadence data is resampled to 240-sec (green), 10-min (orange), and 30-min (blue). The curves show the binned averages while the shaded, hatched, and dotted regions indicate the corresponding one $\sigma$ standard deviation. The parameter differences on the y-axis correspond to the parameter estimate of a given resampled observing cadence subtracted by the corresponding original 2-min cadence estimate, and are plotted as a function of the 2-min cadence data results shown on the x-axis. Bin sizes are $0.275\,{\rm dex}$, $4.21\,\mu{\rm Hz}$, and $0.21$ for $\log \alpha_0$, $\nu_{\rm char}$, and $\gamma$, respectively.}\label{fig:cad_lorentzian}
\end{center}
\end{figure}

Figure~\ref{fig:cad_lorentzian} shows the differences in the estimated $\log \alpha_0$ (top), $\nu_{\rm char}$ (middle) and $\gamma$ (bottom) parameters obtained when the 2-min cadence data are resampled to 240-sec (green), 10-min (orange), and 30-min (blue) cadence. As an example, the y-axes on the top panel corresponds to $\Delta \log \alpha_0 = \log \alpha_{0,{\rm resampled}} - \log \alpha_{0,{\rm 2min}}$, where $\log \alpha_{0,{\rm resampled}}$ is the estimated value for, e.g., the resampled 240-sec cadence data and the $\log \alpha_{0,{\rm 2min}}$ is the corresponding estimate for the 2-min cadence data. Instead of showing the individual parameter estimates for all sectors and all stars in the Cyg~OB and B20 sample with 2-min cadence data available, we calculate the averages in 20 equally sized bins (full connected lines) as well as the corresponding $1\sigma_{\rm std}$ indicated by the shaded green (240-sec), hatched orange (10-min), and dotted blue (30-min) regions in Fig.~\ref{fig:cad_lorentzian}. For all three parameters, the estimates approach the 2-min cadence values as the cadence increases as expected. In the 10-min and 30-min cadence cases, the high amplitude SLF signals are underestimated at longer cadences. In the case of $\nu_{\rm char}$, both the 240-sec and 10-min cadence results are close to the 2-min cadence estimates, while a clear trend is seen for the 30-min cadence data with the differences increasing towards higher characteristic frequencies. The bottom panel of Fig.~\ref{fig:cad_lorentzian} shows that while the curves representing the average differences are generally flat the standard deviations of the differences become smaller when the cadence is increased. 

\begin{figure}%[ht!]
\begin{center}
\includegraphics[width=\linewidth]{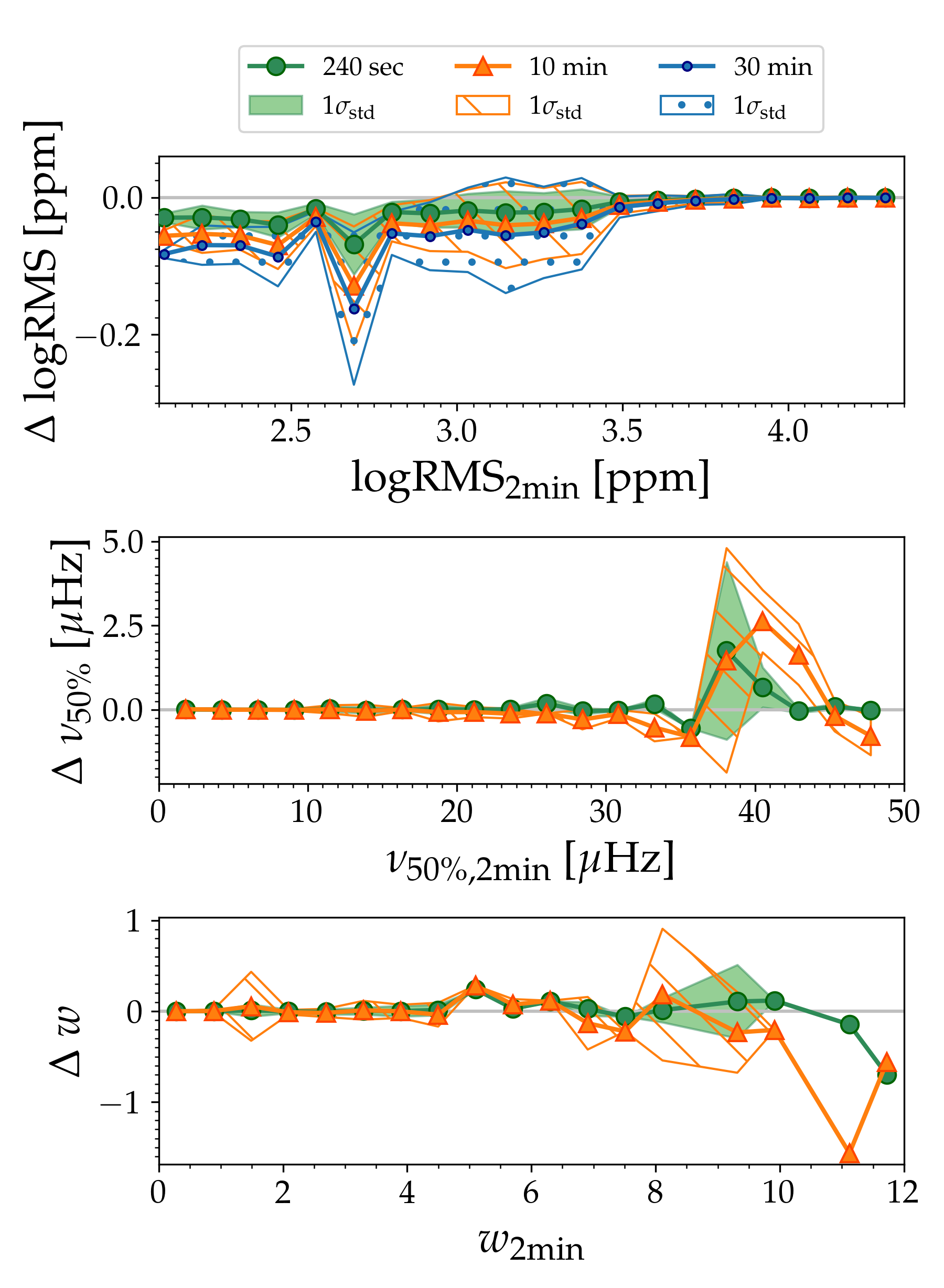}
\caption{Same as Fig.~\ref{fig:cad_lorentzian} but for RMS, $\nu_{50\%}$ and $w$. As we have used the Nyquist frequency of the 10-min cadence data as $\nu_{\rm norm}$ in the calculation of $\nu_{50\%}$ and $w$, we do not include the 30-min cadence comparison in the two bottom panels. Bin sizes are $0.115\,{\rm dex}$, $2.42\,\mu{\rm Hz}$, and $0.60$ for $\log {\rm RMS}$, $\nu_{\rm 50\%}$, and $w$, respectively.}\label{fig:cad_rms}
\end{center}
\end{figure}

Figure~\ref{fig:cad_rms} is the same as Fig.~\ref{fig:cad_lorentzian} but for the three parameters RMS, $\nu_{\rm 50\%}$ and $w$. Because $\nu_{\rm norm}$ is set to the Nyquist frequency of the 10-min cadence data, we excluded the 30-min cadence comparison for the two bottom panels of Fig.~\ref{fig:cad_rms}. We limited the comparisons to stars and sectors with $\nu_{\rm 50\%} < 50\,\mu {\rm Hz}$ and $w < 12$ as only few data points are available above these values. An opposite trend with observing cadence is seen from RMS compared to $\alpha_0$ where the differences in the RMS estimates decrease for increasing RMS, contrary to the increasing differences found for increasing $\alpha_0$ values in Fig.~\ref{fig:cad_lorentzian}.

A summary of the impacts of changing the observing cadence on the estimated parameters is provided in Table~\ref{tab:cadence_dep_2min}. The values in the Table are the total one $\sigma$ standard deviations of the differences between the estimated parameters for the resampled data and the original 2-min cadence residual light curves, and not just the binned standard deviations shown in Fig.~\ref{fig:cad_lorentzian} and \ref{fig:cad_rms}. As in Fig.~\ref{fig:cad_rms}, we excluded stars and sectors with $\nu_{\rm 50\%} \geq 50\,\mu {\rm Hz}$ and $w \geq 12$ in the derivation of the total standard deviations of these two parameters and provide the corresponding values for the full sample in the parentheses. A comparison between Fig.~\ref{fig:cad_lorentzian} and \ref{fig:cad_rms} as well as of the $1\sigma_{\rm std}$ values in Table~\ref{tab:cadence_dep_2min} shows that the RMS, $\nu_{50\%}$ and $w$ parameters are generally less sensitive to the observing cadence than $\alpha_0$, $\nu_{\rm char}$ and $\gamma$ at least for $\nu_{\rm 50\%} < 50\,\mu {\rm Hz}$ and $w < 12$.

\begin{table}
	\centering
	\caption{Standard deviations of the differences in the parameters $\log \alpha_0$, $\nu_{\rm char}$, $\gamma$, RMS, $\nu_{50\%}$ and $w$ (first column) obtained when resampling and binning the original 2-min cadence light curves to 240\,sec (second column), 10\,min (third column), and 30\,min (fourth column) cadences.}
	\label{tab:cadence_dep_2min}
	\begin{tabular}{lccc} % four columns, alignment for each
		\hline
		 & {\rm 240 sec} & ${\rm 10 min}$ & ${\rm 30 min}$\\
           & $\sigma_{\rm std}$ & $\sigma_{\rm std} $ & $\sigma_{\rm std}$\\
		\hline
		$\Delta\log\alpha_0$ [$\frac{{\rm ppm}^2}{\mu {\rm Hz}}$]&  0.079  &  0.19  & 0.256\\
		$\Delta\nu_{\rm char}$ [$\mu$Hz] &  1.731 & 4.347 & 8.542  \\
		$\Delta\gamma$ &  0.157 & 0.471 & 0.724  \\[1ex]
        $\Delta\log {\rm RMS}$ [ppm] & 0.045 & 0.096 &  0.144 \\
        $\Delta\nu_{50\%}$ [$\mu$Hz] &  0.302 (1.286) & 0.481 (2.937) &  \dots \\
        $\Delta w$ & 0.089 (0.143)  & 0.257 (0.307) & \dots  \\
		\hline
	\end{tabular}
\end{table}

\begin{figure}%[ht!]
\begin{center}
\includegraphics[width=\linewidth]{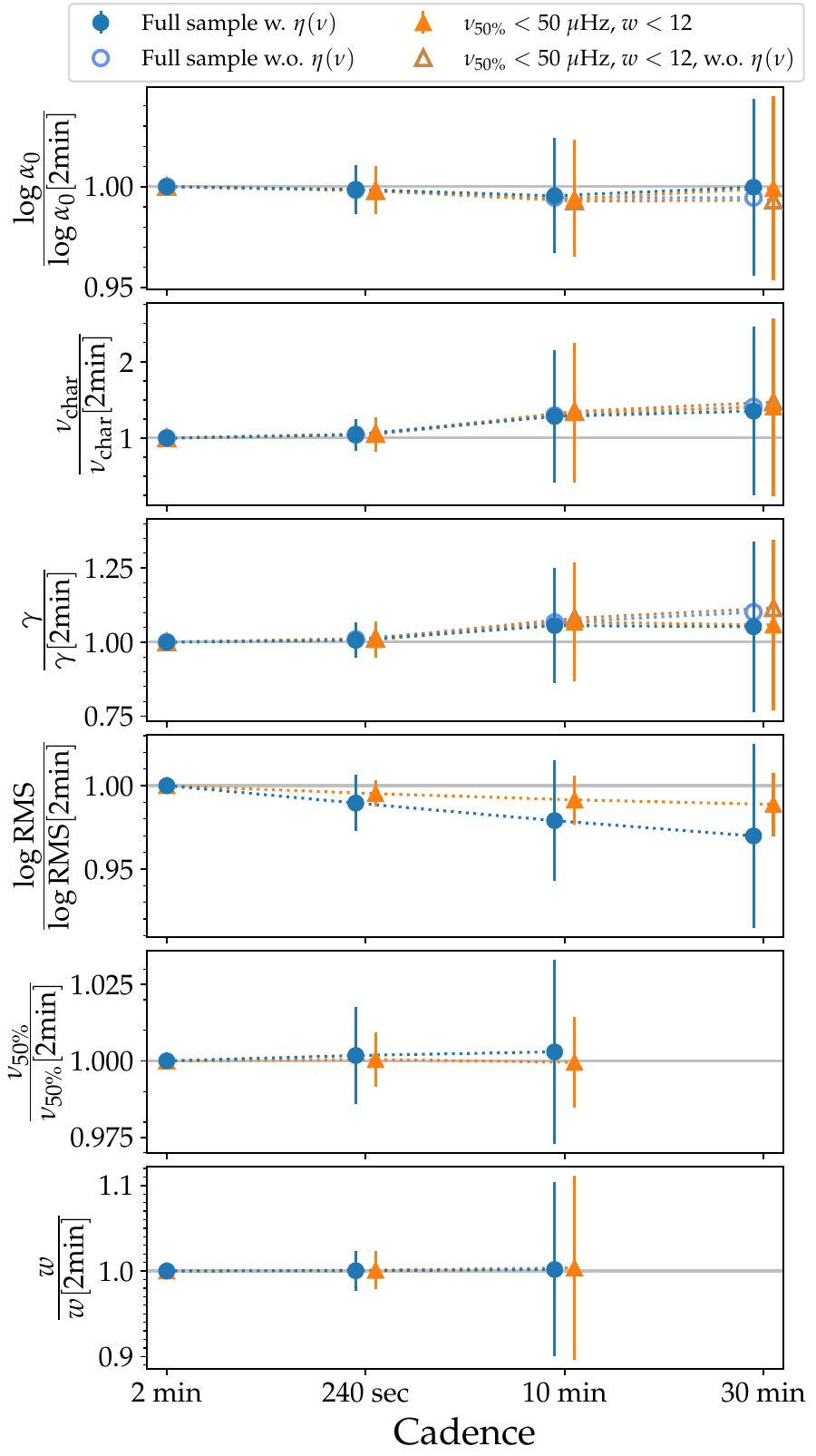}
\caption{Comparison between the average normalised estimated parameters derived after resampling the 2-min cadence data to longer observing cadences. The estimated parameters (y-axes) are normalised by their corresponding 2-min cadence estimates and their averages and $1\sigma_{\rm std}$ calculated across all stars and TESS sectors are shown as a function of observing cadence (x-axis). Results for the full sample of Cyg~OB+B20 stars with 2-min cadence data are shown in blue, whereas the corresponding results after excluding stars and sectors with $\nu_{50\%} \geq 50\ \mu$Hz and $w \geq w 12$ are indicated by orange triangles. For comparison we show as open blue circles and open brown triangles the results obtained if the $\eta (\nu)$ term in Eq.~\ref{eq:model_L} is excluded when fitting the PDS.}\label{fig:cad_frac}
\end{center}
\end{figure}

In their study of SLF variability in Galactic Wolf-Rayet (WR) stars, \cite{Lenoir-Craig2022} also considered the impact of longer observing cadences on the estimated $\alpha_0$, $\nu_{\rm char}$, $\gamma$ and $C_W$ parameters derived from fitting Eq.~(\ref{eq:model_L}) to the amplitude spectrum rather than the PDS. To do so, they looked at the averages and standard deviations of the ``normalised parameters'', e.g. $\alpha_0 {\rm [new\ cadence]} / \alpha_0 {\rm [2min]}$, obtained from resampling the 2-min cadence TESS data to 10-min, 30-min and 100-min. For the sake of comparison, we repeat this exercise here and show in Fig.~\ref{fig:cad_frac} the change in normalised parameters for the resampled 240-sec, 10-min and 30-min cadence data for the combined Cyg~OB and B20 sample. The filled blue points show the averages and associated standard deviation obtained for each cadence; these values are repeated in Table~\ref{tab:frac_cadence_dep_2min}. The orange triangles in Fig.~\ref{fig:cad_frac} and values in parenthesis in Table~\ref{tab:frac_cadence_dep_2min} show the same results after limiting the sample to stars and sectors with $\nu_{\rm 50\%} < 50\,\mu {\rm Hz}$ and $w < 12$. As seen in the figure, the effects of limiting the sample are negligible for $\log \alpha_0$, $\nu_{\rm char}$ and $\gamma$, whereas the standard deviations significantly decrease for the normalised $\log {\rm RMS}$, $\nu_{50\%}$ and $w$ parameters while their averages generally move closer to one, especially for $\log {\rm RMS}$. The open blue circles and open brown triangles show the corresponding result obtained if $\eta (\nu)$ is excluded from the first term in Eq.~\ref{eq:model_L} used to fit the PDS, demonstrating that while including $\eta (\nu)$ improves the results it does not fully solve the issues with the longer cadences. For this reason, we chose to exclude the TESS Cycle 1 and 2 data from our analysis to minimize the impact of possible cadence dependencies on our results. 

\begin{table}
\scriptsize
	\centering
	\caption{Averages and $1\sigma_{\rm std}$ of the normalised $\log \alpha_0$, $\nu_{\rm char}$, $\gamma$, $\log {\rm RMS}$, $\nu_{50\%}$ and $w$ parameters obtained after resampling the 2-min cadence data to 240-sec, 10-min and 30-min. The parameters were normalised by their corresponding 2-min cadence estimates before their averages and standard deviations across all stars and sectors were calculated. Values given in parentheses are the same results obtained after a) excluding $\eta (\nu)$ from Eq.~\ref{eq:model_L}, and b) limiting the sample to stars and sectors with $\nu_{50\%} < 50\ \mu$Hz and $w < 12$.}
	\label{tab:frac_cadence_dep_2min}
	\begin{tabular}{lccc} % four columns, alignment for each
		\hline
		 & {\rm 240 sec} & ${\rm 10 min}$ & ${\rm 30 min}$\\
		\hline
		$\frac{\log\alpha_0}{\log\alpha_0 {\rm [2min]}}$&  $0.999 \pm 0.012$  &  $0.995 \pm 0.029$  & $1.000 \pm 0.044$ \\
                &  ($0.998 \pm 0.012$)$^a$  &  ($0.994 \pm 0.028$)$^a$  & ($0.995 \pm 0.038$)$^a$ \\[1.5ex]
		$\frac{\nu_{\rm char}}{\nu_{\rm char} {\rm [2min]}}$  &  $1.042 \pm 0.212$ & $1.285 \pm 0.87$ & $1.354 \pm 1.103$  \\
                &  ($1.047 \pm 0.212$)$^a$ & ($1.298 \pm 0.87$)$^a$ & ($1.410 \pm 1.113$)$^a$  \\[1.5ex]
		$\frac{\gamma}{\gamma {\rm [2min]}}$ &  $1.007 \pm 0.060$ & $1.056 \pm 0.194$ & $1.052 \pm 0.287$  \\
                &  ($1.010 \pm 0.061$)$^a$ & ($1.067 \pm 0.193$)$^a$ & ($1.101 \pm 0.272$)$^a$  \\[2.5ex]
        $\frac{\log {\rm RMS}}{\log {\rm RMS} {\rm [2min]}}$ & $0.992 \pm 0.014$ & $0.984 \pm 0.029$ &  $0.976 \pm 0.042$ \\
                &   ($0.995 \pm 0.008$)$^b$ &   ($0.991 \pm 0.015$)$^b$ & ($0.987 \pm 0.024$)$^b$\\[1.5ex]
        $\frac{\nu_{50\%}}{\nu_{50\%} {\rm [2min]}}$ &  $1.002 \pm 0.055$ & $1.006 \pm 0.054$ &  \dots \\
                &   ($0.996 \pm 0.031$)$^b$ &   ($1.000 \pm 0.043$)$^b$ &  \dots \\[1.5ex]
        $\frac{w}{w {\rm [2min]}}$ & $0.999 \pm 0.041$  & $0.998 \pm 0.121$ & \dots  \\
                &   ($1.002 \pm 0.031$)$^b$ & ($0.994 \pm 0.086$)$^b$ & \dots\\[1.ex]
		\hline
	\end{tabular}
\end{table}

For their sample of WR stars, \cite{Lenoir-Craig2022} found the amplitudes and slopes to be most affected by the change in observing cadence, with $\alpha_0$ ($\gamma$) showing a negative (positive) trend and larger standard deviations for longer observing cadences (cf. their Fig.~5). The same albeit smaller trends are observed for our sample of OB stars. However, in contrast to our results they find $\nu_{\rm char}$ to be least affected, while in our case it is the normalised parameter that changes the most with its average value becoming almost twice as large for the 30-min cadence sampling. We attribute this in part to be due to the WR stars having a smaller range in observed $\nu_{\rm char} \in [1.2, 36.6]\ \mu$Hz. As shown in Fig.~\ref{fig:cad_lorentzian}, it is above $\approx 30\ \mu$Hz that the largest deviations in $\nu_{\rm char}$ between the observing cadences occur, with our considered full sample of OB stars reaching $\nu_{\rm char}$ values more than twice as high as the WR stars studied by \cite{Lenoir-Craig2022}. 

Finally, we find again that both $\nu_{50\%}$ and $w$ are much less affected by the change in observing cadence than $\nu_{\rm char}$ and $\gamma$ also when the normalised parameters are considered.

%%%%%%%%%%%%%%%%%%%%%%%%%%%%%%%%%%%%%%%%%%%%%%%%%%

% Don't change these lines
\bsp	% typesetting comment
\label{lastpage}
\end{document}